# Insights into the effects of indicators on knowledge production in Astronomy

## Pilot Project Report for PhD Project

Julia Heuritsch

Conducted and written between Autumn 2016 and 31$^{st}$ August 2017.

julia.heuritsch@gmail.com
Centre for Science and Technologies Studies (CWTS). Faculty of Social and Behavioural Sciences, Leiden University, P.O. Box 905, Leiden, 2300 AX (The Netherlands)

## Abstract

Nine interviews were conducted with astronomers from Leiden University, and a document analysis was performed on relevant institutional (self-) evaluation documents, annual reports, and CVs of the interviewees. The aim was to perform a qualitative study about the relationship between the research behaviour of astronomers and how their science is being evaluated. This study encompassed the funding and publication system as well as the indicators used to measure the scientific output, its quality and the research performance. This report sheds light on how astronomers define high-quality research and how they think that creating knowledge of value is encouraged or hampered by the evaluation processes. We found that astronomers are realists who define scientific quality on the basis of "truth" and are driven by curiosity. These two factors make up their intrinsic values and motivation to perform Astronomy. Publication pressure, arising from the requirements of "the system", creates an extrinsic motivation to perform. This results in premature publications, low readability and replicability, risk aversion and a focus on quantity rather than quality. Hence, indicators do not merely represent quality, but also co-constitute what counts as good research. While we observe such constitutive effects of indicator use on research behaviour and content, we do not see that the astronomer's intrinsic values are co-constituted. This gives rise to a discrepancy between what is being measured by indicators and what astronomers define as scientific quality; the so-called "evaluation gap". Findings on constitutive effects and the evaluation gap in Astronomy lays out the conceptual groundwork for further empirical research and for policy advice on alternative evaluation practices and innovative indicators with the aim of bridging the "evaluation gap".

# *Contents*

## *Introduction*

This project is about guaranteeing scientific quality. Researchers, the government and the taxpayers whose taxes fund research want to be able to produce and access high quality research. Several developments, such as the growth of the scientific system and the difficulty to monitor and reproduce all research, have led to the use of indicators of research (e.g. Abramo & D'Angelo, 2011). While indicators have become sophisticated proxies of quality, by definition they are "just" that – proxies. Yet, they are often confused to be representative and because "scientific quality is hard to define, and numbers are easy to look at" (Benedictus & Miedema, 2016) funding and career decisions are made based on those indicators. If funding and job opportunities are dependent on indicators it is only natural that researchers respond to their use in that they want to achieve what the indicators are measuring. This may pose a problem for maintaining research quality, precisely because indicators serve as proxies and not as an accurate representation of quality.

These developments are the reason why the question how to measure and ensure high-quality knowledge production has become a controversial and challenging topic in the sociology of science. Indicators are not merely representative measures of scientific quality. Instead, it has been argued that indicators are political means, as they define categories that are collectively significant in our society and thus come with constitutive effects (Peter Dahler-Larsen, 2014). "Constitutive" here means they rather *shape* than represent what is considered to have value in knowledge production (Verran, 2001) and therefore have an effect of research behaviour and content. Evaluation criteria in many scientific fields include publication and citation rates to assess the quality of research. According to sociologists of science, this gives rise to the "evaluation gap", a term coined by Wouters (2017) to acknowledge a discrepancy of what is being measured by indicators and the quality of the scientific content, as perceived by the researchers of the field. But because researchers respond to indicator use, they may be encouraged to focus on quantity rather than quality (Weingart, 2005). Goal displacement, gaming, information overload, questionable authorship practices, unhealthy competition and aversion to risky/ innovative projects are only a few examples of possible consequences (Rushforth & De Rijcke, 2015; Laudel & Gläser, 2014).

While research has been performed on the role of indicators in evaluating research and publication patterns (a rich literature review can be found in Fochler & De Rijcke, 2017), we lack empirical knowledge about constitutive effects of research assessment, and the interactions between evaluation criteria and the content of academic research (Kaltenbrunner & De Rijcke, 2016). We do not know how the gap between what astronomers see as research quality and what is measured by indicators shapes knowledge production and day-to-day research practices. If we want to ensure research quality, we must first get insights from this uncharted territory.

But why Astronomy? Astronomy (synonymous with Astrophysics) is dedicated to basic research, strongly instrumentalized, involving large collaborations, and the use of (open) archives and huge datasets. Astronomy asks highly fundamental questions, which inspire both scientists and the public at large. Yet, advancement in space research is not only dependent on curious minds, but also on the conditions under which those minds need to work. To which extent those conditions are shaped by performance indicators is subject of this study. The Dutch Astronomy community is fairly large (365 FTE researchers; NOVA, 2016) with four institutes in Amsterdam, Groningen, Leiden and Nijmegen, which together form NOVA (Nederlandse Onderzoekschool voor Astronomie). 1998 was a pivotal year for Dutch Astronomy, when NOVA was rated the top research school by the Netherlands Organization for Research (NWO) and as a result has been receiving funding from the Dutch Ministry of Education ever since. This put Dutch Astronomy into the position where it can actively shape the construction of world-class astronomical instrumentation facilities of the European Southern Observatory (ESO). A close collaboration between NOVA and the two other Dutch Astronomy institutes, SRON for space research and ASTRON for Radio-Astronomy, fosters research at the forefront of Astronomy. Today, Astronomy is a top-science in the Netherlands, where research output meets or even exceeds international standards (e.g. the publication and citation rate are a factor of 2-3 above world average; NOVA self-assessment$^{2010\text{-}2015}$). Hence we reside in a particularly interesting location here at Leiden University to study evaluation effects in Astronomy. Considering the Matthew effect that occurs alongside with success, the focus on basic research, and its trends towards open science, big data and big collaborations, Astronomy is a very interesting field to study the relationship between knowledge creation and the evaluation system.

This report is the result of a pilot-project for a PhD study to explore these constitutive effects in the field of Astronomy. Insights are gained through a combination of secondary and primary research. The secondary research segment provides a qualitative assessment based on the analysis of documents such as CVs, annual reports and (self-) evaluation reports. The latter are particularly interesting as they compare the Leiden Astronomy institute (Leiden Observatory; Sterrewacht) with their national and international counterparts. The primary research segments provide insights based on interviews with (international) astronomers at different career stages from the Sterrewacht. Future research on this topic is proposed and will involve surveys, document analyses, ethnographic research and interviews with international astronomers.

The aim of this study is to contribute to a better understanding of the constitutive effects of the current science evaluation system on research agendas and knowledge production in Astronomy. Studying how astronomers experience the evaluation gap and how it affects the quality of their work, will provide us with valuable insights in how to allow for scientific quality. This includes providing a basis for policy advice to ensure the quality of the knowledge produced in the field of Astronomy. Hence, astronomers, the public and the government will benefit from these results. Depending on the extent of the found evaluation gap, a next step could be an investigation of alternative evaluation practices (Duffy, 2017) and innovative indicators ("re-configuring evaluation"; Fochler & De Rijcke, 2017) which utilize their constitutive effects to bridge the evaluation gap and enforce scientific quality, diversity and societal impact.

The report is structured as follows: The method section is followed by six chapters. *Chapter 1*

provides background information on the Sterrewacht's policies, missions and evaluation practices. It is based on a analysis of annual reports and evaluation documents. *Chapter 2* to *4* are based on the interviews. *Chapter 2* investigates what astronomers value in their work versus what funding, publication and career system value. Chapter 3 studies epistemic restrictions in Astronomy. *Chapter 4* analyses what effects the discrepancy in values has on research behaviour and knowledge production. *Chapter 5* reflects on the results and describes the evaluation gap in Astronomy. *Chapter 6* puts all findings together and draws conclusions on what constitutive effects indicator use have in Astronomy. The six chapters are followed by a discussion, policy implications of the findings and a suggestion for future studies.

## *Methods*

Our research consists of interviews and a document analysis. Semi-structured interviews were conducted with (international) researchers (4 faculty members, 2 postdocs, 1 PhD) and 2 Master students from the Sterrewacht. The Master programme at the Sterrewacht is very research intensive, requiring the student to write two Master theses in total, which is the reason why they are also interesting subjects for this study. In order to shed light on the evaluation gap, questions were developed such that an astronomer's definition of quality versus what is measured by indicators can be studied. Topics include career steps, project funding, exposure to assessments, research evaluation, the publication and funding system, different stages of the knowledge production process – from planning, via doing the research to publishing – and the meaning of quality. Each topic was introduced by one overarching question, followed by several potential follow-up questions. The interview questions can be found in the Table **S-1** of the supplementary material.

The participating researchers were invited via email and all names are anonymized. All interviews, 80-100 minutes in length, were fully transcribed into electronic form, summarised and coded (codes can be found in Table **S-2**) according to Grounded Theory. These codes represent themes which emerged by combining sensitivity towards existing literature on constitutive effects of indicator use with insights from our data. *Chapter 2 to 4* are structured according to those themes are the result of the qualitative interview analysis.

The interview data were complemented with a document analysis of materials collected online or made available via our informants, including CVs of the interviewed researchers, annual reports (1998 to 2015), (self-) evaluation reports of the Dutch Astronomy institutes and their umbrella organisation NOVA. *Chapter 1* is dedicated to this document analysis.

In *Chapters 5 & 6* we analyse the evaluation gap and constitutive effects based on our findings. We draw conclusions as to what extent indicator use shapes knowledge production, research behaviour and research agendas in astronomy.

# *Chapter 1: Background Information*

In order to understand the circumstances under which science is performed at the Sterrewacht, one must not only understand the funding and publication system, but also the history of the institute, how funding is allocated, how evaluations are performed and institutional rules and guidelines for the Master & PhD programme and staff appointments. This chapter on background information on Leiden Observatory is based on annual public reports from 1998 to 2015 and authored by the director of the institute (Annual report[1998]- Annual report[2015]), institute evaluation protocols (evaluation period 2010-2015; Evaluation protocol[2010-2015]) authored by an external committee and self-assessment protocols (same period; NOVA self-assessment[2010-2015] & LU self-assessment[2010-2015]) which were written by NOVA and the institute as a preparation for the evaluation.

## *1.1 History of the Sterrewacht – Maintaining its success*

It is apparent from all documents that the Sterrewacht is proud of its success and remaining its reputation of being a world-leading institute for Astronomy is important. "*During the reporting period 2010-2015 Leiden Observatory thrived; its scientific production, measured in terms of number of papers, citations, PhD candidates and postdocs and the amount of grant money awarded, has never been so large*" (LU self-assessment[2010-2015]). Leiden Observatory is one of the largest and top astronomical research institutes in Europe. The Sterrewacht is ranked among the best Astronomy institutes in the world (Evaluation protocol[2010-2015]). In order to give a brief insight into the history of this success, let us start with the history of the institute.

The Sterrewacht was founded in 1633 and is now the oldest operating university observatory in the world. It has a long list of eminent astronomers. Prof. Tim de Zeeuw is currently the Director General of the European Southern Observatory – the largest observatory in the world, and Prof. Ewine van Dishoeck is the president elect of the International Astronomical Union (2018-2021).

In the "pivotal year" 1998 the national Astronomy proposal "Astrophysics: unravelling the history of the universe" was rated first by the Netherlands Organization for Research (NWO). This proposal was submitted under the umbrella of the Nerderlands Onderzoekschool voor Astrononomie (NOVA). It is the alliance of the four university Astronomy institutes in the Netherlands – the Universities of Amsterdam, Groningen, Leiden and the Radboud University Nijmegen and was rated as top research school in 1998. As a result of the proposal NOVA was guaranteed baseline funding from the Dutch Ministry for Education for 1999 to 2005. Since then this "NOVA grant" has been renewed every 5 years. The grant has been the basis for support of "normal" research activities and the participation in numerous programmes for the construction of astronomical instrumentation. This enabled Leiden to build on its long tradition of radio-interferometry, by getting heavily involved with instrumentation for European Southern Observatory's (ESO) facilities, securing priority access for conducting observations. Additionally Leiden hosts the world famous Sackler Laboratory that bridges Astronomy, physics, chemistry and biology.

The Observatory is an international environment; many students, postdocs and staff come from abroad. The institute has close collaboration ties with other Astronomy institutes in Europe and the U.S. and hosts visitors from across the globe.

Since 1999 an *advisory board* has overseen each of the 9 institutes within the Faculty of Mathematics and Natural Sciences, including the Sterrewacht. This board consists of members with experience in science management and science policy in Astronomy and adjacent fields. The board meets twice a year to advise the director on strategic and funding issues.

## *1.2 Funding*

Leiden Observatory experienced an increase in budget from 8.3 Meuro in 2010 to 12.5 Meuro in 2015. Parts of this budget comes from base-funding from Leiden University (ranging from 3.6 to 3.9 Meuro) and from NOVA (ranging from 1.7 to 1.1 Meuro). The Sterrewacht attributes the budget increase to the success of staff members to secure funding from European and national sources in form of external grants. The main funding agencies include the NWO and the EU European Research Council (ERC). The observatory reported already in 2009 (Annual report$^{2009}$) that "*university funding is changing as a result of external pressures. There is more and more emphasis on temporary, project-based funding, threatening the structural long-term funding that is needed as the basis of a healthy scientific institute. Keeping up our success in funding applications is therefore vital.*" This will become especially true during the next years when the continuation of the NOVA grant is running out in 2023 and NOVA needs to find a different source of funding. The Evaluation Board (EB, hereafter; Evaluation protocol$^{2010-2015}$) emphasises that a lack of funding and the prospect of the opening of too little faculty positions are opening up, will perpetuate the current gender and age imbalance in the institute. Most (90%) staff members of all NOVA universities are above 40 years old. With limited funding for the future, the increasing amount of PhDs and Postdocs (83% and 50%, respectively, between 2010 and 2015) will have low chances of getting a tenured position, which could cause intellectual stagnation.

## *1.3 The Sterrewacht's missions*

NOVA's objective is to "*ensure a front-line role in the next generation of astronomical discoveries*". NOVA intends to maintain this objective by its mission, which is to "*carry-out front-line astronomical research, to train young astronomers at the highest international levels, and to share discoveries with society.*"

Leiden Observatory's three missions are well-aligned with NOVA's:

### 1.3.1 Education & Staff appointments

Leiden Observatory offers bachelor and master programmes in Astronomy of 3 and 2 years, respectively. All staff members are involved in BSc and MSc education, which includes supervision of projects and teaching. The education programme was not part of the latest institute evaluation, but the Sterrewacht reports that it was successfully reviewed in 2013 with particularly positive comments on the close links between education and research benefiting the staff as well as the students. Involving BSc and MSc students in research is one of the Sterrewacht's key principles in teaching.

The "Director of Education", an appointed staff member, is formally responsible for the entire teaching programme of the Sterrewacht and a "Committee for Education", which consists of staff and students, provide regular evaluation. An "Exam committee" is responsible for the (quality) of the curriculum and the exams and advises the Director of Education.

The Sterrewacht's educational mission is to "*to provide excellent education at the bachelor and master level, not only to prepare students for PhD projects, but also for the general job market.*" NOVA and its institutes see this as "human capital development" as according to the NOVA self-assessment$^{2010-2015}$, Astronomy-trained students are attractive to a wide range of sectors in society, from ICT companies to consultancy firms, industrial and environmental labs, and the public sector (teaching, governments). That is because they are trained to handle problems with a high degree of complexity with state-of-the-art instrumentation and software in large national and international teams.

**(i) Master programme**

In 1996 the former 4 year curriculum was changed to a 5 year one with equal content. A gentlemen's agreement between the Dutch Universities (VSNU) and the Ministry was reached in 1998 ensured student allowances for that 5th year, under several conditions: an increase in the pass rate of first years students, the recruitment of more female students and a division of the curriculum of the last year into 3 specialities. In 1998, the answer to those conditions were 3 tracks a Master student could pick from: research in Astronomy, management (today: Science-Based Business) and the educational speciality. A further division into an instrumentation and cosmology track followed later and a track in Astronomy and data science is currently being set up.

The MSc programme is taught fully in English, attracting also many foreign students. In 2015 53 Master students enrolled, including 11 (20%) women and 18 (33%) of foreign nationality. The Sterrewacht reports a dramatic increase in number of students following the MSc program.

Admission to the Master for students without a BSc in Astronomy from a Dutch university requires a recommendation by the "Toelatingscommissie" (admissions committee). Every Master student has their own tailored study plan. A dedicated "Master student Advisor", also an appointed staff member, advises the students on study habits and the choice of curriculum on an individual basis.

**(ii) PhD programme & Postdocs**

The selection of new PhD candidates is coordinated by pooling all project grants and announcing the open positions in one common call. Competition usually is very strong (in 2010, there were over 300 applications for about 15 positions). The final selection is done after a 2-day visit of the 30 top candidates, where potential supervisors and students can see each other in action and rate each other. At the end of 2015, Leiden Observatory consisted of 67.6 FTE PhD candidates, compared to 46.2 FTE Postdocs.

It is NOVA's policy that each PhD candidate has 2 advisors; one from their own institute and ideally the other one from another NOVA university. This can be "extremely helpful in rectifying potential inequities in student support/treatment in their home institutions" according to the EB. However the committee also criticizes that Leiden does not give sufficient importance to that policy. The EB sees the risk of the program not properly serving PhD candidates.

According to the EB, the PhD Program at the Sterrewacht is presumably the largest world-class PhD program in Astrophysics across the globe. NOVA's total retention rate of PhD candidates is over 90%, where nearly all of those graduate within a reasonable time-frame. The requirement for finishing a PhD is 3-4 first-authored paper in a refereed journal; one paper per year. However, by the end of the program the graduates have not only "published a record of innovative research results", but also have acquired strong research and career skills. These include crafting CVs for different audiences and networking in more industry-focused environments. EB recognises that with that NOVA responds to the decreasing likelihood of a tenure track appointment in academia. During the reporting period 2010-2015 54% of the graduates continued as a Postdoc and about 25% with a job in the big data/ software industry. Other jobs outside academia encompass a large range: banks, insurance companies, consultancy firms, and physics teachers.

The EB reports that PhD candidates appreciate the camaraderie that NOVA provides for them in joint conferences and PhD satisfaction is overall much higher than encountered outside the Netherlands.

The EB concludes that the NOVA-led PhD program is one of the "highest quality in all respects and could well provide a model for the rest of the world." This quality is recognised in elite institutions worldwide, inside and outside academia.

However, the EB stated that it is widely and internationally acknowledged that postdocs are frequently not as well-supported academically, personally, or in terms of career development. Hence, the committee suggests the introduction of a similar program for postdocs.

**(iii) Hiring Strategy**

Leiden University appoints staff members on the basis of an US-style tenure-track system. A tenured position can be granted after 5 years of research based on research, teaching, management, and grant-acquisition performance. After a similar period one can be promoted to full professor. According to the LU self-assessment$^{2010-2015}$, tenure-track staff receive guidance, mentoring, and teacher-training. All staff (from PhD to the head of the institute) are reviewed annually by their supervisors where they set personal development goals for the next year together. At the end of 2015, the Sterrewacht comprised 19.3 FTE faculty, 7.3 FTE scientists working on instrument development and 5 active emeritus faculty.

According to the EB, the Observatory's hiring strategy has contributed significantly to its strength. The observatory identifies "targets of opportunity" – candidates of very high scientific standing who may be happy to join the department. At the same time, the Observatory carries out broader searches, putting emphasis on recruiting the best scientists who come forward. However, the EB emphasises that there is an element of risk in this strategy. Because of a shorter publication record, younger candidates may look less attractive. Both, younger candidates and women are less likely to apply if they are not encouraged personally, as the age and gender profile of the Sterrewacht suggests.

*1.3.2  Research at the forefront of modern Astronomy*

As part of NOVA's **research strategy**, education and research focus on three major themes – the "NOVA network science programmes":

- Network 1: Origin and evolution of galaxies from high redshift to the present
- Network 2: Formation and evolution of stars and planetary systems
- Network 3: Astrophysics in extreme conditions

These address major questions in modern Astronomy. Each network has developed a roadmap for the coming decade, including strong cross-network collaborations. The Sterrewacht claims to carry out diverse research, that includes observing programs with access to a large arsenal of telescopes on the ground and in orbit, laboratory Astrophysics, data analysis and interpretation and purely theoretical work.

Research is closely interlinked with technology and hence the observatory's **Research & Technology strategy** consists of 3 pillars: observations, theoretical and astrochemical modelling, and development of key technologies.

Observations play a central role in astronomical research, and as they are performed with instruments, they comprise a big fraction of those key-technologies, and hence, NOVA has an **instrumentation strategy**: NOVA astronomers are actively involved in building instruments. A close collaboration between NOVA and the two other Dutch Astronomy institutes, SRON for space research and ASTRON for Radio-Astronomy are crucial for this development of key-technologies that will enable future astronomical discoveries. Further collaborations with Dutch partners such as TNO Delft, Dutch Space and the Sterrewacht's vicinity to ESA's ESTEC (Technical facility of the European Space Agency) foster those activities. NOVA's instrumentation programme is focused on, because ultimately, most instruments are built in international consortia under the umbrella of ESO or ESA. Those collaborations mean that Leiden astronomers play important roles in the development and operation of national and international instruments. This brings access and guaranteed observation timeslots to Dutch astronomers, and hence NOVA proudly reports that their "*astronomers are among the first to use the instrument, thus reaping the hottest early science harvest.*"

The EB acknowledges that "*these project partnerships place Dutch researchers in a position to take optimal advantage of the scientific exploitation of the resulting facilities.*"

*1.3.3   Outreach & Societal relevance*

Leiden Observatory's astronomers feel privileged to be able to spend their time following their passion of doing research, but they recognise that there is increasing pressure on such curiosity-driven sciences to demonstrate societal relevance. NOVA allocates funding for that, and the Sterrewacht has an Outreach Committee dedicated to ensure the societal relevance according to NOVA's *outreach strategy*. This consists of 3 main pillars: (i) outreach and education (of the public); (ii) pushing technology boundaries and spin-offs with industry, and (iii) human capital development.

**(i) Outreach and education (of the public)**

As "Astronomy has a strong appeal to the general public" (LU self-assessment[2010-2015]), all staff and students "*spend considerable time and effort to explain the exciting results of Astronomy to the general public, in the form of lectures, press releases and newspaper articles, courses, public days and tours at the old observatory complex, and input to television and radio programs.*" These efforts are very successful and play an important role in fostering enthusiasm in school children, not only for Astronomy, but also for science in general. Maintaining high visibility in the media and having a wider participation of the general public (e.g. via citizen science projects), is important for the Sterrewacht in order to "give something back to society".

Additionally, Leiden Observatory hosts the biggest Outreach program for Astronomy world-wide, the UNAWE (Universe Awareness) program. Since its start in 2004, UNAWE has grown to a thriving network of more than 150 UNAWE volunteers and experts active in 17 partner countries worldwide. UNAWE's goal is to "*broaden the mind and awaken curiosity in science, at a formative age when the value system of children is developing by raising awareness about the scale and beauty of the universe and stimulating their development into curious, tolerant and internationally minded adults.*"

For that purpose, UNAWE runs workshops, educational programs and designs educational material.

Another example for the observatory's outreach efforts is "Contact.VWO", which is a liaison between pre-university high schools and Leiden's Departments of Astronomy and Physics. It supports both teachers and their students with various activities.

**(ii) Pushing technology boundaries and spin-offs with industry**

As outlined in *Chapter 1.3.2*, collaboration with technical institutes and industry is vital to exchange technical knowledge and to create industrial spinoff projects. Participation in instrumentation projects brings more benefits, that enhancing the effectiveness of the research program. It creates an expert team of instrument scientists and engineers that complements the expertise already resident within ASTRON and SRON and strengthens links with industry.

**(iii) Human capital development**

As outlined in *Chapter 1.3.1*, Astronomy-trained students are attractive to a wide range of sectors in society, because they are well-known for being able to solve problems based on incomplete and poorly-controlled data. In that sense, the observatory gives something else back to society than research results and technological spin-offs: human capital.

*1.4 Institute Evaluation Practices*

*1.4.1 Evaluation according to the Standard Evaluation Protocol (SEP)*

The Evaluation Board (EB) does not only assess Leiden Observatory, but at the same time also NOVA and the other three Dutch Astronomy institutes that belong to NOVA. The committee's review is part of the assessment system for all publicly funded Dutch research organizations, according to the SEP. The SEP consists of three criteria: (i) Research quality, (ii) Societal relevance, and (iii) Viability.  The committee judges the unit's performance against the SEP assessment criteria, whereby current international trends in science as well as society should be taken into account. The scope of the assessment is set by the Terms of Reference (ToR), which in this case is the information provided by the self-assessment documents of the individual institutes and NOVA as a whole. These documents are a description of the institute's mission, objectives and results. In addition, the EB conducts interviews with management, the research leaders, staff members, PhD program management and PhD candidates. The assessment task does not only include evaluation as such, but also recommendations for possible improvements and a reflection on the research integrity of each institute. Some of the EB's assessments and recommendations we have already encountered in previous sections, however in this section we will elaborate on how the Sterrewacht assesses its performance according to SEP criteria itself as compared to the EB's judgement:

**(i) SEP Criterion 1: Research quality (the level of the research conducted)**

Both, the EB and Leiden Observatory rate its scientific output in terms of quantity, quality, impact, and innovation of research as exemplary. It is a world-class institute which is reflected by the number of high-impact, refereed publications (376 in 2015) across widely distributed fields. Further evidence for this are citation statistics, personal prizes & awards, prestigious fellowships and staff positions at excellent universities obtained PhD candidates and postdocs. The EB also praises the strong synergy between research and instrumentation. The only recommendation would be securing more long-term funding instead of being too dependent on external grants. That would enable the department to mitigate its age imbalance.

Leiden Observatory receives the score 1.0 (on the SEP scale, ranging from 1 to 4).

**(ii) SEP Criterion 2: Societal relevance (social, economic and cultural relevance of the research)**

Both, outreach and interactions with industry was considered by the EB, which is in line with the several ways of outreach efforts that the LU self-assessment$^{2010\text{-}2015}$ outlined:

1. The EB recognises that the Sterrewacht's *outreach program* has a set of well-defined target groups and provide an impressive personnel pool for its "ambitious program". Measured by the large audiences for all of its activities, the Sterrewacht claims that their national and international outreach efforts "*have increased not only the awareness of the public globally of the beauty of the universe, but are also an important reason why the number of students in our Astronomy programme has increased so dramatically*". According to the EB, the UNAWE program "*has matured into one of the most important educational outreach activities in Astronomy with scope to act on a worldwide level*".

2. As described in *Chapter 1.3.1*, Leiden Observatory considers its *training programmes* of PhDs and postdocs as world-class, which was confirmed by the EB. That way the Sterrewacht claims that "*we are providing the Netherlands with well-trained people. Thanks to their background in problem solving, mathematical, physical and statistical modelling, handing big data sets, end to end understanding of complex (observing) systems, computer skills, international collaboration and team work our students are in high demand in the wider job market. The rapid emergence of 'big data' embraced by many companies is making our students even more employable than before.*"

3. Leiden Observatory claims that Astronomy constantly pushes technological development, because it is driven by new observational facilities. In agreement with that, the EB finds the Sterrewacht's "*valorisation activities* in collaboration with industry and the subsequent *knowledge transfer* outstanding". However, the EB criticizes that

"*valorisation seems to be opportunity driven, rather than to derive from pre-determined strategy.*"

The EB evaluates the observatory's attention to "Relevance in Society" with a score of 1.5.

**(iii) SEP criterion 3: Viability (strategy, governance and leadership)**

The EB found that Leiden Observatory is in a very strong and robust position and that, for a faculty of its size and prominence, its cohesion and team spirit are exemplary. The Sterrewacht believes that its success in winning international research funding demonstrates the high profile of its staff. The EB agrees to that, but points out that the considerable growth those grants allowed for, can also bear the risk of being dependent on external funding in the future, if Leiden Observatory cannot ensure more base-funding from the university. The EB reminds the Sterrewacht at this point about its worrying gender and age profile and that taking this risk would make future tenure track positions even more sparse as compared to temporary appointments. The committee also emphasises that this growth will require a serious increase in supervisory efforts, as for example pointed out in *section (ii) of Chapter 1.3.1*. While the EB does not include those points in the judgement of this SEP criterion, Leiden Observatory additionally mentions their a couple of points, that in their view adds to their viability. First, their strong outreach programme attracts the attention of the public and hence provides visibility, which is important for future funding. Second, their strong educational programme trains PhDs to acquire unique skills to be highly demanded on the job market, at prestigious universities or outside academia. Third, their involvement with many new revolutionary observational facilities provides them with the opportunity to continue to do important fundamental research. Fourth, collaborations and interdisciplinary research, crossing the boundaries of their own NOVA networks, ensures qualitative and ground-breaking research.

The EB scores the viability of Leiden Observatory with 1.5.

**(iv) Research integrity**

NOVA considers research integrity of highest importance, which is why PhD candidates at all four universities attend mandatory lectures on scientific integrity. The universities also have data management protocols. By standard, all raw telescope data in Astronomy become public after one year through the archives of the observatories or through the Centre Donnees Stellaire in Strasbourg, where researchers can publish their data when the paper has been accepted. Each university, which includes NOVA, follows the Code of Conduct of the Association of Dutch Universities (VSNU). All universities have more specific codes of conduct for sexual harassment, aggression, violence and/or discrimination. The EB compliments the Sterrewacht's Integrity policy and states that "*all research activities appear to live up to current international standards of research integrity and transparency.*"

*1.4.2 Performance indicators*

In addition to the self-assessments prior institutional evaluations and those evaluations, Leiden Observatory measures its scientific productivity with certain "performance indicators". The Sterrewacht calls them "objective" as they are quantitative and they include:

- **Publications:** The Sterrewacht reports a steady increase of the total number of publications from 288 refereed papers in 2010 to 376 in 2015.
- **Citation rates:** In a citation study, 24 citation parameters (e.g. number of citations, number of normalised citations, number of normalised first author citations) were considered. It shows that the citation statistics for Leiden Observatory are comparable with, or even significantly better than, the average of the 10 top institutes in the USA and the 4 top institutes Europe.
- **PhD theses:** 81 PhD theses were produced between 2010 and 2015. Out of the 44 (54%) of PhD graduates that continues as a Postdoc, 7 received prestigious fellowships.
- **External grants and prizes:** 39 Meuro in grant money was awarded during the reporting period. The Sterrewacht recognises that "the number of honours, prizes and others forms of significant recognition by peers totals 19".
- **Instrumentation & outreach activities:** The performance of its outreach programme is measured by the large numbers of press releases, articles, attendees, teachers and children reached through its various activities.
- **International leadership:** Leiden astronomers are highly visible internationally. Examples are the current ESO Director General (de Zeeuw) and the former ALMA Director (de Graauw). For the International Astronomical Union, Vice President Miley was the architect of its strategic plan and Van Dishoeck takes over as president in 2018. Top-level international committees chaired by Dutch astronomers. Leiden astronomers have organized numerous international conferences inside and outside the Netherlands.
- **Instrumentation programme:** A key indicator of the success here is the on-time, on-budget and within specification delivery of instrumentation (co-)built by NOVA. Another positive measure is the frequent invitations for international collaborations. Many successful spin-off projects can be listed.

In addition to the performance indicators, "excellence" is a rising buzzword to measure the success of institutes and researchers (Sørensen et al., 2015). In the following section we will describe what excellence means to the Sterrewacht, how the EB judges the observatory's excellence and what measures the Sterrewacht takes to remain "excellent".

*1.4.3 Excellence*

The Sterrewacht "*believes that their success in winning international research funding demonstrates that their staff is of* high calibre *and has the drive and commitment to continue excelling. […] Their staff and the faculty board agree that excellence will be the most* *important hiring criterion.*" For that purpose, the Sterrewacht introduces the JH Oort Scholarship to attract excellent students for its Master programme. As described under *section (iii) of Chapter 1.3.1*, the observatory identifies 'targets of opportunity', which are candidates of very high scientific standing, to hire them.

Next to hiring and training excellent researchers, the Sterrewacht has been part of the European Association of Research in Astronomy (EARA) since its inception. It is a vehicle to stimulate collaboration between excellent astronomical institutes: Cambridge University, Institut d' Astrophysique (Paris), Max Planck Institut für Astrophysik in Garching, Instituto de Astrofisica de Canaris and Leiden University.

The Sterrewacht commits itself to all the strategies described so far, because the institute's key goals is to "*maintain the present high level of achievement and to continue to score very well in international competitions for observing time at space observatories and on the ground, as well as for research grants.*" The EB certifies that the Sterrewacht's research and education program are excellent. The institute is one of the leading Astronomy departments in the world and so are the Sterrewacht's graduates. Leiden Observatory scores best for the SEP criteria in the EB's evaluation of NOVA and its 4 universities.

*1.5 The (hidden) meaning of evaluation practices*

We have elaborated on Leiden University's various strategies on how to maintain its "success". We have discovered how that success is evaluated and measured, both in qualitative and in quantitative terms. What we are missing from the reports is an answer to the question who defines quality and if the described measures can satisfy that definition. NOVA claims "*the first part of its strategy [to ensure a front-line role in Astronomy] is to foster an intellectually rich and vibrant scientific atmosphere which allows astronomers to pursue their ideas and push scientific boundaries, and in which young scientists can develop and grow*." This sounds great in theory, but we question, whether individual researchers feel that "success" as defined in the protocols actually allows such a "vibrant scientific atmosphere" and out-of-the-box thinking in practice. That is why we interviewed 9 astronomers from the institute and the remainder of this report is dedicated to an analysis of what evaluation practices and quality measurements really mean for the researchers and their work.

## Chapter 2: An investigation of values

### 2.1 Motivation for becoming an astronomer
### (– The "Aha-Erlebnis")

In order to understand whether indicators shape knowledge production, the first step is to understand the motivation for becoming an astronomer.

*Curiosity* was everybody's answer to the question what drives the interviewees in their research and/or what their initial motivation to become a scientist was. It is the "curiosity for the things we can learn in our field" (PhD Candidate), the most fundamental questions about our universe, which all of the interviewees have in common.

> Faculty Member 1: "Science for me is curiosity driven. I think when your curiosity is awaken … So I think the most magical moment for me was when I first heard about that the universe was expanding and I had no idea what that really meant, but I thought it was totally mysterious and magical and of course [asked myself] all kinds of questions like 'What is it expanding into?'"

Being driven by curiosity means wanting to "know and to understand better" (Postdoc 1). A "passion for knowledge" that answer those questions is found to be synonymous to curiosity.

> Faculty Member 2: "What drives me? I enjoy it. I just get up every day and get to go to work and do my hobby, so …That is just really the *passion for the knowledge* and I just really … I have fun working with images and working with data."

> PhD Candidate: "I think I really like the idea of always trying to learn more and *push your knowledge*."

> Master Student 1: "Well the fact that you can win prizes, was I don't think a big motivation. Because I just wanted to – yeah I think to *understand more* about the philosophical groundwork that underlies physics and Astronomy."

> Faculty Member 3: "Then I have always been a scientist, in the sense that I always wanted to *know how things work*."

Curiosity expresses itself through a variety of other forms. One of which is the love for ***"Puzzle solving"*** (Stephan, 2012). We found that interviewees frequently mention the love for maths and the pleasure of working with methods. Astronomers enjoy the process of discovery.

> Faculty Member 3: "I mean it's the journey and not the arrival, basically. […] It's just simply that it feels good. And in German they have a word for that, they call it the 'Aha-Erlebnis', the 'Oh, is that so'-feeling."

> Master Student 2: "I have an interest in what's going on, but I have […] an interest and an enjoyment, and kind of waiting to see what would […] happen if I did certain things and what results I would get out."

> Master Student 1: "Research now I really like it. And the reason is that you are busy intellectually and you are doing a different thing every day, addressing new problems. You are learning new things."

> Interviewer: "What drives you in your research?"
> Faculty Member 4: "Curiosity. Absolutely. 100%. [As in] 'This is weird, I want to know what's going on'."

The drive to find answers to how the universe works often leads to a passion for working with telescopes. For some astronomers telescopes are not only the instrument to collect data, but also their subject of curiosity and hence might choose projects which include technical work on the telescope.

> Postdoc 1: "I want to know how a telescope works and what's behind the data."

> Faculty Member 1: "You know, pointing the telescope and looking at the sky. That is magical to do that for the first time. So you are dealing with very mundane issues at that moment, but it's all in the service of that problem that you are trying to solve."

Surprising observations or new arising questions may result into a **"hype"** (defined as **"sexy topics"** hereafter) that catches the community's curiosity quickly. The attention that is gathered by sexy topics can be used to attract funding. In this way, hypes can accelerate knowledge production.

> Postdoc 2: "I have had my own curiosity, reading up about some questions. And then I will also realise that, for example, a new question has arisen in the community. Let's say someone has made an observation that is surprising and you want to understand, is this actually true and what are the consequences? So this is more of the hype kind of science 'Oh there is this new result that you want to read upon, because people seem to be interested.' That will also be part of my decision [what to study], so both, my own curiosity, but also, trying to understand how it will be perceived by others."

In Astronomy curiosity is firmly tied to the **notion of truth**; Astronomers are realists who assume a reality, described by physical laws, which is independent of human beings. Objects of scientific knowledge in realist terms mean that "they can be seen as revealing a deeper order [the truth], which is absent in surface manifestations of nature. Scientists tinker to reveal structure, not to impose it. Science is an activity that discovers worlds beneath, or are embedded in our ordinary one" (Sismondo, 2004; p.161). Pushing for knowledge is equivalent for wanting to know the truth and discovering structures of nature.

> Faculty Member 1: "But that moment of – you know – mystery, that is a scientific experience in the sense that there is only one thing that you accept in that moment, that's the truth, you want to know the real answer. And no excuses, only the real answer matters. And that is what drives science, we only want to know the real answer."

The notion of truth and the strive to push knowledge forward, both resulting from the astronomer's curiosity to understand the universe, feeds into the astronomer's definition of quality as we will see in the next chapter.

*2.2 The astronomer's values*

"What is scientific quality" is one of the most pressing and controversial questions in the sociology of science. So far, an explicit theory of quality that is able to capture its multiple meanings and attributes across fields, culture and time has remained elusive. The philosophy of science regards "the scientific method" as the basis for high quality research and has designed a number of normative theories around that assumption, but are not supported by empirical tests. In the meantime other aspects of science have become an important consideration in the question what projects should get funding, such as the relevance for society and potential to result in applications for industry. In order to analyse to what extent indicators relate to scientific quality, we need to investigate what that means to the individual astronomer.

Interviewees were directly asked what high research quality means to them and the answers are in line with their motivation to become an astronomer: Research in Astronomy is of high scientific quality, when it helps them in their endeavour to discover, to ***"understand better"*** (Postdoc 1), to unravel the mysteries of the universe. The research has to "be something new" as in adding to knowledge as opposed to "redundant information". As such it must "push knowledge forward". This ranges from "trying to solve a problem, no matter what the problem is" (Postdoc 1) to "asking an important question" and having the means to solve the problem like Faculty Member 1 describes:

> "Okay, high quality research. I think that's a combination of things. It has to ask an important question. [...] I mean, everybody can see that the nature of dark matter is an important thing. But, you know, solving it is a different thing. [...] So, asking an important question is not enough. I think seeing an interesting avenue to solving that question, that is high quality research. [...] And if you take that important question and combine it with some physical insight, for which you have to know your Astrophysics and turn that into an observational program, that I think is really good research. [...] The key there is the astrophysical knowledge. Because the big questions are obvious, everybody can ask them, but finding a way to the solution is the art, right!? [...] And in present day Astronomy does it provide, not a guarantee, but at least possibilities of making real progress on an important issue? That is different from being innovative."

Two aspects of this answer to the question what scientific quality means to the interviewee stand out. First, the assumption that "the big questions are obvious, everybody can see that the nature of dark matter is an important thing". This brings us back to the ***notion of truth*** about a "reality out there" that astronomers hold in very high esteem. The realist natural scientist assumes a reality independent from the observer, which arises from certain (physical) causal laws. As Astronomy is the study of the universe and its building blocks it, at least from the realist's point of view, seeks to answer the most "fundamental" questions to set a basis for "truth of everything". This is how an astronomer argues the high relevance for society:

> "The inspiration that Astronomy brings and the fundamental questions it raises about the nature of the everything and the place of humanity in the universe, makes it natural for us to engage with fellow intellectuals in seeking connections between arts, humanities, and science." (NOVA self-assessment$^{2010\text{-}2015}$)
>
> Faculty Member 3: "Science that drives the [knowledge] forward, is science that serves society."

The interviewees display consensus about the importance of Astronomy with respect to this mission. However, when asked for a more objective definition of what an "important question" is, the astronomer admits controversy:

> Faculty Member 1: "That is […] difficult to answer, because if you have 5 referees, they will all have different preferences for what is important and what is not important."

Second, the interviewee makes a ***difference between "making progress on an important issue" and "innovation"***. This difference becomes clear when asked about whether the definitions of value and academic quality have changed over time:

> Faculty Member 1: "Well, academic quality I think has always been relatively clear. It has to be verifiable and clear, unbiased etc. I think that is academic quality. But there is these days … a tendency to look at the value of science in terms of economic output, it's called 'valorization'. And I am totally uninterested in that I have to say. It is nice if you can […] use some things… It is always nice if you find applications that are useful and that can actually make you profit even. Why not? But that's not why we do it. And the importance of that is overstated these days. And I don't think that is actually productive."

Here we can see again the high value of "truth" for an astronomer. Truth matters for its own sake. Applications are opportunity driven, but not the goal of the research. Hence scientific quality in the eyes of the individual researcher is independent of its potential to lead to applications for industry. Societal relevance however, in the eyes of an astronomer, arises self-evidently from the relevance of the truths that Astronomy discovers.

The last quote also brings up another aspect of scientific quality as perceived by the researcher: using ***sound scientific methods***.

> Master Student 2: "I guess it's if you followed the methods as best as you can – like to the best ability and take everything into account and thoroughly test your results and outcomes to make sure that they are as concrete and solid as they can be before even

> throwing them out to the general populous … Part of it also is, if you have high quality data, it can be easier to do high quality research, so erm, that too."

> Master Student 1: "So if the conclusion is clear, if it is based on some clear mathematical structure and assumptions and you can do some double checks to see that there are no conflicts."

Asking "important and new questions" that are "useful" in that they add to a better understanding of the universe is the astronomer's first criterion for high quality research. The second one is to then try to answer those questions by robust and careful research. This involves thorough methods, which take ideally all possible factors, assumptions and biases into account and sufficient testing of the methods and results before publishing. High quality research thus, for an astronomer, means to obtain results which resemble "truth" as closely as possible. However, those criteria are not yet enough to satisfy an astronomer's account for high quality research: Conclusions that push knowledge forward must not only describe "reality" but also be "rememberable" and *communicated well*:

> Postdoc 2: "[Quality science] will be a research that is both, careful, so that you know that a sufficient amount of time has been spent on testing the methodology and understanding the biases and limitations of the research. […] And a research that has a broad reach in the field. So that the conclusions you take out of it are rememberable. So you can remember the research in 1 or 2 years from now. And useful."

Interviewees emphasise the importance of a "good writing style" to communicate results well and Faculty Member 4 summarises all three of the astronomer's criteria for high quality research:

> "[*Criterion 1 (Asking important question):*] You have a new science idea. [*Criterion 2 (Sound methodology):*] You have asked the question clearly and well, with a well-defined "Yes, if this works …", "No, if it doesn't". [*Criterion 3 (Clear communication of the conclusions):*] And you have written a paper which demonstrates you have answered that question, one or the other. […] And you have written it in such way that a non-expert in that field can read it and understand what you have done. They may not understand the details, they may not understand the algorithms, but I think high quality research is: You can pick up – a good paper – any Astronomy paper, read the abstract, read the introduction, read the conclusions and know what they did. And why they cared. And you may not know the shear statistics of galaxies of redshifts 2, but good quality research will give you the background and give the context which you should be able to understand. As a scientist you understand it. If it's a crap written paper, then that's crap research – I don't care how brilliant the answer is, if they can't communicate it through a paper or through a presentation, then that's bad research. […] Yeah, I'd say that means high quality. They are able to write and present a compelling scientific argument from start to finish, that any reasonably trained human being can read and think about, you know."

Hence we observe that in Astronomy "the legend" of the superiority of "the scientific method" (Ziman, 2000) is not a legend, but what is actually valued as an important criterion for high quality research. We can explain this again by astronomers holding truth in high esteem, which arises from the realists' attitude. High quality research is mostly dependent on the correspondence of reality with the scientific theory. From which follows that the realist assumption and the assumption that scientific quality is objective go hand in hand. Research is of high quality when it is performed in the service of truth. This is why several interviewees describe the values that define high quality research as something **"clear" and "absolute"**:

> Faculty Member 1: "Academic quality I think has always been relatively clear. It [the research] has to be verifiable and clear, unbiased etc. I think that is academic quality."

> Faculty Member 2: "I think in terms of what constitutes good science and what is academic integrity, all those things don't change – they are pretty close to absolute values I would say."

This quotes also show that from an astronomer's point of view *integrity* is implied by need for clear and verifiable methods that constitute high quality research. Integrity and the quest for truth go hand in hand, hence astronomers see integrity as an "absolute" value for the pursuit of science.

> Faculty Member 1: "But you know, a scientist always has to be responsible in his research methods. So in that sense there is nothing new and there cannot be anything new."

> Faculty Member 3: "[…] Society may well ask 'What do I get for my money?' Society will not get a guaranteed Nobel Prize. Society will not even get an agreed upon product. But the one thing that society will always have to get – *absolutely always* have to get from a scientist – is their *dedication and their honesty*. That, you see – if a scientist is not dedicated to the work, if a scientist is not perfectly straight forward and honest about the work, the *scientist is cheating*. And *cheating kills the subject*. There is no such thing as cheating in science, okay? Now, taking this thing on a personal level. Yes, I tell lies. Yes, I have cheated people. Yes, I am just as – shall we say 'normal' as other people, but not in science. I have never ever, ever, under any circumstances, written or done anything in the sciences that was not fully and totally true, as far as I could tell. I have never, under any circumstances, and I will never in the future, present things in such a way that they look better than they are. That's cheating. You must not cheat. […] The reason that the scientist has to be honest, is not only because it is contrary to the spirit of science to cheat, but also because you may damage society."

How highly astronomers value integrity – both, their own and that of facts – is illustrated by the example of Faculty Member 3, who even makes explicit that he has no problem with waiving his right to stay anonymous:

> Faculty Member 3: "Yeah, but you know, for a scientist it's important to sort of stick to what you say. Even if it's stupid. [...] Even [staying anonymous] is not necessary. As long as it's about the facts. I have no problem with that."

While there is consensus about what is valued about research, when asked how to *measure* this quality, some interviewees do admit that quality is hard to define and often intuitive and no clear distinction between "important" and "unimportant" (other than "redundant") questions can be given. Astronomers seem to have a consensus of what *is* scientific quality, but not a clear idea of how to *measure* it. What quality means in the context of the funding, publication and assessment system and what measures of quality that system introduces will be shed light upon in the following chapter:

### *2.3 Perception of the system's values: Funding, Publication & (Career) Assessment*

*2.3.1 Funding System as experienced by the astronomer*
*(– "Darwinian" or risk-enabling?)*

Money to pay Sterrewacht's employees mainly comes via three different ways: via the Sterrewacht (the University) itself, NOVA and outside grants. How NOVA received its funding is explained in *Chapter 1.2*. NOVA, together with the institute, provide salary for permanent staff. How much money the university allocates to each institute is depended on a formula, which is called ***"Allocatie Eerste Geldstroom" (AEG)***:

> Faculty Member 2: "So what we get from the university is determined by how well we have done over the last few years in terms of how many grants, how many PhD candidates, how much teaching we have done. It's kind of an arrhythmic model that determines how much you get over the next year, it's kind of a 3 year average."

The amount of money that the university receives from the government is based on a similar formula. This model makes the institute very autonomous, but at the same time responsible for paying their staff. The Sterrewacht "is doing very well financially" (Faculty Member 1) and leftover money after those "baseline" expenses can be allocated freely. The Sterrewacht "strategy club" decides what for the money will be used, which could be for example overlap positions.

The most prominent sources for ***outside grants*** are the NWO (the TOP programme, Veni-Vidi-Vici), EU grants such as the ERC and other Postdoc fellowships for example from ESO. Those outside grants are "needed to fund graduate students" (Faculty Member 4) who are the ***working horses*** of the system:

> Faculty Member 1: "[...] if you have money that means that you have also, that you have labour, that you have the effort available. And of course in exchange we need to define a little piece of science that that student can do as part of his PhD."

Grants are personal, however interviewees report that the Sterrewacht is a "very collegial institute" where a lot of collaborations are possible, in order to share money and effort. How holding a grant often leads to collaboration will be explained further in *Chapter 4.3*.

Grants are limited and very competitive. Money available for research is finite and so proposals need to fulfil certain criteria in order to be successful in acquiring grants. As we described in *Chapter 1*, helping build and therefore having access to new observation facilities is a bonus for securing grants. But criteria for a successful are generally not always transparent to the astronomer:

> Faculty Member 3: "[…] it is the sort of work that is surrounded by a lot of almost mystery, uncertainty – you don't really know what you are doing, okay? If you want to build a house, you have someone there that tells you, as to how it should be constructed. If you apply for a grant it is like, I don't know, it is too fluid, there is no … anyway, I find it even hard to formulate what the difficult is, but – on the one hand it is science, on the other hand it is advertising. And it is making certain that your research gets into a popular newspaper."

Advertising the so-called ***"sexy topics"*** are highly valued when the government and funding agencies decide which research proposal to fund. The interviewees frequently report that "the funding system is very much oriented towards the fashion of the day" as opposed to also "extremely important" topics that are "more pedestrian/ basic" (Faculty Member 1).

> Postdoc 1: "[…] the government that actually gives you the funding, they don't really understand. And they don't really care. They just want more papers and they just want to find life on other planets. So as long as you sell it to them like [what] they want to hear – I think it's really bad, but it's the way it is. They don't really understand what we do at all. And you have to make it sound sort of super *sexy*, otherwise they won't give you the money, so if you are …  Like, if I am trying to analyse the chemistry of a star, they would be like 'Why do I care?' But if I say 'Oh, but there might be exoplanets around it and I mean we need to know what's gonna happen to Earth when the Sun dies', then you tailor it to something that they can sort of relate to and then they will be like 'Oh, maybe this is interesting' […] I think the more we try to sell it *sexier* and more – ah, I don't know, sort of it has to make a huge difference, like a massive *impact* – the more we are actually damaging ourselves, because research doesn't work like that."

In what ways favouring of "popular" topics "damages" research is elaborated further from *Chapter 4* onwards. Promising "impact" is important to acquire funding and having impact has an influence on the AEG:

> Faculty Member 2: "So the impact is very important, because if [the evaluation committee] had said that we are doing so-so or it is a field in decline or an institute that are not doing things right the university can start reallocating their priorities [as in funding]."

While there is consensus amongst the interviewees that research topics that are "very much the flavour of the day" (Faculty Member 1) attract the funding, it is not clear if that preference stems from the funding agencies themselves or the referees that review the proposals:

> Faculty Member 1: "And that is of course the referees who do that. I mean that's not the funding agencies that decide we are only going to fund research that is fashionable. It is somehow the referees […] probably because they don't get any instructions at all. […] Or maybe the wrong instructions. For instance if I referee something for NWO or ERC I always have to look, there is a number of criteria to fill out. Of course, is the research high quality? And also does it break new ground and is it innovative?"

As described in *Chapter 2.2*, addressing an important issue, a quality criterion, is different from targeting innovation. The latter leads to a discrimination of basic research in favour of sexy topics. However, the line is fine and as described earlier what is an important issue also often lies in the eye of the referee and so "somehow it's always, it ends up with the fashion of the day" (Faculty Member 1) and **luck** due to the "lack of funding":

> Postdoc 2: "But … I think the biggest problem of the funding system how it is now, is that there is so little money available, that the selection problem is … I would say almost random, not completely random, but you could have a very good project and very robust project but not been given the money, because there are just too many."

> Faculty Member 4: "Funding-wise … We are in such a money poor position – the funding rate is less than 10% now. In nearly all Astronomy fields. And what that means is, yeah, one out of three excellent proposals will get funded and the other two, which are equally excellent, will not get funded. […] The funding is bad now, to the point where if you know you have got an excellent proposal, there is a reasonable chance, you still won't get funded."

This randomness in allocation of funding is what the astronomers call the "TAC-Shot-Noise", which stands for "Time Allocation Committee"-Shot-Noise. The word "time" here instead of "funding" indicates that committees that grant observing time base their decisions on the same values as funding agencies. The interviewees make no difference between grants in terms of observing time and research money when talking about the funding system. In Astrophysics, having been granted observing time is as prestigious and important for ones career as funding as we will see later in *Chapter 2.3.3*.

That grants are often based on luck generates a lot of (psychological) stress for the applicants. On the one hand consequences of this "rolling the dice" (Faculty Member 4) technique include tense competition, favouring quantity over quality and risk aversion. We will come back to those consequences in *Chapter 4*. On the other hand, because the Sterrewacht is in a financially very healthy position, situations, where applicants have been "unlucky" can easily be mitigated and so interviewees report they have received money from the Sterrewacht or NOVA to fund their graduate students, when their proposal when you fall below the threshold of the lucky ones.

The threshold of getting funding is arbitrary and (Merton, 1968 & 1988) demonstrates that "non-winners" of a grant are considerably less successful in acquiring grants in the future than "winners", even though their skills are comparable. Already the way the AEG works gives a hint that the allocation of funding involves the ***Matthew effect***: how much money the institute will be given by the university is "determined by *how well we have done over the last few years in terms of* how many grants […]". The prevalence of the Matthew effect is also frequently reported by interviewees:

> Faculty Member 2: "The funding system is mostly … Mostly looks at your *past achievements*, right? So much of what determines whether you get your next grant is what you did with the previous one, so … how that's evaluated or viewed, or judged, or measured is key."

> Faculty Member 3: "Erm … the funding agencies have a tendency of – where the money follows the reputation. And strangely enough, it's not totally inappropriate that money follows reputation. Erm, it is in a sense *Darwinian*, I mean something has success and therefore you should feed it, you should support it."

While the latter quote argues that the "Darwinian" aspect of funding system makes sense, it is questionable, whether "successful" research, which is the basis of further funding, is equal to being of high quality:

> Faculty Member 2: "And for some projects people – well it depends on who does the reviewing – but sometimes people look at how many papers came out. Sometimes people will just measure the citations and they don't check whether all the citations say that this paper is wrong or whether the citations say whether that paper is brilliant. They all count the same way."

> Postdoc 2: "[…] from what I have seen, funding applications cannot go too much into the details of how high quality the research can be. And most of the time, because the people, who read the application don't have the expertise to judge that … So people will judge about the quality of the research proposal, based on *previous research* of the person that is proposing it and that would be assessed mostly through the citations and *how influential the research is in that field*. So it's not obvious that they take into account the same standards or criteria I care about."

Another interviewee calls this the "chicken and egg problem" and a "winner takes in all" dilemma:

> Faculty Member 1: "Well, for instance for the ERC research grant and also NWO TOP you have to … it's a big part of your application to describe everything you have done so far and … that's a bit strange, because it means that you have to be already famous essentially in order to become famous, right!? I mean you have to be established already. The things that matter there are obviously things like *publications, publication references and impact of your papers*, so that means *citation rates* obviously. […] And how well you have done in the past with PhDs and Postdocs …

> But it's a bit of a *chicken and the egg problem – you have to show that you have done all that already before you actually get the money to do it* […] Well, I think that it is a bit strange that … that your CV is a large part of some applications – in my case it's to my advantage, because my CV is strong, but still I find it strange. For both of the NWO TOP, 40% of your final score is simply based on your CV. […] I think for the ERC Advanced Grant it's something similar. You have between 5 and 10 pages to write down all the stuff that you did and why *you are so important and successful and rich and famous* and … I don't think your CV should count too much. As long as you can show that you are up to it and you have the technical expertise and the knowledge and if necessary the leadership, then I think that should be sufficient. It becomes a little bit – especially with these large grants – it becomes a little bit a *"winner takes it all"*. […] it becomes easier to get a big grant if you have already been successful."

To summarise our observations of what the funding agencies and referees value as perceived by an astrophysicist, generally, the interviewees are well aware of how funding can be acquired and how funding is allocated to the institute and individuals. Even Master students, while not knowing details, are aware of the importance of past output. We observe that the Matthew effect plays a big role in funding in Astronomy based on the astronomer's experiences that the funding system values past output in terms of **publication & citation rates and impact**. Sexy topics are often favoured over basic, but important research. In summary, those values are often not in line with the value of research quality as defined by the astronomer (*Chapter 2.2*). The resulting consequences of this value discrepancy on the research behaviour are investigated from *Chapter 4* onwards.

*2.3.2 Publication System as experienced by the astronomer*

In the prevailing publication system it is the journals and peer reviewers who decide what gets published. They set the criteria for a piece of research to be "publishable". This section will take a closer look to what, according to the astronomer's perception, those stakeholders value in a paper to give the go to publish it.

According to the interviewees, in Astronomy the main journals are Monthly Notices of the Royal Astronomical Society (MNRAS), The Astrophysical Journal (ApJ), Astronomy & Astrophysics (A&A). In addition to publishing in one of those refereed journals, most astronomers publish their pre-print in ArXiv, which is non-refereed. In Astronomy, the choice of the journal does not depend much on the content and topic, unless the research is specialised in instrumentation, for which one can choose refereed journals or the non-refereed Journal of Astronomical Telescopes, Instruments, and Systems (SPIE).

*Sound methodology and good communication* of the results are some of the quality criteria for an astronomer (*Chapter 2.2*), so it is interesting to investigate whether the publication system encourages those values. Referees are supposed experts in the field, but, in the experience of an astronomer, might not always have the detailed knowledge to judge the correctness of the methodology. They also might not take or have the time for a careful check:

> Postdoc 2: "I think [the publication system] can encourage quality, but ultimately it depends on who is refereeing the paper. And what standard they choose themselves to apply to the paper they are reading. […] But I have never had someone looking carefully at the methodology [section] who said "Okay in this you could have done better and this is wrong" or whatever. Maybe it's because I am very careful in what I do. But I am sure I made some mistakes. And it seems to me that indeed, it's not a … at least in the 3 most common journals there is not a huge emphasis made on the methods, trying to understand them."

Another interviewee describes that this makes it "easier to publish things. The journals are less strict [than in the past] about what they will accept." However, non-careful checking of methodology does not mean that fraud would stay undiscovered. Interviewees report that the peer review process or researchers using the published results would "by large we catch all the bullshitters" (Faculty Member 4). Even half-baked results are often turned back:

> Faculty Member 4: "[…] I think the referees are harsh enough. […] If I am trying to push through a dodgy result, I have had referees push back."

However, at the same breath the interviewee states: "If you want to publish it, you will find a journal – All you need is one person saying 'Yeah it's fine.' And it will get published." Whether premature publications or fraud are an issue in Astronomy will be further discussed in *Chapter 4.1.2 & 4.5*.

The ***paper writing style*** is another aspect of publishing that interviewees brought up. The style in which papers are still written is rooted in the past and outdated. While our state-to-the-art technologies would allow for a more reader-friendly format, papers are still written in the linear way that was developed in pre-computer times. This outdated style doesn't enable for example easy implementation for code. On the one hand, this undermines the astronomer's value to demonstrate sound methodology, as code is a crucial part of the analysis in Astrophysics. Bugs often stay undiscovered. On the other hand, this style doesn't encourage good and easy communication of research results, as it would be possible for example with interlinked and pop-up content. The fact that papers are also not updateable diminishes their communication value.

> Postdoc 2: "[…] the way that papers are currently being written is perhaps too much tied to the way that papers were published in the past. So they were actual papers in a journal, so they had to be sequential. But this is no more the case, now that we have other ways to … read or get information. We can have, not necessarily interactive things, but, at least content that can be separated into different sources. So you can have … you can read on one side about the science of the paper, and on the other side about the technical aspects. And currently the two things are merged into a single file, or work. And even if it's true that you intent to have sections like methodologies and results, so if you are not interested in the methodology, or if you actually want to read about the methodology you can go there or not go there. But people will tend to get take [content] away from the methodology section, because they will consider 'Ah, that's too much […], so let's not mention this or put that into an appendix'. So I think

> there should be the possibility for authors to be *very thorough in explaining the methods* and even, that includes the possibility to show code. Erm, that also means – if that is a published content – it means that you must be able to have referees that are able to read about this and that are able to judge the content of these code. Which is not always the case. In fact in the Astrophysics community, the skills in programming are fairly low in general. Which is worrysome, because I think there are a *lot of bugs running around* that are not noticed. And because we can't look at the code we can't say, or see whether this is happening or not."

That the publication system doesn't encourage easier publication of methodology and code also means that the system doesn't value research **replicability** (read more in *Chapter 4.1.3*), because the details of the methodology cannot be thoroughly checked:

> Postdoc 2: "And I think that's very bad, when for example, almost the entire results come from a code which is not publically available. So you cannot look at this code and see … if they are actually doing what they say in the paper. And also if they – sometimes they make a mistake. […] So in the sense, the replicability of the work we do … is not always very high. And in the sense that you can download for yourself, in principle all the raw datasets from a telescope and you can redo everything by yourself. So in this sense, yes it's replicable, but never fully replicable."

An astronomer's drive is based on curiosity, so one quality aspect if whether results add to a better understanding of how the universe works (*Chapter 2.1 & 2.2).* In order to study a potential discrepancy in values, it is interesting to investigate, whether the publication system values certain results over others or has a preference in what topics to publish on. One interviewee states:

> Faculty Member 1: "The publication philosophy is also different than in other sciences. In Astronomy **everything that is not obviously wrong tends to get published**. […] it very, very rarely happens that papers are not accepted if there is nothing wrong with them but they are purely uninteresting."

This fits very well with the astronomer's commitment to truth – "Everything that is not obviously wrong is publishable". However, this gets relativized by other interviewees: "It's not sufficient to be true. It has to be true and pushing knowledge." (Faculty Member 4) Which is still conform with the astronomer's values. As stated earlier, the results need to be "new" and opening new insights in Astronomy. However, what fulfils that criterion is often open to interpretation, so lies in the eye of the reviewer. Often it also depends on the research field:

> Faculty Member 4: "… I think … it depends on the research field. If you discover a planet and you think it's real […] then you publish that and generally the interpretation you leave to another paper. [That field is] so limited by the number of objects. Each time you publish [a planet], it's brand new, it's adding significantly to knowledge. […] So at the moment [that field is] in that regime. But for example radial velocity planets, there are Hundreds of them, so if you wrote a paper saying 'I have just discovered a radial velocity planet', it'll probably get rejected. Just because they'd say 'Well, who cares? We know that there are radial velocity planets – is there something unusual about this one?' If it's significantly pushing knowledge into a new direction, then you publish."

In the field of exoplanets a detection with the right method (direct imaging as opposed to radial velocity) can be enough to publish already without interpretation or analysis. In the field of Radio-Astronomy that is the same case, as detections through long wavelengths are extremely difficult and so a detection is already publishable. Hence in some observational Astronomy fields a ***sole detection*** is highly valued by journals and reviewers. However, in observational Astronomy, ***non-detections*** are much more frequent than detections and about 90% cannot get published (Postdoc 1). Unless the non-detection can "add to new knowledge" by having been able to calculate upper limits or demonstrate anomalies, they are not publishable:

> Postdoc 1: "[Negative results are not publishable], unless you have a very good, as in for example the way we sort of explained the upper limits with the non-detection. […] *The problem is how to tailor it, right?* […] So, yeah, unless you have … like a good way, I mean there is some research that published non-detection – for exoplanets sometimes they publish it when they didn't detect it, because sometimes you sort of predict that it should be there … […] And it's an anomaly or something like that … […] So there are some ways to publish this, but I think it's very … like 10%. There is a whole 90% that doesn't get published and sometimes, like for example, if you just had bad weather, then it's very difficult, right?"

However, it is arguable whether "adding to new knowledge" does not also encompass knowing where and how not to find an object, especially in a field were non-detections are far more common than detections and hence should also be regarded relevant in the astronomer's perception. We will come back to that topic in *Chapter 4.4*.

Whether or not results break new ground, in which cases non-detections push knowledge forward and are hence worthy of being published and whether a sole detection is publishable is in many cases dependent on the judgement of the journal and peer reviewers. This room for interpretation gives space to a bias towards ***sexy topics*** in the publication system.

> Faculty Member 1: "So there is – this is well known in Astronomy, there is a bias towards what gets published.
> Interviewer: "The sexy topics?"
> Faculty Member 1: "The sexy topics, yeah. But even the data that I do not publish are still in the archive, so it's still available for anybody to look at, if they want. But the thing that they know is that I looked at it already and that I haven't found anything interesting, or sufficiently interesting to write a paper about it."

Another interviewee describes to concentrate on topics that are exciting and new, following curiosity, but when they find something unexpected important, then that could mean a shift in focus:

> Postdoc 1: "'We are gonna focus on this, because it's way more interesting, it's new and it's more sexy' […] So this is the thing: When you have sort of … new … sexy things then it's easier to get it published [even when the research hasn't reached a matured stage yet]."

Journals like *Nature* and *Science* especially concentrate on those sexy topics:

> Faculty Member 2: "I mean a few times I have been involved in a paper for *Nature* or for Science. […] Then you know you have very, very limited amount of space, so you have to make your point quickly. It has to be a high-profile point you are making […]. Which is why I don't like this journal actually, because you don't have space for details. So the way you write those is very different. And in fact the motivation for writing there is also very different usually. […] Just to get … trying to make a splash. And most of the papers in *Nature* in Astronomy are actually wrong. Because they go for spectacular results, which are *a priori* unlikely and so the proportion of things that maybe look exciting, but are maybe on the edge is quite high in *Nature*. So many of the results in *Nature* actually later turn out to be wrong. […] They don't, they don't judge the science, they just judge whether it will sell copies."

To summarise our observations on the astronomer's perception of what journals and reviewers value in a paper in order for a paper to get published, there is a preference for **sexy topics**, even though, in principle, any result that pushes knowledge forward, is publishable. As it is open to interpretation whether a result is important and new, often **non-detections** are not publishable and content needs to be tailored to become publishable. A **thorough check of results and methodology** is not part of the publication system, which could be a result of the limited time of the reviewers and the fact that papers are not 100% **replicable** as methodology sections are often not detailed enough. The publication system does not encourage more up-to-date **paper writing style** and so more user-friendly readability and easy publication of code is not made possible, only adding to the difficulty to replicate and check results and methodology. Bugs and mistakes often stay undetected. While most interviewees demonstrate that they have a good feeling of what to publish and how to publish, the publication system does not fully foster the astronomer's curiosity and need for sound methodology. The resulting consequences of this value discrepancy on the research behaviour are investigated from *Chapter 4* onwards.

*2.3.3 Career system as experienced by the astronomer*

This section elaborates on the factors that, from an astronomer's point of view, play an important role for career development in academic Astronomy. We will see that the career system is closely tied to the funding and publication system. The researcher that explained that the things that matter for funding "are obviously things like *publications, publication references and impact of your papers*, so that means *citation rates* obviously" then concluded "Those things are important to keep your standing – to show that you have made an impact etc." (Faculty Member 1)

Impact, number of publications, citations rates, grants and recognition/prestige seem to be linked together and build the basis of what is being assessed in terms of career progress. Scientific merit is gained through publications of papers in peer-reviews journals. Career paths and progress are dependent on the types of grants acquired and your history of previous appointments. This means that output forms the basis of one's future in Astronomy:

> Postdoc 2: "So it is ... erm, it also seems to be somehow the end of things. You do research to write a paper, to be ... published and assessed and ... You need to write a paper to be able to get the grants or to be able to have a position later."

Let us first turn to the most prominent form of output, ***first-author publications***, which all interviewees identify as vital for career advancement. When an interviewee talks about "having a paper" or "publishing a paper" it is always implied that that person is first author on that paper. Interviewees describe not only importance of publishing, but also the "emergency" to do so:

> Faculty Member 4: "Before you have a tenure job, you've got to make an impression and demonstrate that you can produce papers in a reasonably rapid fashion."

> Postdoc 2: "You always want to be the fastest and want to have your results out. But it's not really a deadline, it's more an emergency."

During one's PhD it is generally expected to publish four papers in total – one per year. At Leiden Observatory this is made more explicit and early career researchers are aware of this requirement. That this requirement is fulfilled is also subject to the yearly "R&O talk":

> PhD Candidate: "I am sure that at this yearly meeting they want to make sure that you are making some progress, like that you can actually [publish 4 papers] by the end of the 4 years."

Interviewees observed that at some other universities there is no such explicit requirement of publishing one paper per year to obtain a PhD. However, as Postdoc 1 argues, this requirement remains implicit:

> "The problem is – this is the main thing, right – if you wanna have a job later on, you are gonna have to have papers, because that's how it works. Even though I don't like the system, I don't like the way it is, it is what it is and you *have to adapt to it*, so I think it would be very – way better, if at the beginning your PhD supervisor is clear on that and says 'Look, if you wanna have a job later on, you have to have a minimum of 3 papers [for a 3 year PhD].'"

This dislike of "the system", which involves – amongst other effects – immense pressure on the researcher, and will be discussed later in *Chapter 4.1*. For a Postdoc that expectation to publish only increases. One paper per year is reported to be the absolute minimum, while in some areas of Astronomy it is common to publish two to three papers per year or even more. While recognising these required publication numbers as far too high, interviewees at the same time justify their importance as publication numbers feed into necessary "*filters*". If not all PhDs can find a Postdoc job afterwards and there is not a tenure position for all Postdocs, *competition* is introduced and filters need to be applied. Publication numbers, citation rates and impact play a role in filtering candidates, whether it is for telescope time or for jobs:

> Postdoc 1: "[…] for [the telescope] the first filter is how many papers do [the applicants] have? Because you have so many candidates, we had 4 positions and there were 80 candidates. There is no way you are gonna read all the papers of all the 80 candidates and choose which one is the best. There is no way, there is no way. So you have to do filters … So sometimes you can look at the impact of the paper, right? So you can say 'Okay, this has 10 papers, but this one has 5, but the 5 that he has, has way more citations that these 10', so you have some sort of fact, like numbers that you can use."

The quality of the papers is thereby reportedly perceived as less important than the mere numbers and scientific quantity:

> Postdoc 1: "It's just because there is so much competition, that the first filter you go into is how many papers you have. Doesn't matter how good or bad «laughing» they don't check this that much."

> PhD Candidate: "You know it doesn't necessarily matter how long each paper is […]. Or like the sort of quality behind it, it's more about that you have specifically 4 [during your PhD], like the quantity."

Hence, the career system does not seem to value scientific quality as highly as the astronomer does. What effects that has on knowledge production will be discussed later, from *Chapter 4* onwards. Astronomers value publishing as a means to disseminate their research results in order "to push knowledge forward" (see *Chapter 2.2*), which is most likely the reason why most of the interviewees had not been aware of the importance of the publication rate before they entered the system. Most of them described they came to that realisation how important output and visibility for one's career is, during their PhD or when applying for Postdoc positions.

The requirement to publish first-authored papers, which one has to "produce in a rapid fashion", is a participation in the race of priority (Merton, 1957). Winning this race, being the first one to publish on a particular piece of research, means to be granted one of the biggest rewards in our science evaluation system, which is recognition (Sismondo, 2004; p.22 and Merton, 1968). ***Prestige*** again, is subject to the Matthew effect (Sismondo, 2004; p.35-36), which means, the more recognition a scientist earns, the more easily the scientist can proceed their research at more prestigious universities or acquire "a name" which enables them to be more productive than without this status. In the interviews, we found evidence for this Matthew effect in the career system in Astronomy and what prestige means for an astrophysicist.

First of all, interviewees highlight the connection between ***publishing as a first author and recognition*** and its importance especially for early career researchers. This implies that collaborations, leading to a co-authorship are only recommendable for later career stages, where one has already established their (tenured) status.

> Postdoc 1: "But if you are beginning your career I think [having a lot of collaborations] is very bad, because if you are not the PI of the proposal, it's very unlikely you are gonna be the first author, so you are always gonna be one of the co-authors. So it's never gonna be YOUR research. […] So then you are never gonna have a name in the community and I think 'What are you doing?'"

Acquiring "a name" in the community is essential for an astronomer's career development and being visible in public is a prestige boost. One's publications are usually not enough to become visible in one's research field, so presenting research at other universities or (the attendance) at conferences is common practice. An interviewee describes his realisation when starting to apply for a Postdoc:

> Faculty Member 1: "And that was the first time when I started to think about career type of things […] I need start to make a name for myself, I need to be known, so I need to publish, I need to give talks and things like that … People need to know me, I need to visit placed and give talks."

Second, *citations* of one's publication is seen as a reward itself, as they stand for the influence a scientist has on their research field. While there have been several studies, that citation rates are no representative measures for impact on the community (Sismondo, 2004; p.35), citation rates do not only remain a measure of an astronomer's productivity and success, but a scientist is also cited more often the higher their visibility (Sismondo, 2004; p.36).

Third, a direct implication of the Matthew effect in the funding system, is that acquiring *grants* is not only important for future funding, but also valued in their career. Receiving grants adds to an astronomer's recognition. In Astrophysics, those prestige-raising acquirements can be of monetary value, but also having been granted observing time counts as a "very successful" currency (Faculty Member 1).

Fourth, *"landing" job appointments* themselves are prestigious. The more prestigious the institution, the more highly valued is the job post in terms of recognition. Some interviewees explicitly mention, that their decision of where to apply for a job position is dependent on where the "best group" of their subfield is located. A job post in combination with a grant, like a fellowship, is particularly prestigious and "helps you to get other jobs" (PhD Candidate). Therefore success in terms of having acquired a prestigious job position, in terms of being affiliated with a famous group or university, and accompanied by a grant, increases the astronomer's chances for further success, which is evidence for the Matthew effect prevailing in the career system in Astronomy.

Fifth, astronomers earn significant recognition for *high-impact or ground-breaking discoveries*, especially when their attempt to make those discoveries was *risky*.

> PhD Candidate: "Yeah, but I mean, my PhD for example, it's like a … It is a pretty risky thing, but if I am able to do it then it has a high reward, and with the system

> being so competitive I thought 'Okay, you might as well go for this type of project and if you don't make it – if I don't make the detections or whatever, then, well I mean it's going to be hard to stay in the field anyways, so …' […] But if you want to really define your own place in academia then this would be a really good project to do that, because there is only 5 people in the world working on it and if you are the first to make these detections, then, yeah … You get so many things to explore."

Sixth, if an astronomer can demonstrate ***technical skills and background knowledge*** for technical work with telescopes and instruments, they may gain a substantial amount of recognition for their rather rare training to understand the scenes behind the data production by the machines.

> Faculty Member 4: "So that was the key, you know demonstrating that you can do the process from one end to the other was probably what got my job here."

Prestige through recognition evidently plays an important driving factor of an astronomer's career. Therefore, as we claimed at the beginning of this section, output forms the basis of an astronomer's future in research, if that output increases their visibility. The more measurable this visibility is, the easier for committees to decide whom to grant monetary aids, observation time or job appointments. The most common measurements include publication and citation rates, number of acquired grants (including observation time) and the ranking of the universities of previous job positions. This makes it very obvious that the future career of an astronomer, in terms of landing a job and receiving grants, is dependent on past achievements of grants and prestigious job posts. Again, this is what Faculty Member 1 described as the "Chicken and Egg" problem in the career of an astronomer. The Matthew effect of the funding system reinforces the Matthew effect of the career system with the effect that a "Golden Child Trajectory" (Faculty Member 4) is laid out. The "ideal" career in Astronomy is a straightforward climb of the tenure track; a PhD in minimum time, a couple of Postdocs, Assistant Professor, Associate Professor and Full Professor by success. This involves committing to a professional life in the "cycle of observing, analysis and publishing" (Faculty Member 1). What consequences this rat race for tenure track positions has on knowledge production in Astronomy will be discussed from *Chapter 4* onwards.

## *Chapter 3: Epistemic restrictions in Astronomy*

The previous chapters give us an understanding of what drives an astronomer in terms of intrinsic (the astronomer's values, *Chapter 2.2*) and extrinsic (the system's values, *Chapter 2.3*) motivation[1]. These socially constructed motivations drive the scientific performance and influence knowledge production. However, to fully understand what influence our science evaluation system has on knowledge production in Astronomy we don't only need to disentangle intrinsic and extrinsic driving factors but also investigate the "natural" limitations in the research process, which drive the science performance. That is the purpose of this chapter.

When directly asked about restrictions in their research process the interviewees mostly mentioned ***natural restrictions*** that arise from the nature of research and technological possibilities, which are independent of evaluation or funding. Some however did address restrictions due to the latter which arise from the expectations of the evaluation system and will be discussed further in *Chapter 6.4*.

One "natural restriction" arises from the uncertainty and open-endedness of research itself. Research is the human's endeavour to understand the unknown and as such research is by definition ***risky***. Astronomy contributes to basic research, which uncertainty is on the one hand exciting as one never can never know what to find, on the other hand it can feel restricting as it can lead to a dead end. On the one hand the ***"very open-endedness"*** (Master Student 2) of basic research about the fundamental questions of the universe is what excites astronomers, on the other hand it can make astronomers feel confined in their research abilities. It is very hard, especially for early career researchers to set up a research plan and know which steps are feasible. Additionally, when risk taking is not encouraged, as we will see in *Chapter 4.4*, and a "golden-child path" is laid out (*Chapter 2.3.3*), an astronomer can run the risk to "freeze", not daring to go into any direction or take a next step. Hence, the very fact that basic research is hard to be confined into a research agenda, can lead to a feeling of restriction and inability to move forward in the research process. This is what is called the "***paradox of freedom***"; An absolute freedom in the sense of absence of any restriction from the other (open-ended basic research), will necessarily lead to absolute restriction from the other and fall prey to absolute chaos[2].

> Master Student 2: "There is no ultimate goal, there is no kind of, you know absolute research.
> […] I think I am just like, I am sick of all of it. I think with research it is the *massive unknownness* of it. The very open-endedness, which is why I liked this year's project more than – for this year's project there was a plan. A plan that was very malleable, but it was like 'this month is this, this month is that' and it was kind of like 'Up until this point we know exactly what's going to happen, after this point comes, well, we expect certain things from the data, but we don't know and we'll deal with it when it comes'. But it was very planned. And I am better able to deal with things when there is a proper plan. But everything in research is like 'Oh, just start here, and we see where it goes!'. And then you are doing all these things and running around in circles and then after a couple of months in, all of a sudden it's like 'Oh, you don't have enough done.' – 'But there wasn't a plan, was there?' So and that side of things, yeah I am a bit kind of sick of … Because it feeds into the whole feeling 'Ah, I don't know what I am doing'. If there is no plan and I am sitting at a screen going 'Okay, what do I do now' and then if you are constantly going back to your supervisor, as I found out last year, it costs you marks in terms of initiative."

---

[1] E.g. https://www.verywell.com/differences-between-extrinsic-and-intrinsic-motivation-2795384
[2] Hobbes's Moral and Political Philosophy, http://paradoxoftheday.com/the-paradox-of-freedom/

"Absolute chaos" in research in Astronomy ranges from stagnation towards – in rare cases – fraud and hence may affect scientific quality negatively and can hinder "pushing knowledge forward". Especially for early career researchers who have not acquired enough experience to know how to deal with the uncertainty and open-endedness of their research, this can be quite frustrating and discouraging to pursue a career in Astronomy. Later in the career the astronomer might have learnt how to better confine their research projects. Postdoc 2 for example states that sometimes a higher quality method must be sacrificed for a more reasonable method, delivering less accurate results, when the accuracy is negligible and the conclusions would be the same. However, he does acknowledge, that "Astrophysics has a lot of uncertainties" and so one will never know if one's evaluation of the potential gain would have been right or if the higher quality method would have indeed lead to more insightful results. However, *restrictions can be a bliss* and too much freedom could be a problem:

> Faculty Member 2: "Sometimes, I mean 'restricted' is not the right word, but you are limited by how fast you can get telescope time or how fast data comes. What sort of things are reasonable to simulate, what sort of things are not. But I don't have the feeling that I am … I think I get to do as much as I can handle. Sometimes more resources would be maybe more of a problem."

The limitations Faculty Member 2 mentions here lead point out another form of "natural restriction", which are technical possibilities. Research in Astronomy today is mostly performed on computers. "Theoreticians" develop computations of theories and simulations to test them. "Observationalists" collect data from telescopes and need to process them on their computers. Every technical component in the research cycle might have **bottlenecks due to technical possibilities**. Interviewees focus more on the restrictions in observational Astronomy. What might sound like a joke is true for Astronomy; observations are dependent on the weather. If clouds cover the sky, one cannot make observations. Even if it is possible to observe the sky, the telescope might not be sensitive enough to detect a targeted object. And even if detection could have been made, processing its data may still represent a challenge. For example, working with radio data is more computationally intensive and less intuitive than other observable wavelengths. It could require the researcher to invest significantly more of their time to understand systematics in the reduction and obtain a result.

That is why Faculty Member 1 states that Astronomy "is largely driven by technical possibilities." Others state that their research is driven by the *access to the telescope*. It depends on contracts of the country and institute, whether the researcher can apply to a telescope. Even if so, opportunities to be granted observation time are very competitive, so one has to make a "judgement call" in a proposal for telescope time where one defines exactly what the target of detection is, when it should be detected and how:

> Faculty Member 4: "You say you want to [make a particular observation]. Great, how do we do it? How are we going to do it? I am going to measure it with a big telescope. How long will it take to measure it with a big telescope? It will take 10 nights. Okay, you make a judgement call – are they going to give you 10 nights of competitive time on a large telescope? No. Then you don't do it. You've got to walk away from it. Or you have come up with a clever way of cutting 10 nights to 1 night or some other possibility. Okay, you've got to have a discussion. [That's what you do in] research proposals. You write the pros and cons."

Some interviewees state that their research is driven by the *availability of data*, because that is what is the basis of their research after all, the proposal writing for getting access and observation time at a telescope. Priorities of projects may depend on when telescope data will be available to the researcher. Once the researcher, who was granted observation time, has received this data, they have one year before the data becomes public and open for any other researcher.

In summary, knowledge production in Astronomy is dependent on natural restrictions, arising from the uncertainty of basic research, technical and methodological bottlenecks, and political restrictions when it comes to access to telescopes. The choices of the research topic, research priorities, data collection and methodology are driven by considerations of what is "possible", which is nicely summarised by Faculty Member 1:

> "Just [choose] wherever your interest takes you, right? But not everything is possible. […] You have to follow the technical possibilities and stay at the forefront […] Stay at the frontier of knowledge with the techniques you have … and then do whatever you like. It's really whatever I like … I am in a very nice position!"

"Staying at the frontier" of knowledge corresponds to the astronomers' values as defined in *Chapter 2.2*. However, Faculty Member 1 has tenure and, as indicated by him-/herself, the freedom to be limited mainly by natural restrictions is a privilege of this position.

## *Chapter 4: Effects on research behaviour and knowledge production*

Comparing the astronomer's values and notion of scientific quality with the incentives of the publication, funding and career system, we observe a discrepancy. This discrepancy between what astronomers values in their work and what they perceive is valued when they work is being evaluated is called the evaluation gap. In this chapter we analyse what consequences spring from this evaluation gap. The gap appears bigger the younger the scientist. Early career researchers face obstacles that, by means of for example the Matthew effect, vanish for established, tenured scientists. That is the reason for the generally more relaxed tone of the older interviewees as compared to the younger ones. Those interviewees who are tenured describe peculiarities of the evaluation system and are aware of their "luxurious position". When asked whether the interviewee feels free to question the system how science works, Faculty Member 1 for example replies:

> "Ahhhh, yes, yeah, yeah, yeah I do. I think that's the job of a scientist."
> Interviewer: "To question?" [laughing]
> Faculty Member 1: "Yeah, absolutely, yeah we are not here to work within a system that is imposed. We are dealing with the truth here, I mean there is no other way about it and … it is important to get funding for PhD candidates, PostDocs, etc, but eeerm the thing that matters most is the progress of science. I mean we are all, or should all be at least, in the service of scientific truth."

While the tenured interviewees generally feel that their work is in line with their notion of quality despite the need to publish, they do admit that for young scientists it is more difficult not to compromise their values of scientific quality in order to survive in the system.

> Faculty Member 1: "No, [ the science evaluation system] is not in line with what I expected. I expected much more what you described earlier: just thinking and then suddenly making a big discovery, but that's a kind of very romantic way of looking at science and that's not reality. Eeerm, so it is a bit different. And I am trying to … Well, once again, I am not that happy with that cycle. It can put a lot of pressure. And I am trying to ignore that pressure now. I mean I am [past 50], so my career is established, let's put it that way, so I don't need to prove myself anymore, so I can safely ignore that pressure. But I think that younger people who still have to make that career have to work according to that system and I am not quite sure that that is actually a good thing."

> Faculty Member 2: "I sometimes wonder whether I would have made it through all the selection procedures that students go through now. I have the feeling that it was easier when I went through this."

Do astronomers adapt their behaviour in order to "hit a target"? What consequences does deviant behaviour have on knowledge production? Those are the questions we aim to answer in this chapter.

*4.1 Output orientation & Publication Pressure*

Output in form of papers and its qualities in question measured by indicators such as publication rates, citation rates and impact create the basis of the assessment of the *value of an astronomer*. Hence, not qualities such as methodological correctness and pushing knowledge define a worthy astronomer, but quantitative indicators.

> Faculty Member 4: "I think I was a bad Postdoc for her, because I didn't produce many papers."

> Postdoc 2: "[…] it's a comment that I received many times when I started applying for jobs, and one of the reply I had was 'Well you seem to be a good researcher, but you only have 1 paper accepted.' Even though that paper was a big one."

As the value of an astronomer determines their chances of receiving future funding and hence ultimately their career and this value is based on quantitative output (see *Chapter 2.3*), output then becomes a main driving factor:

> Faculty Member 4: "I look over the next year and what's the next goal to reach? So for me it would be becoming a full professor and that's a timeline of 5 years. So I say *'What do I need to do to demonstrate that ability?'* And having the conversation with [the director] ever year is a great way of focusing. That's generally what drives what I do […]."

This changes the astronomer's motivation from "truth-finding" to *result generation* and as a consequence, astronomers publish more than they would if they had not felt the need to publish:

> Faculty Member 1: "And ahh … if that was not so important for a, you know for a … I would probably not bother so much … I mean I would still publish my papers because I – it *gives a different motivation* to it, right? As a scientist you just want to publish your papers, because you are a scientist and you think this is important for science: 'This is the result, this is what defines the process of science'. […] Yeah, I have my doubts about the usefulness of that system."

It becomes obvious that this extra amount of publications, in the astronomer's opinion, does not add to scientific quality. The need to publish becomes a pressure to publish and curiosity, which is the astronomer's main driving factor to do research, becomes sacrificed in the daily-life process of research. Yet, Postdoc 1 describes that, when the publication pressure becomes too exhausting, she tries to look at the publication from a curiosity perspective:

> Postdoc 1: "Yeah, sometimes it is [exhausting] and sometimes I just try to look at it like curiosity and trying to understand better and ways to better solve the issue that we …So it's always sort of trying to understand more, and that's the way I look at it. But it does feel – yeah it is exhausting sometimes.[…] Maybe more, because sometimes I don't like that I have to publish. I wish that I could just publish whenever I have data. I don't like the pressure … that I have to have papers."

***Publication pressure***, which evolves from the "emergency of publishing" (*Chapter 2.3.3*), is perceived by all but one (tenured) of the interviewees and varies from a "natural" to a "relentless" pressure (Faculty Member 3); from a light version of "always at the back of your mind" to "a real feeling of ***publish-or-perish***":

> Postdoc 1: "You always have [publication pressure] at the back of your mind. I [don't] have a pressure as in that I have to publish this month or whenever, like a day-to-day pressure, but of course like I always have it in my head that … For example, I am finishing that paper that I have had for 1 year and I have data already for 2 other things and I am already thinking I need to do this research. And I am already thinking to apply for more time, because I need to have more data, because I need to publish more."

> Faculty Member 4: "But yeah, there is certainly far more pressure now … Yeah, there is a real feeling of publish-or-perish. We have to …. be seen to be publishing stuff and I think 20 or 30 years ago the field was a lot smaller so the word of mouth helped your reputation in huge amounts and now there is just not enough time to review everything everyone has done. We don't have the community anymore. So you've got to have some metric, where you can quickly do a first cut for all the people you are selecting for a job, let's say. And I am afraid, yeah, H-index is one of them. […] I do it myself – I freak out about H-index and all these metrics which make me look good."

"Publish-or-perish" means that publishing is crucial for an astronomer's career. In order to ***establish a career*** and follow the path up until tenure, an astronomer is expected to hit the required targets in terms of output. This pressure brings consequences.

First of all, we observe ***psychological effects of pressure***, like feeling nervous, insecure, and not good enough.

> PhD Candidate: "Yeah, I was already like – after being here for 1 year – I am like, I was getting nervous."
> Interviewer: "Okay, and is that because you observed that other people have already published in the first year or is that because you think it's expected from you or because somebody said something to you?"
> PhD Candidate: "Yeah, I think it's a combination of these things, because like, I sort of expect it from myself, because like I want to do this and yeah, then I also see other people who are already publishing. […] And yeah, and then I know that you have to make 4 before the end of the PhD, so it's like … Yeah, I think that also my advisors expect that I am already starting a paper so that I can stay on track. And so, yeah, I feel pressure from different areas."

Second, publication pressure has ***effects on the quality*** of research. An indirect effect stems from the constant worry about obtaining the next (temporary) contract, which occupies the astronomer's time and mind:

> Faculty Member 4: " It felt overwhelming at times I think […] 1.5 years ago […] Yeah, then I was flailing about … erm, what to tick, how to get tenure. […] So that's probably why I was running around like a headless chicken. […] Because that was a big decision point – if I didn't get tenure, then I'd have to find a totally different job. So that's for me … pretty … you know, I am in my 40s, I have only done one thing, what do you do? And that takes up a lot of brain space."

A direct effect comes from the focus on quantitative measures of output:

> PhD Candidate: "Your job prospects will depend on this like quantity rate, with which you are publishing."

Hence, the pressure to publish and the emphasis on quantitative indicators lead inevitably lead to sacrificing scientific quality for quantity. Quality criteria like methodological correctness and good communication of the results are sacrificed for a focus on conclusions which are needed to "make a splash".

> Postdoc 2: "Hmm … My impression is that people care much less about …. What's the word? … The robustness of the results obtained and they will more focus on the conclusions."

The need for impact therefore has an ***effect on the narrative and content of research work***. "Creativity is discouraged" (PhD Candidate) and risky but explorative topics are avoided (we will come back to that in *Chapter 4.4*):

> PhD Candidate: "You [the system] are really enforcing this quantity, like that you know that it's sure that you have a success and like a project, instead of being a bit more exploratory.'

> Postdoc 1: "I think the more we try to sell it sexier and more – Ah, I don't know, sort of it has to make a huge difference, like a massive impact, the more we are actually *damaging ourselves*, because research doesn't work like that. It's exactly what I was telling you – the interesting things are not the things you were looking for. So if you sell it for what you were looking for, or you say you are gonna find something super exciting, you may not and you might find something way more exciting, so you have to have the freedom of trying different things […]"

While this effect on content is generally described as negative for research quality, interviewees do hint on the necessity of strategies to enable impact on the community. A ***focus on sexy topics and "playing it safe"*** (Stephan, 2012) are such strategic measures that one can adopt:

> PhD Candidate: "Yeah, yeah! […] I hear it many times, [people] coming up with [sexy topics]. Well, I mean it's like a good thing to do. Like coming up with a project where you know that the outcome of this will lead to a paper."

We observe that the ***impact factor*** of a journal and the choice of the journal in general is less important in Astronomy than in other sciences (e.g. Rushforth & De Rijcke, 2015). Astronomers do not feel that journals with higher impact factor publish better quality papers. However, strategies aimed at increasing one's impact can involve considerations about what journal to target and about including prestigious names as co-authors:

> Postdoc 1: "There are some people that even think [about] strategies, if you sort of want to target it to a community, then target it to that journal. I haven't done that to be fair."
>
> Postdoc 1: "So on this, my first author paper, erm … So, when I wanna publish I do think what journal has more impact, but then, maybe first which one is free and then, second, which one has bigger impact."
>
> Master Student 2: "If you have a certain name on a paper, it might be easier to get it into a more prestigious journal, but that doesn't necessarily mean, that that specific paper is of higher quality."

Postdoc 1 describes that *Science* and *Nature* have by far the highest impact factor, but are difficult "to get into". Of the "normal" journals ApJ has the highest impact factor, followed by MNRAS, A&A letter, and lastly, A&A normal journal. The researcher has to pay for a publication in ApJ, while Leiden University has a subscription to A&A and MNRAS.

The pressure to focus on quantity does not only affect research content and targeted audience, but, just like the publication pressure, also leads to ***psychological effects***. Interviewees struggle with the balancing act of performing high quality research according to their standards and fulfilling the requirements of the system:

> Postdoc 1: "It's a system problem I think. Erm, I try to do quality research, but I do feel sometimes that I end up publishing because I have to publish.", "I wish we could just focus on more like quality papers instead of quantity papers."
>
> PhD Candidate: "Yeah, I am sure that's like something that many people struggle with, because you know that you have to like get so many papers, but I am sure many people want to do something, that is really like with substance to make an impact on the field."

Not only the personal balancing act is described as ***demotivating and discouraging***, but also the observation that more and more low quality papers are produced:

> PhD Candidate: "[…] reading papers from the 80s up until now, seems like they are becoming less and less of substance. […] You know, they start saying the same thing over again and […] Yeah, but that's discouraging"
> Interviewer: "Yeah? In what sense?"
> PhD Candidate: "In that it's becoming less about the quality."

However, whether "publish or perish" is the reason for this or simply biased access, where for example older papers only get digitalised when they are pf high quality, is subject of further investigation. Despite interviewees criticize quantity focus as a result of publication pressure, some also describe the ***usefulness of a certain type of pressure***:

> Interviewer: "Do you think, that sometimes you have to concentrate more on the quantity than the quality?"
> Faculty Member 1: "Yes, I do … That is a bit sad .. yeah, and I am trying to limit that, but I have to deal with the system. […] And erm, … yeah, as I said … I am not happy with that pressure. There should be some form of pressure – "
> Interviewer: "Mhm, a driving one?"
> Faculty Member 1: "Yeah, as a driving force, but it is a little bit too … people look at just that number. Your number of papers per year, which is a very poor measure of scientific quality I think."

Pressure as a reasonably dosed "driving force" can be positive. While, as described above, astronomers would publish less in absence of its need and focus more on research quality, due to the "massive open-endedness" and exploratory nature of research (*Chapter 3*), artificial restrictions (see also *Chapter 6.4* for further elaboration) and confinements are sometimes welcomed to be able to focus. Especially for perfectionists a decent dose of pressure can have a positive effect on their productivity.

> Postdoc 2: "Yes, in a way. If I didn't have that pressure I would probably take more time to study things. […] Perhaps more than would be wise. Because it's easy to get absorbed in research and refine more and more the details, even if at the end it doesn't change the global results of the data, or of your science. […] I think it's both – I still don't know … I think [pressure] is both a positive and a negative thing. It's negative because it sort of pressurizes me into thinking me in terms of 'Okay, what is the paper I will write this year?' rather 'What's the science I will do?'. But on the other hand it helps me identify – how can I say? A project in a sense, what I am doing. So one paper corresponds to a project, one scientific question, I want to focus on. And this pressure of having one paper published per year, encourages me not to spend too much time on a single project."

Only tenured people can afford to "ignore the pressure" (Faculty Member 1, see above) or even perceive less (external) pressure to sacrifice quality for compared to early career researchers.

> Faculty Member 2: "[…] whether you write a paper and what the quality of that paper is are two different things, so I know some people who are brilliant and who don't write many papers, but they are the people I would go and ask if I have a problem or if I have a question. So they are in some sense at least as valuable in the whole scientific enterprise as people that write 20 papers on …"
> Interviewer: "Yeah, okay … And those people are rather the tenured people?"
> Faculty Member 2: "Almost by definition, because otherwise you don't survive."

> Faculty Member 2: "I mean, certainly – If you are on a permanent contract, then the pressure is less in a sense. I mean it can still happen that you are in a competition and you want to be more visible that you colleague from Stanford or whatever. So that's a pressure, you can be sensitive to or not, but for PhD candidates, for their thesis they need 4 papers. And so that is a pressure, like it or not. And so of course, as supervisors we make sure that those papers are there and that they have what they need to be able to do in the 3 years that they have."

Despite publication pressure and the struggles to comply with quantity requirements, early career researchers also declare that they personally would not compromise on quality too much.

> PhD Candidate: "[…] Yeah, I am sure that's like something that many people struggle with, because you know that you have to like get so many papers, but I am sure many people want to do something, that is really like with substance to make an impact on the field."

> Postdoc 2: "The pressure is not extremely high either. If I don't publish a paper in one year I will not be … I will feel … dissatisfied, but I will not do anything to have paper published each year. If I feel I need to delay the paper, because more work is needed I will definitely do it. Even if some other people tell me 'You should publish, you should publish.' Because to me the quality of my research is more important than ultimately my career."

This attitude arises from the astronomer's values (*Chapter 2.2*), but whether it is sustainable on the steep and slippery climb up the tenure track ladder, remains to be seen. What is clear from the interviews though is that publication pressure and the need for impact affect research content, mainly in a negative way when it comes to scientific quality. Focus on sexy topics and risk aversion, as described above, are two examples for strategies to enable high impact and publishable results. However, the need for high impact output is also the source of salami slicing, premature publishing, low replicability, which all shape the content of the published research. What that means for the quality of the research is subject to the next sections.

*4.1.1 Salami Slicing*
*(– Publishing in instalments)*

**"Salami slicing"** (e.g. Broad, 1981; Huth,1986; Moed, 2005) describes the act of cutting up one's research work with the intention to publish as many papers as possible, solely for the sake of publishing. Interviewees do not classify research which is incrementally building on previous publications as salami slicing. The same counts for the slicing results for scientific reasons or for readability. The interviewees give a few examples for pro-slicing reasons. Our first observation is that the "least publishable unit" is largely dependent on the ***epistemic subculture***:

> PhD Candidate: "Yeah in my case, sometimes I … So I can try to publish just a detection of the object and basically that can yield a nice result in itself, just to, that you make the detection, but you don't do a really deep analysis. […] So yeah there is this possibility to sort of split it into multiple papers. One paper just on the detection and … maybe the implications of that just from being able to detect it. Then you can do another paper where you can compare it with other measurements […] – like really focusing on the science that you can do with it. Or there is different things that you can, erm try to compare this measurement to and if you want to do a proper comparison that can take you a lot of time."
> Interviewer: "And do you think it's because it makes more sense from a scientific point of view?"
> PhD Candidate: "Sometimes it does, but now it just becomes more and more about publishing and then they are trying to do more papers."

Because the PhD candidate's field is in the early stages of development, it is normal to first publish only the detection, and the analysis as a follow-up paper. This is because the detection is already a "nice result" in itself as detections were previously impossible to make and new methodologies need to be developed for the follow-up analyses. Therefore, all those research stages – from detection to analysis – are considered an incremental step in the knowledge of the field. While in those kind of cases publishing multiple papers on these results makes sense from a scientific point of view, the interviewee also admits that in general science becomes more and more about publishing. It is a fine line between the intention of achieving better quality or more quantity, when it comes to splitting up results. Postdoc 1 describes this dilemma:

> "So we wanted to do this distance technique for like a survey of more objects. So we first had the … measured the expansion velocity … And then we had these 2 epochs to measure the expansion. And we thought maybe it's better if we publish the expansion velocities first and then we use this paper to do the distances, but sometimes people don't agree with that. At the end we didn't divide it. But I thought … It would have been better. That was my idea that we should have divided it, because I think … The expansion velocity, you can use it for many other things, not just distance. So I thought it would have been good to have a paper on expansion velocity, so that people could refer to that specifically. Because the main paper will be on the distance and then you lose the expansion velocity measurement. But then it would be somehow hidden in the paper. So if you do a search on expansion velocities, you wouldn't get it as a first topic, because it's hidden in the paper, do you know what I mean? […] But my co-authors didn't want it, because they said that it's gonna look like you are just trying to divide it. […] So I gave in at the end. But so I thought that I had my justification."

In this case, information would have been transmitted in a clearer way, which is a quality criterion for an astronomer. Not publishing an extra paper about the expansion velocity dilutes the importance of that result. Another interviewee emphasizes that a ***good communication*** about research results may require different publications. However, he also points out its hazards, which involve leaving out important assumptions in follow-up papers.

> Postdoc 2: "[…] in the cases I have seen, there was always, a justification behind it. In that you would dilute the message or confuse the readers if you publish those two things at the same time. […] But I am pretty sure, there are some people in the community that will do it, just for the sake of having the papers. […] One trend that is very common today, for example when you have a large survey, you would start to write Paper 1, Paper 2, Paper 3, and each of them would address a different thing. And sometimes, Paper 1 is the description of the survey. Which itself is not really a scientific result. It's more a useful piece of information for other people using the dataset, but not a paper in itself!"
> Interviewer: "Alright, I see, and then, when you would prefer as a reader you would think that that was part of the introduction of the first results?"
> Postdoc 2: "Yes! […] Erm, because it often comes to the point – I have seen this happening perhaps more and more – that people will, in their paper, refer to the work done in another one, and sometimes there are some important assumptions that were made and are not discussed, or even mentioned in this one. But you have to read through the previous one to realise that they made this assumption to reach this conclusion. And I think this is very bad."

Good communication of results only involve a clear presentation of important results, but also *readability*. According to the interviewees, shorter papers are more readable than longer ones. Hence, readability might be a valid scientific reason for cutting up results.

> Faculty Member 1: "I don't think they do it just to produce more papers. […] People do cut up their papers into papers that are let's say no longer than maybe 20 or 30 pages, because if it's longer than that then nobody is gonna read it. But that's just a practical thing. […] I think it's just easier to publish a 20 page paper. So suppose I had a 60 page paper, I would be very tempted to split it into 2 or even 3, simply because it is much more readable. […] And because it would probably be easier to get published. Because, God, a referee who has to read a 60 page paper, that is a lot of work …"

> Postdoc 2: "So [I publish] whenever I have this result that *can be summarised in a sentence* and that is interesting enough so that it can be published on its own. That would define my paper unit. So it will be a shorter paper – 10 pages at most. But a more focused idea. So I have some data I can work on and erm, I can take one conclusion out of it, one main conclusion, then I can write a paper about that conclusion. If I have a second conclusion, I may write a second paper, if the two are substantially separated that they can be studied on their own."
> Interviewer: "And why in one sentence?"
> Postdoc 2: "Because I think it's a communication trick, that if you write a 40 page paper that contains a lot of information, people may be impressed about it, because it presents a large amount of work, but they would have trouble remembering …what's the main message of that paper, and if you have too many conclusions and ideas in there, it can be confusing. While having a shorter, well-focussed paper, makes it easier for people to remember. If you can summarise the content of the paper in a sentence or two, then they would remember your paper and use your conclusions when they do their research."

Shorter papers are *"easier to get published" and take less time to work* on.

> Postdoc 2: "Yes, that's one of the positive, well the bonus you would say, that writing shorter papers would also mean that you are spending less time on a given paper."

Whether astronomers split up their research to improve the quality of their publications, and perceive the simultaneously rising quantity only as "a bonus" is difficult to detect. The author's intention is not always easy to assess. An astronomer may hide behind the claim to slice results for scientific reasons, while their intent is to publish more. It is difficult to determine if salami slicing happens and probably that is why the interviewees' answers to whether they observe salami slicing happening vary from "it rarely happens" (Faculty Member 2) to "it happens all the time" (Postdoc 1). Faculty Member 3 even states that "publishing in instalments doesn't just happen, it's standard practice" in order to fulfil one's bibliometric quotas:

> "You publish everything that is done in 5 papers, right, rather than one….] Well, the name of the game is citation. You wanna get quoted, you want to have numbers. Therefore one short paper that states the basics of what you have discovered is not sufficient anymore. […] I mean you compute things under certain circumstances, you publish and then you change the circumstances, you compute again, you publish again and so on and so forth. You get 5 papers out of one single subject." […] I am absolutely certain that this is totally intentional. In fact, if I were the head of a big research group today, I would not just advise, I would require for the people working in my group to do it that way. As much as possible, as wide as possible, you know publish 2 papers that scientifically speaking are identical, but that are just sufficiently different, that you can put one of them in this journal and one of them in that journal and so on and so forth. I would – I mean the funding agencies, basically demand that we do it this way, right? And one has to become cynical to a certain extent."

The fear of being accused of being a salami slicer, however, can be a hindering factor to split up results, even if that action would be in service of good scientific communication, like in the case of Postdoc 1 and as described by Faculty Member 4:

> Faculty Member 4: "Everybody in the field will look at it and go 'Yeeeeah, that's – we call it MPU, Minimum Publishable Unit.' You spot them. And again, it depends, if it's a one of – if people were desperate to get a particular result or it's a very useful contribution to the field you forgive them for it. But if you do it 2 or 3 years in a row […] then the gossip in the community will go around that you are a salami-slicer."
> Interviewer: «laughing»
> Faculty Member 4: "And the trouble is – because it's unofficial, it's a double-edge sword. Because if somebody gets unfairly accused of it, it's very hard for them to realise they have got that reputation."

In summary, there are more "noble" reasons for an astronomer to split up their research results, namely when that action serves the astronomer's values. When publication pressure supports well-communicated knowledge through giving the astronomer an bonus incentive of publishing different important messages in different papers, scientific quality of the published content is improved. When publication pressure forces the astronomer to publish more papers on the same issue, leading to a reduction of readability – for example when important assumptions are left out – then scientific quality of the published content is decreased. Which of those two factors prevail is subject of further study in the form of a bibliometric analysis of salami slicing.

*4.1.2 Premature publishing*

Interviewees generally agree that ***premature publishing*** happens. They give however, different reasons and describe different implications for the field. Premature publishing often lead to bad quality papers which are characterised as ***failing to be sound in their methodology***.

> Postdoc 2: "[…] they would skip some tests, obvious tests, that they could have done, but that maybe take a bit of time, or that they use a method without properly characterising the biases or the assumptions that are used behind this method. "

> Postdoc 1: "[…] you can see it was written in a rush. The sentences are very difficult to understand. […] I can see it mainly when people are trying to explain how they used the data or how they got to the results, that when you are in a rush, you skip things, important things, or you go on a roll of explaining things that are actually not that important. Or you explain it without giving the big picture what it means, right?"

Those interviewees say that "it happens all the time" (Postdoc 1) and that "people publish this like in the news" (PhD Candidate). Faculty Member 4 even admits:

> Interviewer: "And out of this publication pressure, have you observed that people publish the results at a premature stage?"
> Faculty Member 4: "I think sometimes yes, the pressure to publish has forced us to sometimes push out results, where having another observation or two would make a significant improvement on the current results. […] BUT, that's why we have referees and I think the referees are harsh enough. […] If … If I am trying to push through a dodgy result, I have had referees pushed back and saying "No, go away and do it again." But if you want to publish it, you will find a journal that, where some – all you need is one person saying 'Yeah it's fine'. And it will get published."

Papers that are ***badly written are more difficult to read and to replicate***. Their message might be inexistent or drown underneath undeveloped statements. Therefore such prematurely published papers in a lot of cases are not very useful "to push knowledge forward". That rush is of course due to publication pressure and encourages *quantity over quality*, which can "*harm*" science:

> Faculty Member 4: "Yes, yeah, trying to push out as many papers as is possible. […] Whoever is first past the line will get their paper cited more than the other paper."

> Master Student 1: "[…] it can be harmful. Because, yeah, if I trust what you say then sometimes it can bad [if the paper has errors]. [I] lose time or [I] come with false claims. Errors can propagate."

However, Faculty Member 4 brings up an interesting argument that could attribute some quality to pressurized publishing:

> Faculty Member 4: "Some people say if you publish a lot of papers and half of them provoke arguments, then you are doing the right thing. It's the right level of wrongness in your papers. When you are pushing it right to the limits and people are arguing whether you have done the right thing or not. Some people argue that's a great way to go forwards. Cause then you are provoking a discussion in the community."

In that way, by *provoking discussions* within the community, premature papers may help in the process of truth finding. Quicker publishing also brings newly acquired knowledge *faster to the community* which might be additionally beneficial for the dissemination of knowledge and the research process. Astronomers mention three main reasons for the need of "quicker" publishing. The first is, as described above, publication pressure arising directly from the need for output. The second reason is competition, which we will analyse further in *Chapter 4.2 & 4.3*. The third is *telescope application deadlines*, which are described as "natural" deadlines for a paper to finish. As pointed out in *Chapter 2.3.1*, the "Chicken-and-Egg-Problem" implies that often a paper needs to be finished to be able to show this merit in a telescope application.

> Faculty Member 1: "Okay, well, there are practical things. If there is a telescope deadline coming up. Right? For instance if I want to request Alma observing time, then first thing they are gonna ask is did he get observing time last round and what did he do with it? So it helps a lot to publish your data before the next deadline. It just looks good […] And it's good practice, it leads to a much speedier cycle of observation-publication, gets it out to the community much quicker. I think that's actually not bad …Sometimes … it leads to papers that are a bit half-baked, but even that is okay, because the data is out there and at least people can look at the data.[…] There is nothing wrong. But why would you do that? Well, because there is this deadline [laughing]."

To summarise, the pressure to publish, whether it's because of the need for output, competition or to apply for observation time or other grants may lead to premature publications. Those publications have a high potential to be of bad quality, however they might also be useful to provoke discussions and quicker dissemination of knowledge. To what extent they contribute to prevailing information overload will be analysed in *Chapter 6.3*.

*4.1.3 Replicability*

**Replicability of research results and papers are "a big discussion right now"** (Faculty Member 2). Research in Astronomy becomes more and more based on programming scripts which reduce large datasets to deliver results. It is common to publish details about the dataset and the results, but the so-called "data pipelines" or "reducing techniques" in between are mostly not published. According to the interviewees this is under change:

> Faculty Member 2: "It's happening more and more, but by no means all the code is public, for example. And in [my field] now, we are moving towards … basically making everything publically available so that people can really reproduce what was done. So among the competing projects and what we are doing right now, I think we are actually the most open [group], because we are trying to push this. We are basically trying to provoke the others to do the same. [Interviewer: «laughing»] Which will probably happen at some point, but the debate is ongoing and I think it's going into the direction of being more open."

> Faculty Member 4: "What I have started doing recently is: all my data reduction is documented now. So I now have Jupiter notebooks. I have Python notebooks. […] And then I run everything in a Python notebook, so that anybody can download what I have done. […] With the raw data. Press the go button and now get the output result. So I think that's what [responsible research methods] means. It's demonstrating. […] now you can put everything on a computer. You can do all the data reduction on a computer. And I had been doing that for a long time without being aware that that's what I was doing. […] Because at some point, what you do is when you comment code, or when you write a data reduction procedure, you are not doing it for anybody else. You are doing it for yourself in 6 months' time. Cause you forget.[…] So, I was doing that anyway. But formalizing it by making a declaration of 'I will put this on the web for people to publically look at' really makes you focus and tidy it up. But it's a lot of effort, where you don't get any return for it immediately."

> Interviewer: "And, erm, with this data that is published in the ArXiv, can everybody always replicate the published results?"
> Faculty Member 1: "Yeah, that should be possible. All the data is there, yeah. It is all available for everybody. I think that is the *power of Astronomy*. […] Astronomy is already, you know as science that by its nature relies on archives and data and we need to know about the variability of timescales of tens of years for instance and there it is important that all the data is kept. And the philosophy is that everything is kept as you never know what it's gonna be useful for."

According to Faculty Member 2 "positive peer pressure rather than top-down rules" will encourage people to publish their reducing techniques and ArXiv is a step towards data sharing. Yet, as described in *Chapter 2.3.2*, the current **paper writing style** does not easily enable implementation of code. Additionally, as Faculty Member 4 points out, the current science evaluation system offers no incentive to do so. While astronomers in principle are in favour of replicability, because only replicable results can in the end bring theories "closer to the truth", there is no reward from the assessment system to make one's research results and output more replicable. Even the contrary is the case; the astronomer might find themselves in a situation where *hiding information* might be more advantageous for them with respect of their future research:

> PhD Candidate: "[Publishing code] could be useful for the field, but they don't want to make it public, because they want to have it for themselves to do the science."

> Master Student 2: "And yeah, there is a certain element of 'Yeah, we need to get this out first so we need to not talk about it to other departments and universities until we get this out, and then they can do with it whatever they want.' But I think that's across most sciences, that people are kind of 'This is our data, we'll work with it – and then we'll help!'"

While the astronomer values sound methodology, making *mistakes* is human and happens also in science. However, such mistakes shed a bad light on the astronomer and decrease their market value. As a consequence, replicability is also discouraged, because openness might reveal mistakes.

> Faculty Member 4: "As an observational person you should be able to publish all your data reduction scripts from start to finish. It spits out the output files, which you see in the paper and then somebody else can come along. And I know the reason why, is that, there is a fear that, because you made it easy for other people to check your code, other people can find your bugs more easily and so you may get criticized for having buggy code over somebody who never publishes their code and bugs are hidden for years and years and years. There is no incentive at the moment to publish the code."

The *need for output* also decreases replicability. Astronomers do not have time to thoroughly check their peers' results as that will not lead to a paper, unless you can prove that its content is wrong.

> Postdoc 1: "And to be fair, it's mainly because if you want to have a paper, it has to be something new. Sort of. So you are not going to be publishing, checking that someone else's work is fine. That's not gonna give you a paper. You have to either find that something is wrong on the paper or you have to find the same and something more, right? Like, adding to it. So I don't know how much gets checked. I don't think a lot. But I do think if you read a paper and try to reproduce it, it's not very easy from a paper."

In summary, while replicability is better for science, but it is not necessarily better for one's career. Replicability means openness and openness aids the search for truth. However, the astronomer is career-dependent on what is valued in the science evaluation system and that is output. The need for output decreases replicability in various ways. First, keeping information hidden might lead to future publications. Second, spending time on replicating someone else's result or the sake of checking its correctness is taking away precious time from working on one's own papers. Third, as discussed in the previous section, papers are often written in a rush and no incentive is given for making the paper replicable, hence essential information is left out. This often includes reduction techniques without which results cannot be reproduced. Having no incentive from the system to be more open about one's research process, an astronomer cannot risk their career for replicability. The fear of somebody else discovering mistakes on basis of this openness, only further discourages replicability. The consequence is that most papers are not replicable and a lot of mistakes and code "bugs" stay undetected.

In conclusion, we have observed that output in form of papers and its quality assessments through quantitative indicators such as citation rates and impact factors, defines the value of an astronomer, which is in sharp contrast to the astronomer's own definition of quality. The need for this kind of output to survive on the career ladder ("publish-or-perish") however causes publication pressure, which have psychological effects on the researcher and constitutive effects on the (quality of the) content. The latter may include cutting up publications in order to publish more ("salami slicing"), premature publishing and non-replicable papers. In most cases those effects are described as having a negative effect on research quality. Publication pressure however, can also have positive effects, when it supports the focus and confinement of the research question. Salami slicing can also be beneficial for good communication and readability of research results. We discovered however, that not only psychology, but also Astronomy finds itself in a *replication crisis[3]*. That is partly due to the prevailing outdated paper writing style and partly because of the lack of incentive to publish more information which would make results reproducible. Hiding information which gives a competitive advantage for the next paper and the fear of somebody else discovering one's mistakes additionally prevent transparency. Prematurely published information make output even less readable and reproducible and in most cases involve quality of the content. In summary, because the evaluation system undermines astronomers' values, there is a shift of focus from high quality, robust, replicable and well-communicated research results to a considerable amount of premature results on sexy topics, of which conclusions are neither robust, nor replicable, and written in an intransparent way. As a consequence, an astronomer's motivation also shifts to output orientation where safe and accessible projects become the driver and while "The publication is not the aim. [It is] is a means to showing what your methodology is" [Faculty Member 2] it becomes an aim.

> PhD Candidate: "So yeah, sometimes I see that people are really more driven by the telescope rather by the 'What can we really learn in order to advance our knowledge on a certain topic?' Like from the science perspective, rather than the instrument."

In the sections to come we will investigate further consequences of this output orientation on knowledge production in Astronomy.

---

[3] E.g. https://thenib.com/repeat-after-me

*4.2 Impact on career*

An astronomer aspires to research the truths of the universe (*see Chapter 2.1 & 2.2*) and academia promises to be the environment to do so. Academia's career ladder – the tenure track – has very rigid structure; it starts with doing one PhD, several Postdocs, becoming assistant professor, then associate professor, and finally full professor (e.g. Waaijer, 2016). The timeline for this is very strict, as many grants come with the condition that only a certain amount of years have passed since one has obtained their PhD in order for the person to be able to apply.

As described above, one needs to have established themselves in order to reach tenure. We have seen that this means that first-author papers, impact, citation rates, grants and observing time build the foundations of a successful career in Astronomy (*Chapter 2.3.3*). Due to the Matthew effect (*Chapter 2.3.1*), whether or not the researcher receives a grant (including observation time) can have a substantial impact on the astronomer's career. Grants in turn are dependent on the astronomer's publication rate:

> Faculty Member 1: "And in my case I am ok … I have no problem – my CV is very, very strong and … But I do understand the system of course and I am very aware that my publication rate is important, so I do make sure that I publish enough. Fortunately I have an enormous network, so I collaborate with many people … Faculty Member 1: But those things are important to keep your, to keep your standing – to show that you have made an impact etc etc. Those things are important if you get assessed, right? In research proposals …"

While the interviewees who are in a later career stage describe their career as a fairly relaxed, natural path, that did not involve much planning, they do admit the way how the current science evaluation system turns the climb of the career ladder into a ***"rat race" and "postdoc circus"*** (Faculty Member 2 & Faculty Member 3):

> Faculty Member 3: "[My career was] rather a contrast with students today. I have never had a career plan. And in fact, I have always been offered positions, I never had to apply."

> Faculty Member 2: "Yeah, I think that more and more people are thinking about 'Am I doing something which is going to lead to something publishable. Yes, that's a criterion. And people that don't think about that at all do so at their peril, because people work 2 years on a very interesting project, but if you don't publish you are not going to get your next job as a Postdoc.'

> Faculty Member 2: "So I have the impression that the … career structure for the students and Postdocs is becoming more and more like a rat race, so a little bit more stability would be good. […] Then the phase between that and getting into the permanent position or tenure track positions – that's a very chaotic phase and I have the impression that it's more chaotic than it used to be. Also scaring a lot of people away from even entering. […] Yeah, it's in 2 ways uncertain, because you don't know where you are gonna be in 2 years' time , so it's chaotic, but also you don't know what the result is gonna be. There is no guarantee, far from it. So a slightly more stable career structure is probably something which would benefit the field."

One of the reasons for this rat race is, as described in *Chapter 2.3.1*, the limited amount of financial resources and tenure positions, as compared to the amount of PhDs "produced" by the system. One of the consequences are ***fierce competition and risk aversion***. Interviewees describe that before they have not reached tenure, they are less "free to sort of explore, like not be afraid to fail in all of the projects you do" (PhD Candidate). Choosing a risky research topic that "might fail or doesn't have a guaranteed chance of success" means "you are really taking a chance, erm if you want to stay [in academia]" (PhD Candidate). Taking a step further and changing one's whole line of research is possible in theory, but not necessarily in practice:

> Postdoc 2: "It doesn't mean I should do it [changing research field] … It is true that starting, for example, a completely new research project out of the blue, is a difficult decision, *because you need to make sure that it is a viable decision in terms of your career*. […] That is not going to make you waste time on a project that will not be fruitful in research. […] Erm … and so in a way maybe I feel restricted, erm in that I don't have the time to do erm, just sometimes to look around and say, okay I want to spend, say a month in looking around and see what comes out. For now I have – since I started my position here in Leiden, I have been working on, essentially non-stop, working on the projects that I have started, and erm … Yeah, I feel like I don't have much time to stop and take a step back."

Early career researchers feel ***restricted in their freedom of being explorative***. What consequences this restriction has on the knowledge production in Astronomy will be discussed in *Chapter 4.4*. One also might make their career choices depending on considerations about ***competition***. As outlined in *Chapter 2.3.3*, young astronomers need to acquire ***recognition via first-author papers***, which implies that collaborations where they would be co-author may put them at a disadvantage compared to their peers. In order to gain some advantage in the highly competitive rat race one might choose a niche field.

> Interviewer: "And later on, from when you did postdocs, did you think okay, now I should move towards tenure track or …?"
>
> Faculty Member 4: "Not initially. Because, again, I thought that was something other people did and I wasn't good enough to do it and if you build instruments in Astronomy there are always jobs for you, because instruments need to be build and not many people – it was less competitive – so I recognized early on that if you stay in instrumentation it's far less competitive than just being a pure Astronomy professor."

In what sense competition adds to the knowledge production process will be investigated in *Chapter 4.3*.

All early career interviewees state their passion for research, but are at the same time very well aware of the rat race. Most state that they "***will try to stay in academia*** as long as possible" (PhD Candidate), but they accept that they may have to leave academia. This is not only due to the "publish-or-perish" system, but also due to the fact that the several temporary (postdoc) positions one has to go through involve a lot of travelling, which makes it very difficult to establish a family and to settle down (e.g. Waaijer, 2016).

> Master Student 1: "Because I think in science nowadays it really helps if you have been to several places. If you know people from everywhere. Erm, so I don't think a career is really healthy if you have a Bachelor, Master, PhD at all the same place."

> Postdoc 1: "[…] I don't even know if I will get a tenure to be fair […]. Like right now I am doing it because I love it, but I don't – I think two years ago I think I realised that it's very unlikely. Or not very unlikely, but difficult to get a tenure track position, unless you travel – like you move a lot and I am tired of moving a lot. So I think, I sort of accepted the idea that I might leave the field at some point. […] Which I think is a problem and it's not very nice, that the community – like that you sort of have a goal and then almost at the top, like at the climax of the mountain you realise you are not gonna get to the mountain, right? […] there is a lot of Postdoc there that are frustrated, because you already spend 10, 11, 12 years preparing for this and then you realise *'Well, there is actually not gonna be enough professorships. Sorry, thank you for playing.' So I think it's kind of nasty. But this is the system, right?* There is not many things we can change."

Early career astronomers are not sure for how long they want to compromise their private life and their values for the small chance to reach tenure. When asked what tenure would mean to them, interviewees answered that it would bring "more job security" and with that tenure provides "a safety net" that gives more freedom to choose the research topic and to research risky topics that are not guaranteed to be fruitful (Master Student 2). On the other hand, tenure might bring more pressure in terms of more responsibilities for students, teaching and having to acquire external grants to fund their research group. Being occupied with those duties, tenure may decrease the astronomer's freedom to do science. Hence, early career astronomers face a dilemma: is it worth the fight in the rat race where they are not free to do research which does not lead to publications, to reach tenure, where the duties restrict the time that can be spent on research?

> PhD Candidate: "So I think there is a lot of people in Astronomy, like in academia, they realise that there is problems in the way that publishing is being done now. There are many problems that people recognize as very unhealthy system, but they … It's just so hard because your future really depends on some of these things, like if someone were to make a new way of like publishing paper to … You know if you sort of leave the mainstream journals, if you try to make your own route, then you are really sacrificing your job and your career and, yeah if you really want to … It's hard to – say do you really want to make a difference? Then you can't stay in the field, because you know you'd sacrifice that, or … Yeah, you just stay in the field with its

> problems and so on. […] I think it's a widely recognised problem. Like we have these widely recognised problems but it's hard to actually make a change, because it's so competitive […]"
>
> Interviewer: "Yeah, okay, and erm, in terms of, let's say career steps, does that put some restriction on you in some sense?"
> PhD Candidate: "In terms of career … Yeah, so with the career in general I feel like as I am getting older, I am starting to feel more restricted in where I want to live … Or, yeah, just having the option in choosing where you want to live or the type of research you want to do. Or if you can even do research or if you can even teach, erm, yeah, and so, *I mean, it doesn't feel like you have a say* … What you want to do is just what options are available – that happen to be available at the time."
> Interviewer: "Okay … And, and does that make you feel restricted, or more like, let's say, like relaxed, because you don't have that much influence anyway?"
> PhD Candidate: "[laughing] Doesn't make me relaxed. Yeah, because at some point you might want a family and then people say 'Oh you know, during the PhD it's the best time and so you should really do it.' But then, like *I am doing my PhD now and it's so hard and I just want to be able to establish myself somehow*."
> Interviewer: "So it feels like pressure from all sides? This is expected from you and then people say this …"
> PhD Candidate: 'Mhm! Mhm! Yeah! Yeah! So, it *doesn't really feel like I have the control* … I mean that I know that overall […] overall I can make this decision by myself, but to stay in the academic realm is … It feeels … Like it's sort of restricting you to really explore, or so, the different things that you want to do like outside of academia."
>
> Interviewer: "Is it a factor that you don't want to stay in academia, because you don't want to perceive this publication pressure?"
> Master Student 2: "Yeah, partly, because there is this "publish or perish" thing, where it seems to be like "pump it out", like partly it doesn't … Partly it does seem to matter how good the paper is, but it also seems to matter how many you have out there."

Master and PhD candidates in Astronomy know their fate if they want to stay in Astronomy. They take on the struggle of establishing themselves on the tenure track for the sake of research, but they are questioning whether the whole journey is worth **sacrificing or putting on hold their private life**. They generally feel powerless when it comes to choosing their research and career options and also when it comes to the rigid career ladder. How strict that path up to tenure can be is illustrated by an interviewee:

> Faculty Member 4: "So you said 'Hey, you did your PhD really quickly.' Well, that turned out to turn around and bit me in the bum. Because all the ERC grants say that you can only apply within 10 years of your PhD. […] By the time I came here, it was 5 years later than it should be. And so the obvious stepping-stone grants, which get bigger and bigger – there is a starting ERC [grant], there is a Consolidator [grant], there is an Advanced [grant]. And each level is supposed to be research scientist → assistant professor → associate professor → full professor. And they are all gauged from – they don't gauge you by your job title, they gauge you by how many years after your PhD […] I turned out – I didn't do my research [about career steps] carefully enough, but it really burnt me, coming here. I couldn't apply – there is a grant which is clearly the kind of grant I should be applying for and I can't apply for it. So I had to scrabble around for money in really odd places and I have had to think outside the box […]. I have not had a classic trajectory at all. […] And you just have to think a bit more outside of the golden path, the golden child trajectory. Which is you know, awesome PhD, 1 or 2 Postdocs, preferably fellowships, assistant professor, associate professor, full professor within 15 years. […]: Okay? And the reality is, that actually only a few people hit those marks. There are a few, yes, I would say 10% of astronomers go that course. But in reality it can be a much longer path. People often go on and do other things or get bored and wander off and do something else. […] So I am unusual, I am at the very long end of that distribution. I, I, I got tenure last year. And that's 15 years after I graduated? And it should be 10."

The straight forward career path for an astronomer in academia is what this interviewee calls the ***"golden child trajectory"***, which involves a high achievement of quantitative rates valued by the assessment system (*Chapter 2.3.3*), leading to one career step smoothly following the other. An interviewee points out that this involves "discrimination against age" (Faculty Member 3):

"If you have not made your career money-wise in science before you are 45 or 50, you are out […]. You know, discrimination is against the law, except in the case of age, right?"

While quantitative targets are what is promoted by the assessment system, in reality it also takes some ***luck in the competition***. According to Faculty Member 3, a career in Astronomy is "90% luck and 10% hard work".

Faculty Member 4: "Okay, so I am trying to advise them [the Master students who voice their wish to do a PhD] and see how they are going to do for a few months. If they are really thriving, if they are really pushing me on the work. Then that's the kind of mentality – and even then it's also a bit of luck and timing. I rolled the dice a lot, I applied for – so one thing that is incredibly stressful is applying for tenure track jobs. […] Because it's 200:1."

In summary, it is luck, number of papers, impact rates, citation rates, grants and observing time which build the foundations of a successful career in Astronomy. It is also what sets the foundation for a rat race up the career ladder which leads to tenure. This may put restrictions on the astronomer's private life as the competition for jobs in Astronomy usually involves a lot of travelling, which might make having a family difficult. The impact of the assessment system on the astronomer's career can then be manifold: either the astronomer decides that the sacrificing their private life is not worth the rat race and the astronomer leaves academia. Or the astronomer is willing to compromise on private life and continues the rat race. In that case, during the early career phase, the choice of research fields, lines and topics may be dependent on considerations of career viability. In many cases this involves competition and risk aversion, as we will come back to in *Chapters 4.3 & 4.4*, respectively. Only once tenure is reached the astronomer is free to be more explorative and choose riskier research lines, however still then, as described in *Chapter 2.3.3*, the astronomer needs to meet certain assessment criteria. In addition, because of extra supervision and money acquisition duties, the tenured astronomer might not have much time for actual research. Those are the reasons why the early career interviewees state that they might need to leave academia or will not choose to continue.

## *4.3 Collaboration versus Competition*

As we have explored in previous sections, the requirements of the assessment systems, which are focussed on quantitative outputs, foster competition among astronomers. As described in *Chapter 2.3.3,* competition is introduced when *filters* need to be applied. Because "academia is tough and the success rate for the classic Golden Child Path is very limited, it's 5-10%" (Faculty Member 4), especially early career researchers participate in a highly competitive rat race up the career ladder. Competition and publication pressure are interrelated. The pressure to publish in turn can lead to the publication of premature or salami sliced papers. We have explained what impact this has on the content of research in *Chapter 4.1* and how competition can also influence career choices in *Chapter 4.2*. In this chapter we want to explore how competition and collaboration in general are organised in the scientific field of Astronomy, how the assessment system influences that and what effects collaboration and competition have on knowledge production in Astronomy.

As pointed out in *Chapter 4.1.1*, telescope application deadlines are described as **"natural" deadlines** for a paper to finish. Being granted observation time is not only necessary for observational astronomers to have data to do research on, but is also **prestigious** (*Chapter 2.3.3*). That is why telescopes are generally "always over-subscribed and there is competition for the observing time" (Faculty Member 1).

Hence, deadlines also arise from a competition point of view. Research is driven by the **availability of data** and once data could be gathered from the telescope, the astronomer is given the **priority** to work on those data for one year before the data becomes public (*Chapter 3*).That is the point when competitors can become active in analysing the same data, so one needs to be sure to be done with the research before that. Similarly, if other astronomers work on a similar topic, the goal is to publish first in this race of priority (Merton, 1957).

> Postdoc 2: "Erm, there is the *explicit deadlines*, for example when some of the data I received from a telescope is given to me in *priority*, because I was the investigator of that project. But after a year or 6 months the data become public. And any other astronomer can download them and do the science I want to do and therefore take away my science from my project, so that is in a way a deadline – I need to get the science out before anyone else can do it with the data. And there is this *less explicit deadline*, that is erm … always annoying, that you know that other teams are working on similar projects."

> Faculty Member 2: "So sometimes there is just a *natural deadline* for some project or another. Like for example if there is a telescope application deadline. Or you hear that there is a *competing team* that is about to publish a paper on something very similar. So you decide to 'this needs to go first.'"
>
> Faculty Member 1: "There are some people who want to do exactly the same thing with the same telescope. That is of course head-on competition."

Competing with other teams often imposes an "emergency to publish" (*Chapter 2.3.3*) on the astronomer and **reinforces the pressure to publish**. However, competition for telescope time or competing on the same research topics also opens up the potential for **collaborations with the competitor**.

> Faculty Member 1: "So there is always competition and one way of dealing with it – I mean there are several ways of dealing with it, but one way of dealing with it is collaborating with the competitor."
>
> Interviewer: "So when you have this 'head-on'-competition, would you write to that person and ask 'Hey, we are looking at the same objects, let's just collaborate?'"
> Faculty Member 1: Sometimes. […] Well, I should back up here: The *first thing I am interested in is getting the science done*. Then of course, erm, *money comes in at some point*, because the work needs to be done and somebody needs to hire a PhD candidate for instance or a Postdoc. Well, these things, at least in my experience, usually solve themselves. In the sense, let me give you an example: I have now a research project, using Alma, eeerm, in a very big collaboration, 20-30 people, something like that. Worldwide – it's a very large project and I have a PhD candidate for that. I had that student already. There is so much work to be done, that actually no problem at all to me move him into the collaboration. […] Because if you have money that means that you have also, that you have labour, that you have the effort available. And of course in exchange we need to define a little piece of science that that student can do as part of his PhD. And *that science is important, that science needs to be protected, we don't want competition, you know against poor PhD candidates, who have to write their thesis. That's very normal to ring-fence the science that the PhD candidate does.*"

Because PhDs and Postdocs are the "working horses of the system" (*Chapter 2.3.1*) and supervisors have the "responsibility for their next career steps" (Faculty Member 2), their science needs to be **ring-fenced**. Whether that requirement is met in a potential collaboration is one of the potential criteria for or against collaboration. Hence, whether or not you seek for collaboration with your competitors also depends on whether or not you are already established, because that defines how important the publication is the future career. The ***"goal" of the publication*** for an early career researcher is shifted towards "publish or perish", away from the astronomer's initial values.

> Faculty Member 4: "I think before I got tenure I would have done it by myself. It's … before you have a tenure job, you've got to make an impression and demonstrate that you can produce papers in a reasonably rapid fashion. […] It depends on what your end goal is, if you want to get a publication out and you don't care if you are first author then you probably talk to them, as they have exactly the same point of view – why should they become second author when they worked just as hard as you have? […] So unless you negotiate it out beforehand usually the answer is 'Thanks but no thanks'. […] But it really depends on what the goal is. If the goal is, produce as many first author publications as possible then you go for broke and if you think you are not going to do it on time, you may as well collaborate with somebody and get your name on the paper and then maybe you work with them further down the line."

Hence, an early career astronomer has to evaluate whether a collaboration that does not bring a first author paper damages their career. Tenured astronomers need to make sure, that the collaborations they seek to engage with are beneficial for the early career researcher as well as for their own science and possibly career. We investigate the reasons that speak for a collaboration, and start off with an interesting relationship, that motivates collaborations: the ***relationship between authorship and credit (prestige) versus the relationship between authorship and collaboration.***

Authorship ideally reflects upon the ***contribution*** of the individual researchers in a collaboration.

> Interviewer: "Okay, and are you mostly the last author? Or how does that work?"
> Faculty Member 2: "Well, it depends on – we have that rule, it depends on how much each person contributed to the specific project or to the overall dataset. So we have a system where there is 3 categories of authors. Then there is authors who actually did the work of the paper itself and then there is a category of people that helped to provide the whole dataset. […] And then there is people who helped to motivate the science, that had a smaller role in the paper. […]: So each of those groups are typically alphabetically, so I am usually somewhere in the middle."

Reality however, is more complicated. On the one hand it is not always easy to evaluate the worth of the individual contributions and on the other hand, authorship in the current assessment system becomes a form of ***credit***, a way to grant and acquire recognition. This is because, as outlined in *Chapter 2.3.3*, first-author papers are required to establish a name in the community and are the basis for one's publication rate, and both are determinant for the astronomer's future career.

A simple case is a publication by a PhD or Postdoc, where the role that the individual researchers play in their authorship are rather obvious. Those cases are treated as a "collaboration" between the supervisor and the early career researcher, where the latter becomes first author, because they usually do "all the work" (contribution) and need ***visibility*** (prestige).

> Postdoc 2: "Or the Postdoc, or the [PhD] student themselves. And in that cases there is always a collaboration between the person who does the work and the supervisor. So these people are always listed among the first authors of the paper."

> Faculty Member 2: "So you try to get visibility to the [PhD] students and the Postdocs, who actually are the ones who spend most of the work. Who most of the time do the work, even though we discuss it."

In more complicated cases the role an individual researcher plays in their authorship is less clear and their position as an author *may not reflect at all the value of their contribution*. In many cases, the first author "does all the work" and other "collaborators" contributions are negligible, as these examples illustrate:

> Postdoc 1: "The paper that I have been dragging along for a year, I have 7 co-authors, but I am doing all the work. It's only me! They literally just read it and give me like grammar errors. So it's like arrrrgh. This is the thing: there are collaborators that work very well, and there is collaborators where you have to do all the work and where you only get the 'thanks'-kind-of-thing and the first author does everything. […] So for this paper – I mean, literally we are 7 co-authors. I mean 7 people working on this. I was the one having the idea. So I had the idea, I wrote the proposal, I got the time in the telescope. It was like a survey, so it's a huge thing. I got 52 hours at the telescope.[…] I got the data, I analysed all the data. I wrote all the paper. I did all the analysis, I sent it to the co-authors and then they started saying 'Ah, have you thought about doing this, or what about this, what about this, what about this?' They just gave me more, sort of like expansion and I thought 'Okay, this is good.' And then I did all the things they wanted me to do when I said 'Okay, I am gonna submit in a month, let me know if you have more comments.' And the day before I was submitting, I got comments from my ex-supervisor, saying 'Ah, no, I think you should do this and this and this.' And I got really angry [laughing]. […] So this is the type of collaboration I don't wanna keep doing, because I don't wanna keep doing all the work. Because I don't have time to do – when you also have to teach and you also have to do other [data] reduction and there is no time for this."

> Postdoc 2: "[…] in all the papers that I have been involved, I think the choice of the co-authors have always been motivated by contribution or interest […]. So in general if papers are circulated among a broad audience of people that could be interested and then, there are the people who actually did the work. Which may actually just be one person. And the people who may have supervised the work, or given them the initial idea. The supervisor essentially. And some people who may have … not worked on that specific project, but did some other work, that has been used by this project. […] So I am thinking in particular of the people that make astrophysical images. They spend a lot of time doing that, but in the end they 'just' make an image and we don't publish that. So the way … that the community arranged that *those people get credit* – what they did is that they get invited on all the papers that use this image as co-authors. And then it's up to them if they want to contribute or not, to the actual work."

In large collaborations it is mostly first and possibly the second author who put most of the work into the scientific research. The first author gets the highest visibility, but being listed as a co-author may not be directly benefitting the early career researcher's career, but may also be *valuable* for one's paper record:

> Interviewer: "[…] and if you are 'just' a co-author, does that count in your opinion still, or is that just a way, basically to say 'Thank you' to you and that's it?"
> Postdoc 1: "Depends on where in the co-author list you are, because if you are more like after 4, 5 it's more a 'thanks', yeah! […] In a way [this is also] kind of good. For example, in this paper I am gonna be around 4 or 5 because I only analysed the data, it's not that I did the work. He is doing all the work, so most of the time, right? So I will add it to my paper list and I think it's important, but it's not super important. But for example there is another paper we just publish last … like a week ago, 2 weeks ago, that I am the second author, because we literally, both of us, did all the work – the first and the second author. So this is way more important. So it does depend, where you are [in the list], but I think, being a co-author is not …. I think is necessary as well. I think, is not like you couldn't do all the job all the time."

In summary, workload is usually unequally distributed in collaborations and while authorship aims at reflecting upon individual contributions, in reality this is a complicated state of affairs. This is not only because it is not a straightforward process to determine the value of individual contributions, but also because in the evaluation system authorship is handled as a currency of credit. This has an ***adverse effect*** on knowledge production for two reasons. First, only being first and at most second author brings real visibility and directly benefits the career of that researcher. Second, many astronomers who are at a low-ranked co-author position feel entitled to not contribute much work, precisely because they do not receive much credit for it. They do however still aspire to be a co-author, because, while they do not receive the credit that the first author gets, they can add it "to their paper list", which is picked up by quantitative indicators. Hence, from an incentive perspective, it is ***beneficial for an astronomer to be in collaborations where small contributions earn co-authorship***. Hence, collaborations may be additionally valuable for one's career, also for early career researches, insofar as they produce enough first-author papers in addition.

Other than credit through authorship, we observe the following reasons for an astronomer to collaborate (with a competitor): First, when the astronomer benefits from the ***other astronomer's money or observation time***. Second, an astronomer may chooses to collaborate if the ***"collaboration is fun"*** and beneficial for their own science. Third, large observation facilities require ***large collaborations***.

We start exploring the first reason. Whether or not an astronomer may be granted access to a telescope depends on who paid for the facility:

> Faculty Member 2: "So, sometimes when projects are organized around a facility like a telescope or an instrument then access to that facility is given by the institutes that helped paying for it and then you may have access or you may not have access, depending on if you are in the right institute or in the right group. Other projects, like the survey we are doing is based on a public facility, so you simply apply for the time.

And then the data come and it's up to you to decide who is in the team and who is not.'

If astronomers are "not in the right group" they may seek collaboration with that group to benefit from the access to the facility.

Faculty Member 4: "The cynical answer is: If I don't have access to a telescope, which I need access to, I'll see if one of my friends have access to it and ask 'Hey, do you want to work together on this?' Is bluntly the way to do it when you are resource limited."

Without (monetary) resources an astronomer is limited in their research and may not be able to afford PhDs or Postdocs. Faculty Member 2 describes that the Sterrewacht is an especially "collegial institute" where a colleague with a large personal grant helps another one out by an invitation to collaborate:

Faculty Member 2: "In principle it's [a] personal [grant]. Although, it's also a very collegial institute with natural collaborations where people sometimes say 'Okay, let's do a project together, I have money for the student.'[…] *So the research follows the money a little bit.* In the sense of, where the resources are, that's where you can do things. […] So it's really quite constructive the way we collaborate and build something together. Which is actually something special in that institute, I think."

Hence, collaborations may **benefit the research work by distributing resources efficiently**. Those resources may include different skills sets, for example collaborations between observational astronomers and theorists:

Faculty Member 4: "But then if somebody is an excellent theorist and knows how to run a computer model you ask them, you just email them and if you know them you just cold-call them. And you say 'Hi, I am interested in X'. And that's the purpose of meetings, yeah, you can't do it over Skype, if you go out for coffee in the evening of a conference. You can gauge whether there is somebody you can work with. You work out their personality very quickly.'"

Next to personal contacts from previous job appointments, **conferences** are good occasion to find collaborators. The reason for that is that they open up "natural" environments for astronomers to exchange their research work in formal and informal settings. The interviewees put an emphasis on the importance of getting along well with the collaborator and conferences give the opportunity to "gauge" whether collaborators "match".

Faculty Member 1: "Well, I do say no to collaborations sometimes when I have no confidence in the ability of the person. Faculty Member 1: You know, a collaboration has to be fun, if you don't like the people, then there is not much fun."

Not only does such a "match" support the second reason for astronomers to collaborate – enjoyment of the collaborative work –, but is also important for the collaborators to trust each other that they do not *secretly compete*:

> Faculty Member 2: "So I have done this [collaborations] with some smaller projects I think. Depends a bit on whether you know them [the collaborators], whether you have history with them. With some people you know it wouldn't be a good idea, because some people have a reputation of being a shark and if you tell them you are working on something they would just speed up. So sometimes it's better to just be quiet."

A trustful collaboration may not only be "fun", but also benefit the research work in the sense that scientific discussions can speed up the research process and may lead to more robust results and better quality papers:

> Faculty Member 1: "Now I personally like a lot to collaborate with people. I prefer that very much to working in isolation on my own. *It's much more fun and science is discussion*, so it's much more fruitful and you get more things done."

> Postdoc 1: "The paper [with the 7 co-authors] is going to be better after this [long process of collaborators revising the paper], which is the main point."

A Postdoc describes their motivation for engaging in collaborations from the view point of *curiosity*:

> Postdoc 1: "[Collaborations that] I ended up in, because I knew how to analyse the data for example. And they – there was sort of, I started to talk to people and they were like "Oh, can you check if there is something in the data?". And I checked, and there was something and I was like "Wow, this is really cool". This sort of things, just random stuff. I had at least 2 or 3 collaborations that started like that and, and for example yesterday […] I spent almost the whole afternoon just trying to analyse this. Because I was just like "This is really cool!". It wasn't … It's not my paper – all of this stuff I am not the first author, but these are the sort of things that I think are really cool."

Even though this collaboration is not directly beneficial for the career an early career researcher may take the risk of prioritising their values over the assessment system's values.

The third reason for astronomers to collaborate are the requirement of large collaborations to build large observing facilities required to perform research beyond the earth's atmosphere and to analyse their output. As this endeavour requires not only remarkable technological abilities but also considerable amounts of money and extensive labour to conduct space missions and build instruments (mainly telescopes) with which one can observe the universe, collaboration is needed. This collaboration ranges from a political level (acquiring funding) to a scientific level (sharing the output). Such collaborations may *restrict the astronomer's freedom to explore*, as whatever they use the instrument for needs to benefit all collaborators.

> Postdoc 1: "So in this case, for example, we have to find targets, that everyone can do their own things [with]. So in this case it will be restricted in a way, right? But it's the way it works. Because for JWST you are not going to get a proposal from one co-author saying 'Oh, I wanna observe this target, it's very interesting for my own research.' Cause for satellites you have to have a big collaboration. You have to have a big impact, so it has to be many may people."

In summary, collaborations may benefit knowledge production in Astronomy when it is curiosity driven or when resources in terms of money, observation times or skills are efficiently distributed. Collaborations can be the breeding ground for fruitful scientific discussions and high quality papers. While large collaborations make it possible for many astronomers to share resources, they may not be able to be free in their research.

Competition, on the other hand, may also benefit the knowledge production process. Some interviewees describe that competition can be *"healthy"*:

> Faculty Member 1: "Which is also healthy, as long as it's, you know, kind of constructive competition. […] Yeah, to stimulate."

If competition drives the science and "pushes knowledge forward", so if it agrees with the astronomer's values, it is seen as "healthy" and beneficial for the knowledge production process. An example where an interviewee competes with other groups on the same project illustrates that this process can be healthy as the researchers can learn from each other, because they use different data and methodology. On the one hand, they are driven to find a "spectacular result" first, but on the other hand the other groups will discover if they make a mistake due to the rush. So in that sense the research process is speeded up, but at the same time controlled for mistakes. Results have more credibility and are more robust as a consequence.

> Faculty Member 2: "So you try to hide a little bit in detail, but you learn from each other, and in the end … the idea is, it's very difficult measurements, so we all feel it's very valuable to have different people doing the same thing. If we disagree then, you know, we all have to sort something out. And if all agree then it gives more credibility"
> Interviewer: "But then it's only the first one to be able to publish the thing you agree on?" Faculty Member 2: "No, you can say, I mean all the datasets are different and the calibration strategies are different."
> Interviewer: "So you could all publish?"
> Faculty Member 2: "We can all publish of course. You'd try to be first if there is something spectacular. But you also try to be correct [Interviewer: laughing], so you can't rush too much."
> Interviewer: "And how do you feel about this competition? Does it stress you in a good way, in a bad way? Do you think it's fostering research or do you it's stressing?"
> Faculty Member 2: "Usually it's good. In this particular case it's good. Both, because it gives credibility also to people outside this particular [describes area of research], where they agree mostly, and disagree on some things, which we need to sort out and in the end we will all be better for knowing what the answer is. If there had just been one measurement and that was it, then somebody's mistakes would just stay …"

It is an interesting observation, that despite the prevailing pressure on especially early career researchers and the needs of "filters" which puts them in competition, young astronomers generally perceive Astronomy as a less competitive field than others.

> Master Student 2: "Astronomy is not one that I have seen massive levels of competitiveness between institutions, but that might just be because I haven't been introduced to it. I haven't been high enough in a level to kind of see people hoarding things and not talking about things. Maybe that's just because I haven't dealt with it ... Because I know, there was a conversation we had with another side, where they were giving out about different institutes not talking about other institutes, because they were trying to publish things and stuff. But that wasn't in Astronomy. Astronomy seems to be quite open in dialogue with each other."

In this section we have explored in what relation competition and collaboration stand in Astronomy, how the evaluation system influences that relationship and how collaboration and competition influence knowledge production in Astronomy.

The requirement for credit may lead to competition, as elaborated on in previous sections, where we discovered that competition and publication pressure are interrelated (*Chapter 4.1*) and described consequences of that pressure, which include output orientation, premature publishing, salami slicing, keeping information closed off and so forth.

However, in this section we made an interesting discovery: collaboration (with the competitor) may be beneficial for the astronomer's career and science. The need for (prestigious) observation time or resources in terms of money or human skills may lead to collaboration that are mutually beneficial. Those benefits may include an exchange of resources and credit through co-authorship, which often requires minimal contribution in terms of workload. Collaboration benefits research quality when they are "fun" and driven by curiosity, when they spark scientific discussions, when they speed up the research process and when they add to the robustness of results and better quality papers. Collaborations may distribute resources efficiently in a way that supports those benefits for pushing knowledge forward.

Similarly, competition adds to scientific quality when it is "healthy", which means that it can be seen as a driving factor and stimulator where several competing groups have a competitive interest to cross-check each other's results. Hence this kind of competition can be a motivator in pushing knowledge forward and at the same time leads to more robust results due to cross-checking.

In summary, the requirement for credit to build (on) one's career can support both, competition and collaboration.

> Postdoc 1: "I think you have to have a little bit of both, otherwise you never finish anything."

One needs to be output oriented to survive in the assessment system, but collaboration might benefit that, as co-authorship brings credit without "doing all the job all the time". Hence, while collaboration might spring from a purely curiosity driven motivation, it can also be seen as a *trademarket for resources and credit*.

*4.4 Relation between riskiness, failure & negative results*

In previous sections we have explored that the need for impact and publishable results, required to build one's career and acquire future funding, makes an astronomer averse to risk taking. We have already observed in *Chapter 4.1*, that *focus on sexy topics and "playing it safe"* means that "it's like a good thing to do [to come] up with a project where you know that the outcome of this will lead to a paper." (PhD Candidate) An exception are "high risk – high gain" projects (*Chapter 2.3.3*) that promise high-impact and a career boost. However, choosing a risky research topic that "might fail or doesn't have a guaranteed chance of success" means "you are really taking a chance, if you want to stay [in academia]" (PhD Candidate). "Risky" by definition means open to failure. If failure is not accommodated for in the assessment system, the researcher will try to avoid failure and anything that can lead to it like risky research. To what extent that happens and what this implies for the quality of research in Astronomy is subject of this section. In that discourse we will question what failure actually means for an astronomer and how that definition relates to the requirements of the evaluation system. We will discover that failure in Astronomy has an interesting relationship with negative results. We will explore how astronomers deal with failure and what that means for knowledge production in Astronomy. Let us first turn to the question, what *failed research* means to the astronomer:

What all interviewees agree on is that, to them, research is failed, if it *"doesn't add anything new"*:

> Faculty Member 4: "Failed research? Anything were you are rehashing over an old result. So if your piece of research adds nothing new. And by that I mean *literally nothing new*; you have used the same methodology. You have used the same dataset and you are drawing exactly the same conclusions, then that's failed research."
> Interviewer: "Is that not replication in a good way? To see if research is done correctly?"
> Faculty Member 4: "No, I'd say … But then you'd use a different methodology. If you run the same piece of code on the same data, that's not doing anything new. If you write a new data reduction routine, in a different language with different tests, that's useful! […] A paper *has to advance knowledge*. And advancing knowledge *also includes confirming a contentious result*."

This can be explained through the astronomer's values: high quality research pushes knowledge forward. It is the communication of new insights to the community. New insights can also be gained through *negative results*, such as non-detections, which are the most prominent example of such "null-results" (Faculty Member 2) in Astronomy. Hence they are not necessarily failure. Consistent with the astronomer's values is also that another aspect of failure entails mathematical or methodological errors. Premature publications (*Chapter 4.1.2*) are an example of a research piece, which have a high potential to *fail in providing sound methodology*.

> PhD Candidate: "So when you think that you have reached a conclusion, but you haven't really taken the time to *understand all of the systematics* that are involved and then you want to publish the results – I think that's like *premature publishing* […] And that's like an *example of failed research*. [Giving an example:] They didn't understand all of the systematics of the instruments, but still trying to reach conclusions, based on the data that they observed. […] You know, you are [re-]plotting things and they don't make sense. Just because, maybe this thing is dominated by a systematic that you haven't accounted for. […] they just move on and they are still trying to like extrapolate some things based on that plot. […] if you are trying to publish something that you *recognize can have flaws in it or where there might be some mistakes*, but then you publish it anyway. I think that would be an example of like *failing to do good research*."

> Faculty Member 1: "Failed research? … That's an interesting [question]! Well, you can think of different kinds of failure. If I try to observe something and don't detect it, that can still be an interesting result. *Non-detection can be interesting*. So that is not necessarily failure. *It is failure if I made a mistake in my calculation*, when it turns out that I tried to do something that was totally impossible. […] Failed research … it's very hard to define failure here, because Astronomy, at least the way I do it, is very observation driven science and if you do your observations right then there is always interesting data in the archive that will be interesting for somebody. […] So I would define failed research as research that goes off in … that takes good data, but goes off in a direction of interpretation that is *very artificial or not supported my empirical evidence*. […] Eeerm, in the sense this is like … You know Astronomy is a bit like the journeys of discovery like centuries ago. It doesn't matter where you go as long as it's *new territory*, it's *always interesting*, right? And it's always useful at least. And we are *filling in the blank parts on the map*. That can be uninteresting research, but that is much more a matter of taste."

Another interviewee draws the distinction between "failed" and **"bad" research**. The latter amounts to failing to account for sound methodology, while finding a real definition for "failed" research is trickier, given the nature of research. As we discussed in *Chapter 3*, research is the endeavour to discover the unknown and us such always the "discovery of new territory" (Faculty Member 1), hence research can hardly fail with respect to this venture of "filling in the blank gaps" (Faculty Member 1).

> Master Student 2: "Erm, so yeah, I am not really sure, that I would have a definition of failed research. Bad research is different, but I am … Failed research?"
> Interviewer: "Okay, what's bad research?"
> Master Student 2: "I mean, you know, depending on what you are doing, if you are *not following the correct steps*, or you are *lying about results*, or you are *not fully checking results*, or you are not maybe taking certain *assumptions* that you have made into

> account, […] Like all, because if you are drawing conclusions from something that you – where the methods are not sound – then they are *not necessarily correct* conclusions to draw. […] *But failed research just sounds really harsh …* [laughing] … to me".
> Interviewer: "Why harsh?"
> Master Student 2: "I don't know, because it just feels, if you spend like 4 years of your life doing something and then somebody tells you it was failed research, then that would just be quite cruel, you know, you have just wasted 4 years of your life and have nothing to show for it."
> Interviewer: "Is it maybe because research by definition is quite uncertain?"
> Master Student 2: "I think that's it, yeah! Research is – there is no ultimate goal, there is no kind of, you know absolute research [both: laughing], so I don't know how it could be failed … Yeah, you know, it sounds harsh to me."

Faculty Member 3 on the other hand emphasises that in the case of mistakes "research may be failed because it never sees the day of light", but draws the distinction to wasted research, where "research is never wasted, because it could acquire a second life sometime in the future":

> "He told me and he showed that [mathematically] it's wrong, but said 'Don't throw it away – put it into a drawer'. And that was very good advice, because maybe the idea you had, maybe the mathematics you developed for it , maybe the observations you were trying to explain will acquire a second life sometime in the future. *Nothing you ever do is wasted, but it may have failed*, alright? So your line of research your idea, your mechanism, whatever it is you are after, did not work out. It doesn't fit the facts, it's mathematically inconsistent. All sorts of things that can go wrong. That's failure. […]. Erm, and failure is not the same as not reaching a particular goal, right?"

While the interviewees share their opinions on the value of sound methodology and pushing knowledge forward, they are **less in agreement** on what "something new" actually means and when a piece of research is useful for the community. As Faculty Member 1 stated "What is interesting is a matter of taste", also translates to the question, what outcome of research counts as interesting enough to be valid as "something new".

> Faculty Member 2: "And so, at some point you give up and say 'There is something wrong with this dataset, but we can't figure it out'. And so you may have learned a lot about methodology and this particular data, but there is *no result* in there. […] So that's *failed in the sense that it doesn't yield the publication* that you were hoping for when you started out."

In this example even though the researcher finds out "something new" about methodology and a dataset, it does not count, as no new results have arisen out of this enlightenment. Faculty Member 2 goes even further and entitles this research as failed as it is **not worthy of getting published**.

By contrast, some interviewees emphasise the importance of writing up such pieces of research, precisely because it has failed. This adds to new knowledge in the sense that other researchers know how not to do it. Not publishing such insights poses the danger of ***reinventing the wheel*** which is a "waste of the earth's" resources:

> Interviewer: "Okay, yeah, I see. So hitting a dead end is not failed research?"
> Faculty Member 4: "Correct – no, not at all! It failed and if you don't write it up nobody will know! And that sucks, because somebody might accidentally do exactly what you have done and it's wasted the earth's resources. […] You have just wasted somebody's brain power and somebody's life. The one thing you can't buy is hours. Right? The one thing you can't get more of is the time in your life. And it sucks if somebody says 'Ah, yeah, I knew that 10 years ago, but I didn't bother to write it up!'"

Hence, we observe a discrepancy between the definition of failed research that arises out of the evaluation system ("fail to publish") and the astronomer's definition which is based on their intrinsic values as discussed in *Chapter 2.2*. We observe that this discrepancy causes a ***shift in what counts as new discoveries in the community, from "anything new" to "publishable results"***. This in turn fosters output orientation, away from risky research and away from negative results.

> Postdoc 2: "Hmm, that's true, what I define as failed research and what erm, and what is considered as failed research by the community … are 2 different things."
> Interviewer: "Okay, please tell me both."
> Postdoc 2: "Okay, so what I just told you, so it's essentially negative results are considered as failed research by the community. […] And on that side I disagree. […] So there is always information to be taken from research that is well conducted. Given that the research is using state of the art data, and state of the art methodology, whatever the result is, should be interesting, because that is, that is a research project that I could be willing to do myself."

Because non-detections are the most prominent example for negative results in Astronomy and, especially for observational astronomers, are actually more common than positive results, we are dedicating the following paragraphs particularly on the value and "publishability" of negative results.

Most interviewees are convinced that negative results should "absolutely be publishable" (Faculty Member 4). This is because they are seen as valuable with respect to ***new knowledge about what does not work***. As research is the discovery of the unknown, this kind of information is also essential, "because it either can help [the researchers] discount certain theories, or help them kind of support other theories" (Master Student 2).

> Master Student 2: "[…] I mean even not getting a result doesn't necessarily make it failed research if … because you know, if different methods are not applicable, or getting you a result, can usually tell you something else about what's going on."

> PhD Candidate: "And you can still try [to publish], like when you can extract something from it. Like you can learn something new even from the non-detection. […] Yeah, I guess like, yeah, I guess it really depends on what's being done … Because it could be … a viable method 'It doesn't work for these reasons' and there you could learn something. […] Yeah, I would publish it, but also […] I would also discuss 'What are the implications of that?' So that you can learn something […]."

Other interviewees emphasise however, that negative results are only useful if such *"implications"* can actually be made.

> Master Student 1: "So of course a negative result, so if you try something and it doesn't work, then that's not failed research. […] That should also be publishable. […] But if you don't have anything concretely that you can say [about why it doesn't work], then it's, yeah I am not sure whether you have added something [new]."

> Master Student 1: "[…] We didn't manage to deduce the error in the radial velocity. And it's actually *publishable if we know why that is*, why it doesn't help. Otherwise there is always 'Look, we have this method for reducing errors in radial velocity measurement, but it doesn't work in this example and we don't know why' That's a little bit of a weird paper."

In the case of non-detections, the most prominent implications one can make to make their results useful is providing upper sensitivity limits and cross sections for what one wanted to measure. In that case a negative result becomes a *"null"-result*.

> Faculty Member 2: "Well, you can report this [dead end], but I am not sure, whether it's useful. If you learn something about this particular method, [that it] suffers from problems, which means that you will never be able to use it for 'Blah', then yes that's a useful thing to say, but if it's 'I went to a telescopes, I observed for 10 nights, I came back with data and they weren't good enough' – I don't see why you should publish that. […] So there is a difference. It's not like you say 'Okay, I was trying to measure some signal due to dark matter and I didn't find it'. That's a Null-result or an upper limit. That's very much worth publishing. So it's *whether it's a methodology or a data [problem]. I think that's the big difference*. If the data aren't up to it, then it's not really worthwhile."
> Interviewer: "Right, but I mean, with your example, the data just doesn't yield anything that can prove …"
> Faculty Member 2: "Yeah, but that can be maybe, because there are subtle technical issues in the telescope which means that you cannot trust the images and then it's … Then it's something about the telescope. And in a technical journal about that particular telescope you can publish. If you then actually find the cause of it and analyse it."
> Faculty Member 2: "But then usually, if it's for dark matter, you just forget about the telescope and go with a different telescope. […] So sometimes you have to be pragmatic about what to … *What's going to get you to your goal*. […] No, but that's *[upper limits and cross sections] very valuable*. So that's the *difference between a*

> *Null-result and not being able to get a measurement.*
> Interviewer: "Right, so as long as you can publish about that and can say 'Look, we tried this method and it didn't work.' – "
> Faculty Member 2: "Yes, and I wouldn't say that the emphasis is on being able to publish. The point is that you are hunting for something and make sense of it …"

While Faculty Member 2 emphasises that the goal of research is not to publish, but search for truth and "make sense" of it, as we described in earlier sections, those goals can be easily confused, when one's career is dependent on publishing. As we saw above, the individual astronomers have a different view on failed research and negative results than the system has, which also causes a shift in the community's perception. Because individual astronomers feel judged by the system and the community, they can experience an ***inner conflict***. Even though high quality research is what motivates an astronomer to do science to begin with (*Chapter 2.1*), they might feel failed if they do not "reach the goal" of the publication despite having done proper research work.

> Master Student 2: "Well yes, but that's, again, that's maybe a personal kind of, that I haven't maybe managed as far with it as I could have, or, erm, I might feel like I failed in it, because I didn't necessarily get results. But I know that that doesn't mean that it's failed research, I just feel that I have failed. Yeah you know … but that's because, if there is a goal set and I don't reach it, then that feels like a failure, but I wouldn't say that, to me the research is a failure. More that I have failed whatever the set goal was, if that makes sense?"

On the other hand, even if the publication is ensured by for example providing upper limits, due to their inner conflict of what the goal is, the researcher may still feel ***unworthy*** if they fail in making a detection.

> PhD Candidate: "Non-detections. [...] It's just really hard to work with the telescope and I really want to be able to figure it out and do this thing and I think personally I would feel failed if I wouldn't be able to at least … put some limit, that gives a good sort of low sensitivity to it. PhD Candidate: But, erm, yeah, so if I don't make detections and I just report that I have this lower limit on these objects, I think that can still be valuable and there is still science that we can learn from that. Erm, but personally I would feel a bit failed."

We observe that ***publishability of negative results is a controversial topic*** among the interviewees. While, as described above, upper limits may make the non-detection useful and publishable, some astronomers think, there should be a "journal for non-detections" regardless of whether implications could be found. As some observational astronomers "have way more non-detections than detections, so in a sense [publications on those] would be a lot more relevant" (Postdoc 1):

> Postdoc 1: "I mean when I was at [the famous institution] we said, we should start a journal on non-detections. [Both: laughing» We should really do a journal on non-detections! Because I am really sure that there are people that have been observing the same objects on and on and in without knowing that other people have already done this. Because nobody published when they don't detect anything."

Hence, while astronomers display *a general wish for publishability* of failed research and negative results to avoid reinventing the wheel, in reality their publishability depends on the subfield, whether or not implications can be drawn and on the relevance (sexiness) of the topic. Dark matter is a subfield that is "sexy" enough, such that non-detections which are part of the daily agenda in that field, just like Postdoc 1's, can be published easily while that is different for Postdoc 1's subfield. Another issue when publishing negative results is that the quality of that research piece cannot be assessed easily (Postdoc 2). Does the fact that nothing could be found mean that the research was performed incorrectly, or does it lead to a new insight of what does not work? Hence, it is much more difficult to write a high quality paper on negative results than on positive ones. In addition, the community is less interested in reading about negative results, unless they are *surprising*, and hence these will not add to one's citation rates, nor bring prestige. The following quote summarises the publishability dilemma of negative results and draws the connection between riskiness, failure and negative results.

> Postdoc 2: "One of the projects that was advertised in this application was erm, perhaps risky, because it was difficult to evaluate the results that come out of it – would it have any impact, or would it just reproduce what has been done before? […] and one of the referees of the application criticised this – it's difficult. It's true, but I also think it's not a good way to assess research, because research should be risky and sometimes I feel that in general in the community, that failure is not very well accepted. […] It should be that publishing … Well, it should be that doing research that leads to no conclusion is still research, but the way the publication system is, you have to publish a result, a positive result essentially. And while journals will – I think – if you do your job properly – will accept to publish a paper, in which you say 'We applied this methodology, we have not found a result' – People will not be interested in these papers and they will not benefit you as a researcher."

We can conclude that, while research is risky by nature, which implies that positive results are not guaranteed, researchers generally do not feel negative results are as valuable for a publication or their career. The result is that the astronomer feels that the community regards negative results as failure and hence may feel devalued in their worth as a researcher when they cannot show positive results. Because negative results are then perceived as having a negative impact on one's research life, risk aversion is the consequence. *"Playing it safe"* is a consequence of an assessment system, where the pressure to publish does not account for uncertainty, which is so intrinsic to research. As we saw in *Chapter 4.1*, this is "damaging" knowledge production in many respects. As astronomers feel restricted in their freedom to be explorative, creativity and "out-of-the-box thinking" are discouraged. But "Creativity is what in essence research is all about." (Faculty Member 2). If that is taken away novel and innovative ideas to discover "the truth" are not giving any breeding ground to grow. What

Pallota[4] says about charities is all the more true about research: When the evaluation system prohibits uncertain outcomes of research projects which may include failure, ***innovation is "killed"***, which means "you can't possibly solve large problems [in Astronomy]".

Some astronomers find their way out of this dilemma by employing a ***"mix of safe and risky projects"***, the riskier the better for the reputation:

> Faculty Member 1: "You have to do a mixture of those things! If you play it safe all the time that would be uninteresting. But you have to do the right mixture, if you only do very risky stuff, then most of the time you won't detect anything and that's gonna be very boring, that's gonna be bad for you …"
> Interviewer: "Yes? In terms of grants?"
> Faculty Member 1: "Yes, you have to show judgement. […] And a mixture of safe, solid, but still interesting stuff and some high risk – high gain projects."
>
> Faculty Member 4: "[…] it's a safety net and I am very cautious in that I always have a backup plan. If something falls through, there will be something which I can back up to. And that's true in research as well. I don't have one main research project. I have two main research projects. Because with research one thing can just fall flat, through no fault of your own. Something can just stop working or a telescope breaks down."

However, as elaborated in *Chapter 4.2*, Faculty Member 4 emphasises that an astronomer who has not reached tenure has to think twice about whether they can afford (such a distribution of) risks:

> Faculty Member 4: "So, yeah, I have the freedom to do that now [take on riskier projects]. And I probably wouldn't have done it 2 years ago when I was waiting for tenure. And, so people are averse to do risky projects. I think I still would have done it, because it's too good to miss. I am a real believer in going for it. If you are … If you thought about it, if you slept on it for a few nights and you are still waking up in the morning and still excited about it then 'You know what? Go for broke!' I'd be kicking myself. I don't want to be sitting on my death bed […] going: 'Ohhhh I wish I had done X.' So this could probably fall through. It could still blow up and fail horribly, but at least I gave it a go. I wouldn't be able to sleep at night … […] Yeah, absolutely, you have to [play it safe]… if you have enough time you do a bit of both. And this is me with plan B. I always do a risky project, and then I do a safe project."

Master students, nevertheless, might be an exception from that rule due to the fact that they do not have to publish yet.

> Master Student 1: "So that [proposed project] was more of a certainty project. [The other project] would be very nice, if it works, but it's also very risky. So then I chose the risky thing."
> Interviewer: "Alright, okay, okay! And why did you choose the risky thing over the

---

[4] Dan Pallotta addresses the relationship between risk, failure and innovation in his TED-talk "The way we think about charity is dead wrong".

certain thing?"

Master Student 1: "Erm, because I think it was more exciting."

To summarise this chapter, we have explored the astronomer's definition of "failure" and "negative results" and how this relates to riskiness and the pressure to publish. Failure to an individual astronomer is equivalent to "bad" research and is defined in relation with mathematical or methodological mistakes, hence a violation of the astronomer's quality criterion. Failed research can also mean a low quality paper that does not communicate "anything new" to the community. Premature publications have a high potential of being of low quality. Negative results, on the other hand, are not considered as failed research by the individual astronomer. Because Astronomy contributes to basic research which can lead to a deadlock, negative results are a natural occurrence and, as long as knowledge is pushed forward (an astronomer's quality criterion), they should be publishable in the astronomer's opinion to avoid the risk of reinventing the wheel. However, their publishability is a controversial topic among the interviewees. That is because, whether or not negative results are "valuable" for publishing depends on the subfield, the sexiness of the topic, whether the result is surprising and whether implications in form of upper limits and cross sections can be drawn. Hence, a project is perceived as "risky" when it has the potential to fail to get published, which is not one of the astronomer's quality criterions. Because publishability of negative results does correspond with the astronomer's definition of quality, the astronomer may experience inner conflict about the goal of research and feelings of unworthiness. The community may adopt a new definition of failure which is related to publishability. Risk aversion is a consequence, which has a "damaging" effect on research insofar that there is no place for innovation and novel ideas. At best a compromise of a mix between safe and risky topics can be found. However, especially, early career researchers who have not reached tenure need to be careful with risk taking, which can harm their career. If they do not also have safe projects and are too dependent on the risky ones, they may be tempted to commit fraud, which we will explore in the next section.

## *4.5 Effects on Integrity*
## *(– Brutal honesty versus "local beer house")*

As we saw in *Chapter 2.1*, integrity is implied by need for clear and verifiable methods that constitute high quality research. Integrity and the quest for truth go hand in hand, hence astronomers see integrity as an ***"absolute" value***. As Faculty Member 1 quoted "I mean we are all, or should all be at least, in the service of scientific truth", he continues:

> Faculty Member 1: "And that includes of course the way you do things and the way you answer questions, including scientific integrity, which I think should never be an issue, because it should be at the forefront of everybody's mode of working, but apparently it is necessary to remind people sometimes of that, sadly."

That means that the opposite of integrity, such as fraud or corruption is the ultimate opposite of scientific quality. As we found in the previous section, "lying about results" is regarded as ***failed research*** by an astronomer. Fraud ***damages the knowledge production process*** as it is a setback for the search for the truth. The study of the effects of the evaluation system on knowledge production in Astronomy would hence not be complete without an investigation whether corruption is one of those effects.

While all interviewees agree that integrity is fundamental to research and that this is an absolute value, even constant over time (*see Chapter 2.2*), they do admit that "the pressure maybe means that people […] push the envelope maybe a little bit", which can go as far as fraud.

> Faculty Member 3: "The situation for young people, particularly in getting published, in getting funding, making a career […] is terrible, it's really very, very hard. Now they are *made to compete in unscientific ways*. They are made to compete in terms of personality, they are made to compete in terms of gender. Erm, and all sorts of things, that are basically unscientific. It is not a surprise then that they respond by the usual unscientific ways that we have in society. *You lie, you cheat, you pretend, you bluff*, okay? You do all the things you might do in the *local beer house*, so to speak, okay? […] If you demand that attitude of your students and your graduate students in the competition, you will teach them the wrong things. And I think that young people are definitely being taught the wrong things in that respect … I believe – really coming back to this – I am not saying that we are some sort of priesthood in which you are supposed to be holy. Okay? I am not even going to begin to tell you the unholy things that I have done in my life, some of which I am badly sorry. But that's a different story. This small part of one's existence – in the sciences – that is sth where you must be honest. […] And also of course because then, automatically, you are responsible to society, the people who pay you. And the people who trust you. Because it is easy to lose trust and it is very, very difficult to gain trust, so once you are caught cheating, it's basically all over. You know, society will have the tendency to say 'Ah, sciences are just an opinion.' We earn our respect and we gain our trust, by being scrupulously honest. Okay, no just simply because we have to, but because that is the trade, that is the way how science works."

As we saw in the previous chapter, non-detections often cause the astronomer to be scared that those negative results will not be publishable. An inner conflict of what the goal is, may cause a feeling of unworthiness and hence a need to gain recognition in the community, which can lead to fraud. Especially when an astronomer relies on only risky projects, with no sight of a possible publication, the boundary between integrity and corruption might be overstepped.

> Faculty Member 4: "Some people go for just the risky projects. And it makes me bite my nails and fear for them. […] Because they … That's when you get stressed and you make it – That's the point when people tend to fake results."

Faculty Member 4 continues with explaining that the importance of publishing blinds the corrupt researcher in a way that they can justify fraud to themselves, while they are still in the best hope for the results to turn out to be right in the end.

> Faculty Member 4: "And I think you see it in pharmaceuticals – the stuff where there is money on the line you see people stressing out and they think 'What if I just fake this result? And you know, I am sure it's right! I am sure it's right! 'You know, the next research grant I get' … And that's where you get. … I think, part of the reason is you get people are so … It has to work! It has to be an amazing result. And people get into that train of thought and that's deadly, cause some things just don't work."

Faculty Member 4 uses the pharmaceutical industry here as an example and not Astronomy as generally the interviewees don't feel that fraud and corruption are as big of an issue in Astronomy than other sciences, where **"there is a lot of money on the line"**, paid by someone with a particular interest.

> Postdoc 2: "Maybe [responsible research methods] a term I would most likely see in erm, say other sciences, not necessarily Astrophysics, but perhaps more social sciences, or medical sciences, where erm, researchers can be pushed towards reaching one conclusion rather than another, given who is paying for the research. And in that sense using responsible methods would mean that they use biased indicators to make sure that regardless who is paying for the research, to make sure that the results are what the data, what nature is telling them."

Interviewees also emphasise, that "cheating" in Astronomy is rather difficult, given the comparative "openness" of raw data and papers published in the freely accessible ArXiv.

> Faculty Member 1: "And that, let's be fair, not everybody plays by the rules in that sense, so you have to be responsible in the sense of … being honest, first of all, of course, but also, and that is easy for Astronomy – your data needs to be available for other people to check. […] And, but, so the things, as far as we know, reliability and not faking data is concerned, I think in Astronomy we are in an extremely good shape. Because all data is archived and it's publically available for everybody and that makes cheating very, very difficult. You can always argue about interpretation, but the data is always there."

> Faculty Member 4: "I think the publication system is pretty good [in the sense that] I think, by large we catch all the bullshitters. […] *It's relatively easy to cross-check and validate*.

That "***cross-checking*** and validating is relatively easy" is one side of the coin. The other is that yet, as we discussed in *Chapter 4.1.3*, only raw data and finished papers are open, which in the majority of cases do not include reduced data and the code that lead to them. An interviewee mentions that the unopenness of such data reduction procedures, feels

> PhD Candidate: "Yeah, it does feel like it's a bit cheated or like they are trying to hide something or … Yeah, I think it does make me more sceptical of their results they are getting."

We described earlier that mistakes, such as code bugs often stay undiscovered. If that is the case, no real prediction can be made yet, how often intentionally faked data stays undetected.

Hence, to what extent it is really that easy to spot fraud, will have to be a topic of further investigation. For now we have observed that competition and the pressure to publish in Astronomy rather leads to premature publications rather than fraud.

## *Chapter 5: The Evaluation Gap in Astronomy*

We have conducted a qualitative investigation on how astronomers' perception of their own values and their motivation to perform science relate to the incentives given by the evaluation system and how their science is actually assessed. We have observed that astronomers are generally intrinsically realists, driven by a curiosity for the "truth". Their main motivation is to "push knowledge forward" that approximates the truth, confirming realists' implications described in Sismondo (2004; p.6): "*According to realists' intuitions, there is no way to understand the overall increase in predictive power of science, and the technical ability that flows from that predictive power, except in terms of increase of truth. That is, science can do more when its theories are better approximations of the truth, and when it has more approximately true theories. [...] When it discards theories, it does so in favour of theories that better approach the truth.*" Hence, it becomes clear that astronomers start off their career holding an "ideal view of science" (Sismondo, 2004; p.7) and even as they climb the tenure track ladder, they struggle with the balance act of compromising their ideals for surviving in "the system".

Interestingly, **two opposite notions of science** emerge from almost every interview. One notion is the "ideal" image of science, where astronomers are driven by their curiosity and the search for truth, which is only limited by epistemic restrictions (*Chapter 3*) such as technical possibilities. The other notion is the image of "the system" of science, comprising the funding, publication and evaluation system, and whose values are not in line with the astronomer's intrinsic values. Interviewees put those two notions in direct opposition. Faculty Member 1, for example talks about "a system that is imposed" versus "dealing with the truth", and taking action "to get funding" versus for "the progress of science". Faculty Member 1 emphasizes that "science is curiosity-driven" and "you want to know the real answer", while at the same time committing to the "cycle of pressure". Other examples are:

> Postdoc 1: "It's a system problem I think. Erm, I try to do *quality research*, *but* I do feel sometimes that I end up publishing because I *have to publish*."

> Postdoc 1: "It's difficult, because then it's the same problem, right? *You end up doing papers just to publish or you actually wanna do research?* […] Because I don't think – it's almost impossible to have, during your PhD 3 amazing papers. It's almost not even doable."

> Postdoc 1: "I wish we could just focus on more like quality papers *instead of* quantity papers."

> Postdoc 2: "Hmmm … Well, I discovered that there was a bit more of politics and some boring aspects that I didn't consider existed, but that have be there in order for the job to work."

> Postdoc 2: "I guess some people had an idealist view of the job – we do research all day and then everybody is happy. It's not always the case."

The reason for the discrepancy between the "ideal" and the "system" notions of science is that scientific discovery is naturally unpredictable, which makes it difficult to determine what "being a successful scientist" actually means. In general, effort cannot be monitored efficiently, so assessment cannot be based upon it. Therefore, a scientist is rewarded and funded for quantitative achievement instead (Rosenberg & Nelson, 1994). Quantitative measurements of achievement, so-called performance indicators include bibliometric data, such as publication & citation rates, H-indexes, impact factors etcetera. Quantitative targets supersede scientific quality as defined by the astronomer, for which we found three main criteria:

> **Quality-Criterion 1:** Asking an important question for the sake of understanding better and to push knowledge forward.
>
> **Quality-Criterion 2:** Clear, verifiable and sound methodology.
>
> **Quality-Criterion 3:** Clear communication of the results in order for the community to make use of them.

As we have observed from the interviews, these quality criteria often stand in opposition to what "the system" requires, summarised by Faculty Member 3:

> "It has to be moral in a sense, it has to be reproducible, it has to be honest, it has to be straight, it must be free from cheating. Even free from pretending. Erm, as long as the research has those qualities, erm it is quality research. Now … what people mean mostly when they are talking about high quality research – they do not mean these qualities. They rather mean successful research. That's a different story. Because success depends on people's opinions. Success depends on *whether you are at the right place at the right time*. Erm, and many other things. Erm, success sometimes depends on whether something is applicable in a practical sense, you know *if you can make a product out of it*. Erm, and so, that's really different. Sometimes people use the term "*high-impact*", okay? Erm, that gets a little closer to the daily scientific practice. But still, [quality] is not something that you can really define."

The fact that the two notions of science were mentioned in contrast to each other because of the way the question was phrased (e.g. "How does the funding system encourage scientific quality as you defined it"), does not undermine the importance of the opposition; if there was no opposition, the astronomer could have replied that their values are in line with the system's. However, the fact, that those values are often either in opposition or at least not in line, gives rise to the evaluation gap. We studied the ***effects of this gap on research behaviour and knowledge production in Astronomy***.

We observed that astronomers adapt their ***research behaviour*** in order to "hit" the required targets. This is what we call ***"strategies" or "gaming"*** in this report. Astronomers explain that at the beginning of their career "they were completely oblivious to the idea that I was gonna have to" plan out the career ladder (Faculty Member 2):

> Faculty Member 4: "And I got the impression I was the worst person in my year for having no career ladder. I was just like I wanted to do a PhD, because it was interesting and it was Astronomy and I wanted to go to telescopes and I got to build things and I got to plug them in and I enjoyed it for the moment instead of looking beyond."

> Faculty Member 2: "Well, I guess, when you are applying for Postdocs, that's the ultimate assessment, right? […] 'Will you hire me – yes or no'? That's a pretty harsh assessment. […] So there – it's really the first time I felt I had to write down what I have done, what my plans were and give people a chance to judge me. […] Ah, okay, so I was completely … oblivious to the idea that I was gonna have to do this. […] Well, you have no choice, right, at some point? If you want a Postdoc, then you have to apply …"

As astronomers climb the career ladder they realise more and more "how the system works" and what "boxes to tick" in order to "demonstrate their ability" (Faculty Member 4). Master Student 1 describes that he observed senior researchers having a "paper filter", by which he means that they recognise publishable topics and develop strategies giving priority to the question "How can we work towards publishing?" instead of quality. We have discussed strategies regarding a range of targets – from the choice of topics and research priorities, to the timescale of publishing, boosting one's impact, ensuring recognition and considerations about competing versus collaborating all the way to "resorting to stupid tricks" [Postdoc 2]. It is interesting to note, that even with respect to strategies we observe astronomers putting those in opposition to their "ideal" notion of science:

> Faculty Member 1: "And ahhh … of course it's also *very easy to play the system*, right? Because there is … these days … large collaborations. […] and there is a reason for it, collaborations are getting larger, because projects are getting larger, more expensive and more complex. *It's not just people playing the system, it is also a real thing.*"

> Postdoc 2: "And I don't know … I think also if there was less pressure … Financial pressure to conduct research, people *would not have to resort to stupid tricks*. And *trying to make themselves appear more … high quality researchers than they are* by for example *publish too many papers or publishing wrong things or hasty or too quickly* without taking too much care. I think indeed, the lack of funding is … is *hurting the research quality*. Not really in the sense that we don't have enough money to do all the research that we want, but it's affecting the research that is being done, by *sacrificing quality for efficiency*."

We will come back to the effects of the application those strategies in *Chapter 6*. However, we would like to first explore one more observation regarding gaming: having to adapt one's research behaviour and making use of certain strategies can lead to psychological struggles, and. A third notion of science arises out of the balancing act between the two notions already discussed.

Early career interviewees in particular struggle with the balancing act of performing high quality research according to their standards and fulfilling the requirements of the system:

As discussed in *Chapter 4.1*, publication pressure leads to struggles directly, but also indirectly as pressure leads to a focus on quantity, which is not in line the astronomer's quality criteria. For some researchers, this discrepancy is unacceptable. As a consequence, they wish to leave academia:

> Master Student 2: "Yeah [I don't want to stay in academia], partly because there is this 'publish or perish' thing, where it seems to be like 'pump it out'."

This image arises when astronomers decide to accept "the system" as a "fact of life" (Waaijer et al., 2017) and to "deal with it". This notion can be described as a synthesis or mix between the other two notions – the "ideal" and the "system". When astronomers master the balancing act between staying true to their own value's while at the same time fulfilling the quantitative requirements, when they are being practical with respect to their work, they find a middle ground where psychological struggles are minimised as the astronomer accepts "the system" and adapts to it. We can observe this in interviews, where especially tenured astronomers describe how they practically "deal with the system" in terms of funding, telescope times and publishing. They emphasise how their science is "observation-driven" and explain how artificial deadlines are "natural" deadlines to them. As a result of practically working according to this third notion, tenured astronomers feel that their work is generally in line with their criteria of quality. While having to "adapt to the system" which they do not like (Postdoc 1), early career researchers also declare that they wouldn't personally compromise on quality too much, because research quality "is more important than ultimately my career" (Postdoc 2).

Because we have observed that in practice research quality is harmed in many respects, either the amount of astronomers who manage the balancing act without sacrificing research quality is extremely low, or there is a fine line between working according to the third notion of science and a **bouncing between the "ideal" and the "system" notion** of science, where quality is sacrificed at least occasionally and justified by having to survive in the system. We find contradicting information from interviewees that could be evidence for the latter case. While for example, almost all interviewees – even those tenured astronomers who feel that their research is in line with their notions of quality – acknowledge problems of the "publish-or-perish-system", Master Student 2 observes that "different people talk about the publish-or-perish thing and how it hurts. And then other people seem not to have much of an issue with it." On the one hand, astronomers know they need to play along with the system. On the other hand they know what really matters (Faculty Member 2).

> Annual report[2014]: "With 16 PhD theses and 318 refereed papers, the scientific 'production' was fantastic. However, in 2050 it will not be those kinds of facts that count, it will be the true discoveries that have stood the [test] of time that will be remembered."

Whether astronomers manage the balancing act and work according to this third notion of science, or they flip between the two other notions, the majority of astronomers seem to indeed accept the pressure to publish as a "fact of life". Waaijer et al. (2017) find that being able to cope with the system enhances the early career researcher's sense of autonomy and independence. That is probably why so many early career interviewees state that they "try to stay in academia for long as possible" (PhD Candidate) and why pressures can even be party self-enforced (Waaijer et al., 2017):

> Postdoc 2: "If at all possible, yes, I would like to continue in academia. And in a way this rule I have – 1 paper per year – is the standard I have posed on myself in order to have a good chance to continue."

On the one hand, this would be consistent with Waaijer et al. (2017) who claim that, while many PhDs (from different fields) state that publication and grant pressure is too high and had made them hesitant to choose a career in academia, it has not been a decisive factor in their actual job choices. On the other hand, early career interviewees are aware of the fact they might have to leave academia and are working on accepting that. Thus, to what extent a third notion of science, can be held by astronomers in practice, and early career researchers in particular, is subject to future investigation. This notion would basically be another ideal – the "ideal way" to deal with the system. It would be interesting to see whether or not such a third notion implies a bias towards the positive aspects of "the system" in order to guarantee one's survival on the career ladder, which would give justification for sacrificing scientific quality.

In any case, we observed consequences of the evaluation gap. Because grants are needed to secure a subsequent job contract, publication pressure, grant pressure and career prospects form a complex network of factors that influence the scientific environment and foster competition that is often quite fierce. Due to this, the system's effects are ***constitutive***. We dedicate our final chapter to these constitutive effects on research behaviour and knowledge production in Astronomy.

## *Chapter 6: Constitutive Effects*

We have found that performance indicators in research in Astronomy do not reflect the astronomer's definition of research quality. That gives rise to an *evaluation gap*, which we found have consequences (i.e. *formative effects*) on research behaviour and knowledge production. Because those consequences are formative, they are *constitutive in their effects*. The indicator then stands in "a constitutive relation to the reality it seeks to describe" (Dahler-Larsen, 2014). Meaning is being constructed (e.g. citation rates equals impact) and practices are being established (pushing publications). Dahler-Larsen (2014) identified five prominent constitutive effects that indicators can have on science and researchers. We will introduce them and elaborate on their validity and occurrence in research in Astronomy.

### *6.1: Indicators define interpretive frames and world views*

When scientific performance is measured by bibliometrics of the output "a view is enhanced where scientific activity is similar to industrial production. Scientific inquiry, reading, collecting data, serving as an editor or reviewer, giving advice or engaging in debate are not counted" (Dahler-Larsen, 2014). In that situation "indicators become the way through which the world is defined."

Constitutive effects of indicators include **political effects**, in "how they define a strategic landscape in which practitioners must navigate" (Dahler-Larsen, 2014). We have observed this happening to a large extent. As outlined in *Chapter 1*, the Sterrewacht developed numerous *strategies* and defined *missions* to "maintain their success". They defined the landscape in which individual astronomers must navigate, even though they may not agree that these strategies necessarily lead to higher scientific quality:

> Faculty Member 2: "The students are also heard, but again, it's like an external visiting committee that evaluates the program, the way of examining, the processes, making sure that conflicts of interests are properly dealt with and this kind of thing. So much of that is also very formal and process oriented. Some of it, and to my taste not enough, is about the content. […] They just want to know spreadsheets and things like this. They don't care so much about the quality of the teaching. […] The answer to every possible issue is another layer of bureaucracy or another form or another document or another registration."

> Faculty Member 2: "The government here … every few years they come up with something new, right? So every few years they invent something new and then we all have to jump through new loops and then we can all collectively give the feedback that that is a crazy idea. So we are allowed to give the feedback, but it doesn't help."

The latter quote illustrates what an immediate impact government decisions on funding can have. If indicators define the **landscape of success**, they also define its inverse; the **landscape of failure**, which is also what we observed in *Chapter 4.4*. Individual astronomers have a hard time defining "failed research", due to the very risky nature of research. They were only confident to describe what bad research is – the opposite of good quality research in their definition. The community and the system do have a definition of failed research however, which is the opposite of successful research as measured by indicators. The constitutive effect of this indicator then is a shift in what counts as new discoveries in the community, from "anything new" to "publishable results". Hence, as long as negative results can't be put in a context (i.e. "tailoring"; Postdoc 1) where they publishable, they are regarded as worthless: the research project failed and the researcher feels like a failure. This is the point where we can see that not only do indicators have constitutive effects on world views on what is success and failure, but also on identities.

*6.2: Indicators define social relations and identities*

"People know who they are through the figures that describe them. Statistics help define populations and collective identities" (Porter, 1994).

The afore-mentioned strategic landscape defined by the indicators is "political, as it consists of policy-related categories by means of which [researchers] seek to manage themselves and thereby represent themselves and their values" (Vestman and Conner, 2006 & Rosanvallon, 2009). We have observed this happening with the institute. The Sterrewacht defines itself, its success and its researchers as excellent on the basis of the number of grants received, the number of publications produced and the scores on categories received by the EB. In this section, we will elaborate to what extent we observed this with individual astronomers.

Astronomers need to survive in a system where "filters" are necessary due to a limited amount of financial resources and tenure positions, and where they need recognition to establish themselves in the community. We have observed that this turns the climb of the career ladder into a highly competitive "rat race" and "postdoc circus" (Faculty Member 2 & Faculty Member 1). The Matthew effect of the funding system reinforces the Matthew effect of the career system: "Funding agencies are much more likely to support scientists who have established track records of successful research. The *cycle of credibility* (Latour & Woolgar, 1986) is the cycle that allows scientists to build careers and continue doing research. Continuing research is *central to the identity* of most scientists." (Sismondo, 2004; p.112)

As we discovered in *Chapter 4.1*, performance indicators based on output define the ***value of an astronomer.***

> Master Student 1: "But as a PhD candidate [or at a later career stage] you are being judged for the results of your research. If don't have a success, then you are really judged by that and that affects your possible career options."

Hence, if research is central to the astronomer's identity, and what is seen as 'good research' and 'a good researcher' is dependent on indicators, then those indicators have constitutive effects on the identity of an astronomer. "Publish-or-perish" can become quite literal for an astronomer's identity. The extent of those constitutive effects on the identity and values of an astronomer is the subject of this section. They can be disguised in several in several ways and we will concentrate on two of them: psychological and motivational effects.

First, in *Chapter 4.1* we observed that pressure to focus on quantitative output has ***psychological effects***. These include not only demotivation and discouragement, but also feelings of unworthiness; an 'attack' on one's identity. While the university promised to provide a "vibrant scientific atmosphere that allows young people to develop and grow" (*Chapter 1*), Master Student 2 describes what that actually entails in practice:

> Master Student 2: "From looking at people who are doing PhDs, erm, you know there is still, they are in on weekends, they are doing more than 8 hour days, they are doing more than 40 hour weeks. You know, they don't take the full amount of holidays allocated to them, which I didn't realise. […] As much as I have been told, that this university really encourages you to have a life outside your PhD, I see very few examples of that. And the examples of that, that I see, are people who […] basically don't let themselves be *bullied by their supervisor* into feeling that they have to do all of this additional work. Some people are happy with this, but I don't want my entire life to be one thing …because it causes me too much stress for my entire life to be academia. […] I think I figured out that it would be *constantly proving that I was good enough*. Constantly proving that I was worth the money, constantly proving that, you know, I was worth the time and the energy and all of that and that sounded exhausting before I even started it. And sounded like I would *constantly be battling with feeling I am not good enough*, while trying to tell other people that I was good enough. And I kind of went 'No' – and I am not – I know, there is gonna be an element of that in jobs as well, but I feel a little bit, in a job, at least there should be a break, like this is 9-to-5 or whatever. And then I can go home and I can leave it there. Whereas with academia, it's kind of like, yeah you can go home, but then you are getting emails, until maybe 20:00 or 21:00 in the evenings and still doing things."

Second, we find constitutive effects on the ***motivation of an astronomer***, which may affect their identity. When the value of an astronomer, which is defined by indicators, determines their chances of receiving future funding and hence ultimately shapes their career, output becomes a main driving factor. This changes the astronomer's intrinsic motivation of "truth-finding" to an extrinsic motivation of "result-generation". This in turn has constitutive effects on research agendas, content and time frames as we will describe in the sections to come.

As we saw above, Waaijer et al. (2017) find that being able to cope (by acting out of extrinsic motivation) with the system enhances researchers' sense of autonomy and independence, especially for early career researchers. Therefore, effects on motivation are also constitutive with respect to identity.

It becomes clear that *"[…] science consists of interacting social worlds. […] Participants are invested in their worlds, and attempt to ensure those worlds' continuance and autonomy. They strive to continue their own world and maintain identities"* (Sismondo, 2004; p.153). We have seen that, if an astronomer wants to maintain their identity as a researcher, they must not perish, but publish and hence adopt extrinsic motivation to survive in the system. Which raises the question: do these constitutive effects on motivation and identity reach as far as to also (re-)shape the astronomer's values? That question is of high importance as in this case, constitutive effects would gradually adapt astronomer's motivation and values until finally the evaluation gap vanishes and the two notions of science, the ideal and the system image (*Chapter 5*), become more aligned.

Intrinsic motivation is based on astronomer's intrinsic values, which we described in *Chapter 2.2*. Similarly, their extrinsic motivation is based on the values of the system, as defined by indicators, and the kind of rewards and incentives promoted as a consequence. The interviewees express an opposition between "things that matter [in the system] like publication rates", and a "different motivation" an astronomer would have without the need of publication rates (e.g. Faculty Member 1). The astronomer would still publish, but "to progress science". We have also seen that young astronomers do adapt their actions to survive in the system (extrinsic motivation), and yet they claim they wouldn't sacrifice their values (e.g. scientific quality for quantity). Hence it seems that, while "indicators become the way through which the [scientific] world is defined", the realist character of astronomers remains stronger than the quantitative character of the system in respect to what an individual astronomer values.

> Postdoc 2: "Because to me the quality of my research is more important than ultimately my career."

This shows that, the astronomer's intrinsic values are generally stronger and they try to maintain them. This is despite the fact that research is central to their identity, and astronomers adapt their motivation to maintain this identity. We observe that the initial values (intrinsic) and motivations of an astronomer remain as their ideals, while in their daily research life they need to adopt different motivations in order to survive the system (extrinsic). Hence, while indicators give an extrinsic motivation to an astronomer, the constitutive effects on identity do not reach as far as to alter the astronomer's intrinsic values to a noteworthy extent. As a consequence the evaluation gap remains. This conclusion is illustrated in *Figure 1*.

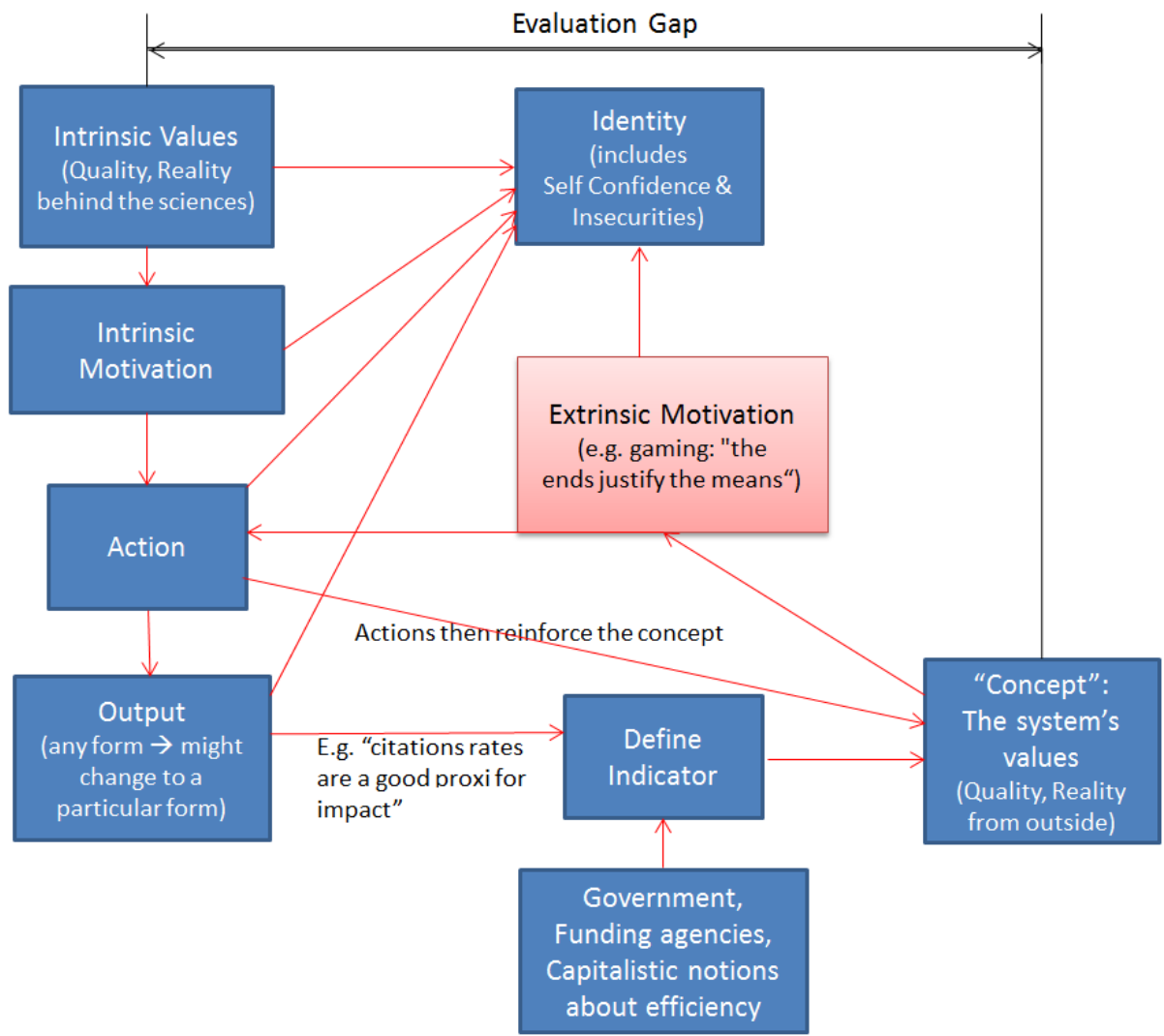

**Figure 1: This figure illustrates the evaluation gap and constitutive effects on the motivational cycle behind research in Astronomy. The cycle starts with the astronomer's intrinsic values and shows what constitutive effects (red arrows) each element has. Because the system does not have constitutive effects on the astronomer's intrinsic values, it also does not influence their intrinsic motivation. Hence, there is not only an evaluation gap between what an astronomer values and what is actually measured, but that gap also results into a gap between intrinsic and extrinsic motivation. This has constitutive effects on the astronomer's identity in form of psychological effects, for example feelings of unworthiness.**

### *6.3: Indicators define content*

As *Chapter 2.2* explains that astronomers have an ideal of what scientific quality *is*, but acknowledge that it's less clear how to *measure* it, because quality is more a matter of intuition. In the previous section we confirmed that "through indicators, practices which are otherwise more intuitive and implicit are exposed to explicit criteria of success" (Munro, 2004). Dahler-Larsen (2014) explains: "Indicators define what is central in work. They help determine what actors should strive to accomplish in a given activity. [When indicators are based on output] some observers predict that researchers will slice their work up in its smallest publishable parts and focus on research questions that are not risky." We have observed the constitutive effects of indicators on the (quality of) content of research in Astronomy and will discuss them in this section.

The constitutive effects of indicator use on research content in Astronomy start with the ***effects on research agendas***. These effects range from task prioritisation, to the choice of research topics and national research agendas.

As outlined in *Chapter 4.1*, indicators influence the decision of which tasks are to be prioritised according to the imperative of "What do I need to do to demonstrate that ability?" (Faculty Member 4). The choice of research topics depends on whether it is a viable decision in terms of [an astronomer's] career" (Postdoc 2; *Chapter 4.2).*

> Master Student 2: "I have heard a lot of people give different opinions on [publish-or-perish] and give out about this kind of thing, because in their opinion it leads to substandard work. Erm, it's not something I have dealt with very often – I think it leads to a very high pressure environment. And it kind of, you know, where research is kind of for research's sake, it feels like it's kind of 'Pick this, because you can get more papers out of this.' or 'If you do something like this you will *definitely* get a result that will be publishable and then you can ...'. So it feels like you end up narrowing down what can actually be researched until you get to such a point, where you have enough of a name, where you can then go like 'Ohhh, I could go into this area ...'"

Research topics have to promise success (in the landscape as defined by the indicator; *Chapter 6.1*), and are ideally "sold as ***sexy***" in that they fit the current "fashion of the day" and national research agendas (i.e. "what the government wants to hear", Postdoc 1). Sismondo (2004; p.112) explains this phenomenon and its (constitutive) effects:

> "To gain funding scientists typically write grant applications to public and private funding agencies. The success of grant applications depends upon evaluations of the likelihood of concrete results coming out of the research project. That in turn depends upon issues of credibility and direct evaluations of the do-ability of the project itself (Fujimura, 1988). Scientists have a large stake in finding doable problems, both to gain immediate funding and to build up their base of credibility for the future. A standardized package is attractive because it helps to routinize research, research that is intrinsically uncertain. […] One of the effects of this ***standardization*** is that members of a field can see each other's work as meeting standards of objectivity."

Quantitative measures are meant to be "more objective" than qualitative ones. By design, they make researchers and research projects easier to compare to each other. At the same time however, Sismondo (2004) emphasises, that standardization and standardized measurement is not necessarily suitable for research: "There are always cases that, because of crucial ambiguity, instability, or novelty, do not fit neatly into the categories set out by formal rules. Experts may be able to recognize these cases and respond accordingly. Therefore, formal objectivity is not always desirable. This is especially true, given that real research operates at the edges of knowledge, rather than merely reiterating knowledge – by definition research is supposed to create novelty, and not merely maintain cultures" (Sismondo, 2004; p.118).

Because funding rather mostly "targets guaranteed, short-term goals" (Loeb, 2015), and because an astronomer wants to avoid failure in terms of non-publishable research results (as discussed in *Chapter 6.1*), the consequence is **risk aversion**. As we elaborated extensively in *Chapter 4.4*, this "damages" knowledge production in that it leaves little space for innovative and novel ideas, which is what would take astronomers "out of their psychological comfort zone, but closer to the truth" (Loeb, 2013).

> Faculty Member 3: "These young folks are scared! They are afraid! […] And you know what, that is *ultimately bad*. This is ultimately bad, because in such a science where you know, ignorance is so big, being scared is not the right thing to be. […] You get results by your brains, your hands, by the collaboration with your colleagues and stuff, but you have to have a sort of *courage*. And it is *bread out of the young people*. Because they are *not rewarded* for their courage. And I find that very, very, very bothersome. That generation – people growing up like that. *How are you ever, ever, ever going to understand the universe if you don't have courage*?"

Creativity, which is so valuable to the research process, is discouraged. Hence, since the use of indicators leads to "playing it safe" and favouring sexy topics, it has constitutive effects on the content of research in Astronomy. These effects are mostly damaging research quality from the astronomer's point of view. As such, those effects go against the Sterrewacht's objective of "[…] fostering an intellectually rich and vibrant scientific atmosphere which allows astronomers to pursue their ideas and push scientific boundaries […]" (*Chapter 1*).

High-risk, high-gain projects promise **big impact and a boost in reputation**, which are needed to establish one's career, however. The most prominent indicator to measure impact and reputation is **citation rates**. While there have been several studies indicating that citation rates are not representative measures for impact on the community (Sismondo, 2004; p.35), citation rates provide "a source of emotional energy" (Sismondo, 2004; p.36), as a reward. One of the astronomer's values is to push knowledge forward in the sense of making contributions to the field. Thus, as the career system values citation rates for their alleged measurement of influence, the astronomer feels encouraged in their drive to discover. However, when this reward is targeted out of the extrinsic motivation needed to climb up the career ladder, citation rates as indicator have constitutive effects on the content of scientific research, and so does the need for impact in general. We found that strategies aimed at increasing one's impact can involve considerations about targeting journals with high impact factors and including **prestigious names as co-authors**. **Sexy topics** also promise a higher impact and **safe projects** are chosen over the potential to fail.

The need for impact and recognition is only one aspect of the wider need for output in general, as their indicators are based on output. **Output indicators** involve all sorts of bibliometric indicators: publication rates, citation rates, impact factors, etc. Apart from the ones already described in the previous paragraph, we have observed other constitutive effects of output indicators on research content in Astronomy. These include **cutting up research** *results ("salami slicing"), premature publishing and non-replicable papers*. In most cases these effects are described as having a negative effect on research and output quality (*Chapter 4.1*). There is no incentive (yet) for astronomers to publish their data reduction code. ***"Hiding information"*** even gives a competitive advantage in the race of priority, which results in numerous code bugs and mistakes staying undiscovered. ***Information overload*** goes hand in hand with increased, questionable quality content due to strategies to increase output. Interviewees describe that they "publish more than they would if they didn't have to" and that "referees have no time for careful check of the massive load of papers." The worst effect an indicator can have is when it drives the astronomer into committing fraud. As fraud is the opposite of integrity, fraudulent research is the opposite of qualitative research.

These constitutive effects of output indicators on content in Astronomy stem from one element that they all have in common: ***publication pressure***. Nevertheless, we observed that publication pressure can also have positive constitutive effects on research content and quality. It can be a "healthy" driving factor and support the astronomer to focus and confine the otherwise "massive open-endedness" of research questions.

We have observed something similar for ***competition*** (*Chapter 4.3*): despite the many negative effects that the "rat race" has on research quality, competition with other astronomers can be "healthy". When it is seen as a stimulator to push knowledge forward and when competing groups are especially critical towards each other's outputs, competition can add to higher quality content as results are likely to be more robust. However, an astronomer may also choose to ***collaborate (with the competitor)***, which is often a strategic move with respect to indicators and incentives. We found that collaborations are effectively a *trade market* for resources and credit. The need for (prestigious) observation time or resources in terms of money or human skills may lead to collaboration that are mutually beneficial. Collaborations benefit research quality when they are driven by curiosity, when they spark scientific discussions, when they speed up the research process and when they add to the robustness of results and the quality of papers. Collaborations may distribute resources efficiently in a way that supports these benefits for pushing knowledge forward.

In summary, one needs to be output oriented to survive in the assessment system, but collaboration might benefit that, as co-authorship brings credit without "doing all the job all the time". Hence, insofar as the system, its indicators and its incentives "push" astronomers into collaborations to avoid competition, they have a positive constitutive effect on research quality.

There is yet another constitutive effect on research content stemming from the evaluation system. As we saw in *Chapter 2.2*, astronomers see a difference between "making progress on an important issue" and "innovation". From the astronomer's point of view, pushing knowledge forward is the goal of research and what serves society (e.g. Faculty Member 3). If that knowledge leads to applications outside science, that is a bonus for the astronomer, but it is not the primary intention. The astronomers criticise the increasing "tendency to look at the value of science in terms of economic output, it's called 'valorisation'" (Faculty Member 1), because the outcome of research cannot, by its nature, be planned for.

> Faculty Member 3: "They think that they can direct science. They think – 'they' is these politicians, okay? – *They think they can order discoveries like you order a pizza. You. Cannot. Order. A. Discovery.* […] You have to work on it, you have to try things, you have to experiment, you have to stick your neck out. […] But since science is funded, mostly, by public funding we are, you know, dependent on the strange conceptions that politicians have on how science works. Science doesn't work just simply by ordering as discovery. It has never been like that and will never be like that – that's the essence of research. You don't know."

Nevertheless, while the EB compliments that Sterrewacht's "valorisation activities in collaboration with industry and subsequent knowledge transfer were found to be outstanding", the EB points out that those efforts are opportunity-driven, rather than derived from a pre-determined strategy" (*Chapter 1*). Because valorisation is an important aspect in the SEP evaluation procedure[5] the committee recommends "that NOVA continues to review its scientific strategy on a regular basis, in order to be able to exploit rapidly emerging scientific opportunities." Hence, the EB encourages a ***strategy for valorisation***. This, in the astronomer's perception, would alter the research content towards an economic output. The intention of an economic output is to sell, while the intention of an astronomer is to find out truths about the universe. Hence, as far as astronomers adapt to any valorisation strategies to come, the evaluation system determines not only the value of, but also research content itself, and is thereby constitutive. In *Chapter 1* we nevertheless encountered an emphasis on a wide range of spin-offs and the applicability of skills acquired through studying Astronomy to jobs in industry. However, astronomers see these contributions to society as a bonus, which helps to secure funding, but not the primary reason behind Astronomy.

In summary, we have seen that the prevailing science evaluation system and its indicators have constitutive effects on research agendas and content in Astronomy. From the point of view of individual astronomers these are mostly negative with regard to research quality. At the same time, self-assessments and the EB report the high-profile character of the institute, measured by those same indicators which the interviewees mostly criticise. This discrepancy in viewpoints between individual astronomers and "the system" is further proof of the existence of the evaluation gap (*Chapter 5*). It arises from the two opposing notions of science: the ideal ("ask an important question and push knowledge forward") and the system ("it has to promise output and needs to be the flavour of the day") image. Interviewees reflect upon the evaluation gap by calling the effects of indicator use "adverse" instead of "constitutive":

> PhD Candidate: "You know, you always have some adverse effects [of measuring quality], that … It's not healthy. […] Yeah, sometimes it's even more an intuitive thing; it's really hard to define how to [measure quality]…"

Some constitutive effects on research quality are direct. For example, strategizing about one's career "takes up a lot of brain space" (Faculty Member 4), which is taken away from accomplishing qualitative research. However, most constitutive effects on research quality are indirect, enabled by gaming strategies followed a publication pressure, which is a constitutive effect itself.

---

[5] More information on SEP's interpretation of valorization: https://www.rathenau.nl/nl/files/definitions-and-policypdf

*6.4: Indicators define time frames*

"Some academics fear that the increased focus on counting publications on an ongoing basis leads to increased pressure to publish quickly. Time is not innocent, however. It is feared that time pressure inhibits deep, innovative and risky research projects. On the other hand, advocates of the bibliometrical system argue that researchers, who used to keep lengthy publications in their drawers for too long, will actually benefit from a bit of time pressure" (Dahler-Larsen, 2014).

This is exactly what we observed. Publication pressure is in itself a constitutive effect of indicator use. In the previous section, we described that, in turn, publication pressure has constitutive effects on research content. In this section, we will elaborate more on which factors cause publication pressure, and define research time frames more generally. ***Publication pressure*** is always "at the back" of an astronomer's mind (Postdoc 1), given that publication rates set the base for of an astronomer's value (*Chapter 4.1*) and their future career (*Chapter 4.2*). This pressure may increase when one or more of the following factors are present:

- First, when the astronomer faces head-on competition, there is a ***race for priority***, which pushes the researcher to "publish as fast as I can … as soon as I get the data" (Postdoc 1).
- Second, we observed that timescales of projects and publishing are "tied to the ***timescale of [PhD] students and postdocs***" (Faculty Member 2), because "they need to get their thesis chapters out. They need to be ready for the job application season".
- Third, ***telescope application deadlines*** are perceived as deadlines for publications, as performance indicators are part of the assessment criteria for observation time allocation.

As it becomes apparent when analysing these factors, defining time frames ultimately means setting restrictions. Thus, a new constitutive effect of indicator use arises as ***man-made deadlines become "natural" deadlines***. In the astronomer's daily research life, these deadlines are indistinguishable from epistemic restrictions (*see Chapter 3*) an astronomer faces.

In addition, interviewees emphasise the ***lack of incentives for reviewers***, which means that the system provides no time frame for the reviewing process. As a consequence, papers are not reviewed properly, which is damaging to the research content. This also means that the process up until an astronomer can make their final submission takes up to half a year, which in turn increases the pressure on the astronomer.

> Postdoc 2: "[…] typically I had to wait half a year, between when I submit a paper and the moment it gets accepted. And … in an ideal world that wouldn't matter, but the thing is that you have to apply for a job, you have to show that you have enough published papers. And sometimes you need the papers to come out quickly."

We once again witness the distinction between an "ideal world" and reality as shaped by indicators. Interviewees oppose the ideal notion of science where research is driven by curiosity and only limited by epistemic restrictions to the reality of "practical" restrictions which arise from "the system". We find that practical restrictions arise as a constitutive effect from the evaluation system and manifest themselves as publication pressure (timely restriction). This in turn leads to restrictions on agendas, topics, riskiness, publishable results and collaborations. These restrictions on research content comprise constitutive and mostly damaging effects on research quality as outlined in the previous section.

### *6.5: Indicators change their meaning as a result of their use as indicators*

"Official indicators are themselves social constructions reflecting the variety of practices and meanings accorded to them by different groups of participants in different social contexts" (Vulliamy & Webb, 2001).

According to Dahler-Larsen (2014), an "indicator superimposes a mindset of productivity and effectiveness" upon research behaviour and knowledge production. As ***Goodhart's law*** states, "when a measure becomes a target, it ceases to be a good measure". The mere use of an indicator puts pressure on its own integrity. A simple illustrative example are citation rates: while initially adopted as an indicator for impact, once they become a measurement citation rates become a target (for manipulation). The very fact that indicators have constitutive effects on themselves proves that no quantitative indicator can ever be truly representative for the qualitative account it is supposed to measure. The indicator is rendered useless with respect to its original intention, unless targeting strategies are accounted for, or unless the indicator's intention is to increase performance by developing gaming strategies so as to reach the required targets and claim the involved incentives – in a "setting the bar higher" fashion.

Even if targeting strategies are accounted for, research[6] has shown that, while productivity can indeed be increased by using incentives (***extrinsic rewards***), the actual quality of the work performed is influenced by intrinsic factors. In the case of the astronomers, these are based on their passion for truth-finding and its intellectual challenge. If they can engage those, they are more likely to have novel ideas and find creative solutions.

In *Chapter 6.2* we explained in detail how the misalignment of extrinsic and intrinsic motivation lead to the evaluation gap. The government's, funding agency's and employer's need for measurement ultimately constitutes the value of an astronomer and which aspects of their research work are valued. Those are the aspects the astronomer strives for to survive in academia (extrinsic motivation). They are different from the values an astronomer would go for if they were free from an evaluation system (intrinsic motivation). This difference constitutes the evaluation gap and the evaluation gap itself has constitutive effects. The constitutive effects of indicator use therefore have the potential to either reinforce the evaluation gap or reduce it.

---

[6] E.g. https://www.verywell.com/what-is-intrinsic-motivation-2795385

*6.5.1 Utilizing constitutive effects*

We found that in the case of Astronomy the former is the case in the current evaluation system. We have elaborated in detail on the constitutive effects of that gap and it's mostly "damaging" consequences on knowledge production in Astronomy. ***Utilizing constitutive effects to close that gap*** seems to be the obvious solution to enable best research quality while being able to perform evaluations. But how? That is the biggest question and challenge Evaluation Studies have to face. The remainder of this section is devoted to a few thoughts on this challenge.

Different indicators could be found, with constitutive effects which could be utilized to foster research quality and enable astronomers in their strive for truth-finding (Fochler & De Rijcke, 2017). One now might assume an apparent solution is to utilize those constitutive effects to align extrinsic motivation with intrinsic motivation. However, research also shows that being rewarded for something that one does out of intrinsic motivation may decrease their intrinsic motivation for it, which is called the ***"overjustification effect"***[7]. In that process, one's extrinsic motivation may take over to keep performance levels high, but the quality of one's work may suffer again. This is the reason why finding alternative evaluation criteria that match with what the professionals themselves value is not only tricky, but a nearly impossible endeavour.

Although it is unclear how feasible this would be in practice, one theoretical solution could be to have a flexible set of indicators and incentives that change in a reasonably timed manner. This would render (long-term) gaming strategies obsolete. These indicators must be close enough to the astronomers' value, so that the evaluation gap is minimised at all times. This would enable astronomers to follow their intrinsic motivation, which would keep research quality high without being too distracted, but still motivated by extrinsic motivation as set by the indicator, which would keep productivity high. That way the overjustification effect could be avoided and indicators would be flexible enough to not become a target for a longer term. This situation could be illustrated by an *asymptotic spiral* (Figure **2**): the constitutive effects of the indicator used at a point (/period) in time spiral around the ideal notion of the astronomer. The distance between those points on the spiral and the centre of the spiral would then amount to the evaluation gap. Future studies will aim to demonstrate the feasibility of such a concept.

---

[7] E.g. https://www.verywell.com/what-is-the-overjustification-effect-2795386

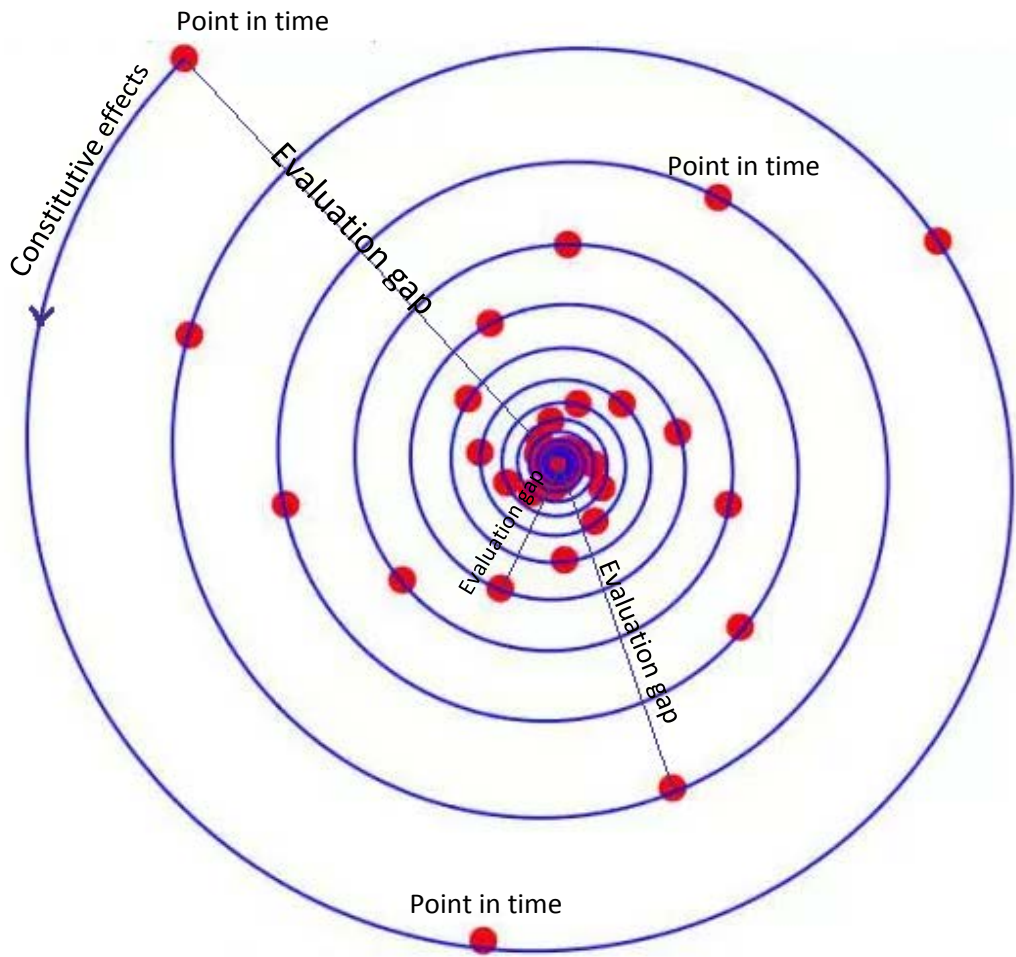

**Figure 2: Constitutive effects of indicators could be utilized such that the "system's values" asymptotically approach the "astronomer's values" (the centre). The difference between those values at a given point/ period in time represents the evaluation gap. Note that the spiral does not need to point inwards at any given point/ period in time; the evaluation gap may also grow again on the spiral when it has become sufficiently small, to then be able to minimize again.**

## *Discussion*

We have analysed 9 interviews with astronomers from Leiden Observatory and a collection of (self-) evaluation documents and annual reports from that institute and the Dutch astronomy umbrella organisation NOVA. We have elaborated on what values drive an astronomer to enter academic research, how they perceive the values of the publication, funding and evaluation system and what epistemic restrictions astronomers face. We then analysed how the astronomer's values relate to the system's values and what effect a discrepancy – the evaluation gap – has on knowledge production in Astronomy.

We found that astronomers are driven by curiosity, truth-finding and "pushing knowledge forward" (*Chapters 2.1 & 2.2*). During discussions of the interviews with CWTS's group for Science and Evaluation Studies, the question was raised of whether these values are based on a folk theory based on the public's enchanted[8] view of how science works. A folk theory is a belief based on received wisdom, rather than concrete evidence and facts. However, while especially young astronomers are likely to hold an enchanted view about science and may become disillusioned by their experience in academia (e.g. Postdoc 2), we have observed that the astronomer's intrinsic values hardly change due to this disillusion. Therefore, we conclude that the astronomer's values are based on the realist account that astronomers generally hold, rather than on a folk theory about scientific quality. Astronomers derive scientific quality from their values, and define quality as "objective" when it meets those values. Therefore, while astronomers agree that quality is difficult to measure, we have nevertheless been able to establish an "objective" account for scientific quality and found three criteria:

> **Quality-Criterion 1:** Asking an important question for the sake of understanding better and to push knowledge forward.
>
> **Quality-Criterion 2:** Clear, verifiable and sound methodology.
>
> **Quality-Criterion 3:** Clear communication of the results in order for the community to make use of them.

While astronomers agree on what quality is, they do admit that it is difficult to measure. Because resources such as funding and positions are limited, proxies for scientific quality – the quantitative indicators – help decide whom or what to fund. Those indicators include bibliometric measures such as H-indices, citation and publication rates. They include the number of grants and how much observation time an astronomer has been granted. The more prestigious the affiliations a researcher had, the better their profile and chance to climb up the career ladder.

In order to survive in the current science evaluation system, which includes the funding, publication and assessment systems as described in *Chapter 2.3*, the astronomer needs to fulfil the requirements of what is valued in "the system", as constituted by quantitative indicators. As there is a discrepancy between the astronomer's and the system's values those constitutive effects of performance indicators on what is valued as quality by the system give rise to the evaluation gap (*Chapter 5 & Figure 1*). We found that the evaluation gap in turn has a variety of constitutive effects on knowledge production (*Chapter 6*), ranging from research agendas, researcher's behaviour and identities to research content.

---

[8] Science in Transition, Position Paper 2013, http://www.scienceintransition.nl/

Interestingly, we observed that the astronomer holds two opposing notions of science: the "ideal" one which corresponds to their intrinsic values and the "system" notion. This means that, while in their daily research life an astronomer adopts an extrinsic motivation to perform science, their intrinsic values and motivations remain as their ideals. Hence, while indicators give an extrinsic motivation to an astronomer to perform, their constitutive effects do not reach as far as to affect the astronomer's intrinsic values to a noteworthy extent. As a consequence, the evaluation gap remains and a $3^{rd}$ notion of science arises: coping with the system. The astronomer always tries to manage the balancing act between their intrinsic values and the requirements of the system.

We conclude that Dahler-Larsen's (2014) finding, "the indicator stands in a constitutive relation to the reality [here: of scientific quality] it seeks to describe" does not apply to Astronomy. While indicators are indeed no "more or less valid representation of the concept [here: scientific quality] they claim to measure", we find that the astronomer's values are hardly co-constituted. However, constitutive effects of indicator use may not shape the reality of a realist, but they do shape research agendas and have epistemic implications on day-to-day research practices. Man-made deadlines become "natural" deadlines.

To summarise, we found that research in astronomy is driven by the $3^{rd}$ notion and includes: First, questioning what is the most important result to communicate to the community (*Chapter 2.1 & 2.2*); second, epistemic restrictions and data availability (*Chapter 3*); third, what brings impact (*Chapter 3*); fourth, what are the explicit deadlines, such as application deadlines and what are the less explicit deadlines, such as the "emergency" due to various reasons such as publication pressure (*Chapter 6.4*).

There is a shift of focus from high quality, robust, replicable and well-communicated research results to a massive amount of premature results on sexy topics, the conclusions of which are neither robust, nor replicable, and written in an intransparent way. Risk aversion discourages creativity in the scientific process which inhibits innovative ideas, while valorisation gains ever-growing importance.

However, because success in science or in the scientific career is not only dependent on quantitative indicators, but also on luck (e.g. Faculty Member 3; "90% luck and 10% hard work"), especially young researchers have psychological struggles with this uncertainty. Van der Weijden (2017) elaborates on this.

We conclude that the institute's goal (*Chapter 1.5*) of "*fostering an intellectually rich and vibrant scientific atmosphere which allows astronomers to pursue their ideas and push scientific boundaries, and in which young scientists can develop and grow*" is not compatible with its strategy to ensure a front-line role in Astronomy if this front line is defined by quantitative indicators. Instead, we propose (*Chapter 6.5.1*) to find alternative indicators, whose constitutive effects could be utilized such that the evaluation gap is minimized. In such a scenario, the institute's goal could be met as the astronomers could act upon their intrinsic motivation, while at the same time being extrinsically motivated to perform at a high level. Dahler-Larsen (2014) needs to be kept in mind on this mission to design policies that enable high quality research:

> "If an evaluation subsequently finds that the bibliometrical system has worked efficiently because it has helped increase the number of publications over time; then that way of understanding the impact of indicators is already based on the world view implemented by the indicator. The constitutive effect of having defined research only in terms of its outputs should be attended to rather than taken for granted. Pre- and post-tests that measure changes in a defined indicator should not be isolated from the (contested) institutionalization of world views connected to that indicator. The observation that one form of indicator influences subsequent forms of knowledge creation would suggest that political effects of indicators may be far-reaching."

## *Policy Implications*

Because political effects of indicators are far-reaching, it is our goal to continue our study and find concrete advice for policy makers in order to enable high quality research in astronomy. Keeping in mind some points that have arisen during the interviews could be valuable for future policies in that respect.

First, utilizing constitutive effects as elaborated in *Chapter 6.5.1* could be a revolutionary way of reconciling quantitative indicators with scientific quality. Such a scenario would take into account that "the [emotionally] disinterested scientist is a myth" (Sismondo, 2004; p.26) and would put incentives more in line with the intrinsic motivation of an astronomer.

Second, more career stability and prospects would enable the astronomer to concentrate better on qualitative research instead of strategies for how to survive on the career ladder. This goes along with adapting the distribution of money: "*Funding distributed on a long-term basis could enable more permanent contracts, and thus improve the career prospects of junior academics more substantially*" Waaijer et al. (2017).

> Faculty Member 2: "So a bit of continuity, and a bit of tranquillity every once in a while is useful. And it also means that [the early career researchers] wouldn't need to think immediately after they arrive 'Am I gonna get enough papers for when I apply again in 1 year or 1.5 years?'"

Third, allocating money to "risky" research that doesn't promise "positive" results or an economic output would foster creativity and novel ideas (e.g. Loeb, 2012).

> Postdoc 1: "Utopically I wish we had … I don't know, a specific amount of money, it doesn't have to be unlimited, but a specific amount, but you can just try things …"

Fourth, adapting the journal templates and hence the paper writing style to what is enabled by state-of-the-art technology instead of writing in an out-dated linear format. Papers would be of higher quality if they included more information on in-between-steps and if there was an easy method to implement code. This would decrease errors due to otherwise invisible mistakes or bugs. Interviewees wish for an interactive way of delivering visible feedback and updating out-dated information. Moreover, they emphasise that co-authors should be listed in a more refined way, which reflects who contributed what more clearly, while at the same time acknowledging all the contributors. Typeset [9] has made first efforts to accomplish this goal of making writing and reading papers easier.

Future studies could provide more information to further develop these policy implications.

## *Future studies*

In our analysis, we stumbled across several questions that require further investigation.

First, astronomers argue that cutting up research results is often due to epistemic reasons and to improve readability, and not a drive to get more papers published. We therefore propose conducting a bibliometric and full-text analysis of astronomical papers to investigate the extent to which salami slicing happens intentionally.

Second, from the interviews we observed that publication pressure and competition rather lead to publishing premature papers that fraud. Moreover, interviewees claim that fraud in Astronomy would easily be detected by peers. At the same time, they admit that because data reduction code doesn't need to be published, mistakes such as code bugs often stay undiscovered. If that is the case, no real prediction can be made yet on how often intentionally faked data stays undetected. Hence, how easy it is to spot fraud will have to be a topic of further investigation.

Third, the 3rd notion of science – the "ideal" way to deal with the system – needs to be studied more closely. Can the balancing act between the two other notions really find equilibrium or will the astronomer always struggle to cope with the system while not sacrificing their values? A possible equilibrium could be found through "embodied objectivity":

> "A more modest form of commitment involves simply making one's position transparent. […] Haraway (1988) argues that it is possible simultaneously to strive for objectivity and to recognise one's concrete place in the world. This 'embodied knowledge' can produce partial knowledge though knowledge that is in every way responsive to the real world. […] researchers can identify the perspective from which they write and act, without relinquishing their status as objective researchers." (Sismondo, 2004; p.109)

---

[9] https://www.typeset.io/

Fourth, because constitutive effects are interactive phenomena, instead of simply "adverse" effects of indicator use, an investigation of alternative evaluation practices and innovative indicators that foster scientific quality needs to be studied further.

Fifth, as this PhD pilot study only concentrates on astronomers in Dutch academia, qualitative and quantitative studies world-wide could give more insight in the diversity if evaluation practices and their constitutive effects on knowledge production and research behaviour. Surveys, field studies of research behaviour and evaluation procedures and interviews with astronomers from other countries are just a few elements which would help complete the picture of this pilot study.

## *Table of Acronyms*

EARA: European Association of Research in Astronomy

EB: Evaluation Board (Evaluation protocol$^{2010\text{-}2015}$)

ERC: EU European Research Council

ESA: European Space Agency

ESO: European Southern Observatory

NOVA: Nerderlands Onderzoekschool voor Astronomomie

NWO: Netherlands Organization for Research

SEP: Standard Evaluation Protocol

ToR: Terms of Reference

UNAWE: Universe Awareness outreach program of Leiden Observatory

VSNU: Vereniging van Samenwerkende Nederlandse Universiteiten

## *Acknowledgements*

First of all, I would like to thank the 9 interviewees for their willingness to participate in this study and for dedicating research time that is even more valuable in times of publication pressure. Next, I would like to thank my supervisor Sarah de Rijcke and CWTS's Science & Evaluation Studies research group for inputs regarding interpretation and structure of the results. Jochem Zuijderwijk's comments were especially valuable and helped me lay down the core for *Chapter 5* and subsequent interpretation. I would also like to thank Cornelis van Bochove for his constant support and trustworthy advice. A big thank you goes to Cathelijn

Waaijer and Andrew Barton for their corrections and comments. Last, but not least, I would like to thank my boyfriend, Enrique Garcia Bourne. Not only did he help me wrapping my head around constitutive effects and the evaluation gap through lengthy discussions, but he was also extremely supportive in very stressful and uncertain times.

## *References*

Abramo, G., & D'Angelo, C.A. (2011), "Evaluating research: from informed peer review to bibliometrics", *Scientometrics*, Volume 87, Issue 3, p. 499–514

Annual report[1998]- Annual report[2015]: Annual public reports, authored by the director of the institute (1998-2015), https://www.strw.leidenuniv.nl/research/annualreport.php

Benedictus, R., & Miedema, F. (2016), "Fewer Numbers, Better Science", *Nature*, Vol 538, p.453-455

Broad, W.J. (1981), "The publishing game: Getting more for less", *Science, 211(4487)*: p.1137–1139.

Dahler-Larsen, P. (2014), "Constitutive Effects of Performance Indicators: Getting beyond unintended consequences", *Public Management Review*, 16:7, p.969-986

De Rijcke, S. et al. (2016), "Evaluation practices and effects of indicator use – a literature review", *Research Evaluation*, 25(2), p.161–169

Duffy, D. (2017), http://blogs.lse.ac.uk/impactofsocialsciences/2017/08/14/rather-than-promoting-economic-value-evaluation-can-be-reclaimed-by-universities-to-combat-its-misuse-and-negative-impacts/

Evaluation protocol[2010-2015]: Evaluation Document, authored by an external committee. Includes the valuation of NOVA and its four institutes. Period 2010-2015.

Fochler, M. & De Rijcke, S. (2017), "Implicated in the Indicator Game? An Experimental Debate*", Engaging Science, Technology, and Society 3*, p.21-40

Fujimura, J.H. (1988): "The Molecular Biological Bandwagon in Cancer Research: Where Social Worlds Meet." *Social Problems 35*: p.261-83.

Huth, E.J. (1986), "Irresponsible authorship and wasteful publication". *Annals of Internal Medicine, 104(2):* p.257–259.

Kaltenbrunner, W. & De Rijcke, S. (2016), "Quantifying 'Output' for Evaluation: Administrative Knowledge Politics and Changing Epistemic Cultures in Dutch Law Facilities", *Science and the Public Policy*, p.1-10

Latour, B. & Woolgar, S. (1986): "Laboratory Life: The construction of scientific facts", 2nd edn. (first published in 1979). *Princeton, NJ: Princeton University Press*

Laudel, G. & Gläser, J. (2014), "Beyond breakthrough research: Epistemic properties of research and their consequences for research funding", *Research Policy 43*, 1204-1216.

Loeb, A. (2012), "Fostering the discoveries we can't see coming" , http://www.pbs.org/wgbh/nova/blogs/physics/2012/10/fostering-the-discoveries-we-cant-see-coming/

Loeb, A. (2013), "Thinking outside the simulation box", *Nature Physics, Vol 9*: p.384-386.

Merton, R. K. (1968), The Matthew effect in science". *Science 159:* p.56–63. Page references are to the version reprinted in Merton (1973).

Merton, R. K. (1988), "The Matthew effect in science, II: Cumulative advantage and the symbolism of intellectual property". *Isis 79*: p.607–623.

Merton, R.K. (1957), "Priorities in Scientific Discovery: A Chapter in the Sociology of Science", *American Sociological Review 22(6)*: p.635-659.

Moed, H.F. (2005), "Citation analysis in research evaluation". *Dordrecht, The*

*Netherlands: Springer*.

Munro, E. (2004), "The Impact of Audit on Social Work Practice". *British Journal of Social Work, 34*: p.1075–95, in Dahler-Larsen, P. (2014)

NOVA self-assessment[2010-2015]: Self-evaluation Document NOVA (2016), including NOVA's evaluation of its four institutes and a joint appendix. Period 2010-2015

Porter, T. M. (1994), "Making Things Quantitative". *Science in Context, 7:3*, p.389–407, in Dahler-Larsen, P. (2014)

Rosanvallon, P. (2009), "Demokratin Som Problem", *Hägersten: Tankekraft Förlag,* in Dahler-Larsen, P. (2014)

Rosenberg, N. & Nelson, R. (1994), "American Universities and technical advance in industry", *Research Policy 32*: p.323-348

Rushforth, A.D. & De Rijcke, S. (2015). "Accounting for Impact? The Journal Impact Factor and the Making of Biomedical Research in the Netherlands", *Minerva 53*, p.117-139

Sismondo, S. (2004), "An introduction to Science and Technology Studies", *Blackwell Publishing*

Sørensen, M.P. et al. (2015)," Excellence in the knowledge-based economy: from scientific to research excellence", *European Journal of Higher Education*, DOI: 10.1080/21568235.2015.1015106

Stephan, P. (2012), "How economics shapes science", *Harvard University Press 2012*

LU self-assessment[2010-2015]: Self-evaluation Document University Leiden, Leiden Observatory (2016). Period 2010-2015

Van der Weijden, I. et al. (2017), "The Mental Well-Being of Leiden University PhD Candidates", https://www.cwts.nl/download/f-x2q213.pdf

Verran, H. (2001), "Science and African Logic", Chicago, IL: University of Chicago Press

Vestman, O. K. and Conner, R. F. (2006), "The Relationship Between Evaluation and Politics" in Shaw, I.F., Greene, J. C. and Mark, M. M. (eds), "The Sage Handbook of Evaluation", *New York: Sage Publications*, p. 225–42., I. F.

Vulliamy, G. and Webb, R. (2001), "The Social Construction of School Exclusion Rates: Implications for Evaluation Methodology". *Educational Studies, 27:3*: p.357–70, in Dahler-Larsen, P. (2014)

Waaijer, C.J.F. (2016), "Quantized careers: origins and consequences of the preponderance of temporary and junior jobs in academia", *Doctoral dissertation (Centre for Science and Technology Studies, Leiden)*

Waaijer, C.J.F., Teelken, C., Wouters, P.F. et al. (2017), "Competition in Science: Links Between Publication Pressure, Grant Pressure and the Academic Job Market", *High Educ Policy*, https://doi.org/10.1057/s41307-017-0051-y

Weingart, P. (2005). "Impact of bibliometrics upon the science system:

Inadvertent consequences?", *Scientometrics 62*, Vol. 62, No. 1, p.117-131

Wouters, P. (2014), "The Citation: from Culture to Infrastructure". In: Cronin and Sugimoto (eds) "Beyond Bibliometrics: Harnessing Multidimensional Indicators of Scholarly Impact", *Cambridge, MA: MIT Press*, p.47-66

Wouters, P. (2017), "Bridging the Evaluation Gap", *Engaging Science, Technology, and Society 3*: p.108-118

Ziman, J. (2000), "Real Science: What It Is and What It Means", *Cambridge University Press 2000*

# Insights into the effects of indicators on knowledge production in Astronomy

## Pilot Project Report for PhD Project

### Julia Heuritsch
julia.heuritsch@gmail.com

## *Supplementary Material*

*Table S-1:* Interview Questions

| Topic | Research Question | Interview Question | Type Respondent (Faculty, Postdoc, PhD, Master) |
|---|---|---|---|
| **Introduction** | Background | E.g. How did you get this position, which career steps were necessary? | All |
| | Topic (How much does the choice of the research topic depend on the need to get funding? (avoiding risk taking?)) | What is the topic of your research? | All |
| | | | All |
| **Project funding** | Conditions of funding | How did you received funding for this project? | All |
| | Institutional conditions of funding | How is funding allocated in your institute in general? | All |
| **Exposure to assessments** | | What role do assessments play in your work? | All |
| | | > Do you have yearly appraisals/ R&O talks with your supervisor? Peer review for funding applications & mid-term reviews for projects? | All |
| | What role do assessments play in an astronomer's (daily) life? | Are you held accountable to the founder/ review panels on a regular basis? | All |
| **Knowledge production - Planning research** | | How do you decide on a topic for your research? | All |
| | | What do you advise PhD students when they ask about how to select a research topic? | Senior - Faculty |
| | | How do you give priority on topics if you have more than one to work on? | All |
| | What is the choice of topic dependent on (e.g. preference of supervisor/ funding/ own interest/ riskiness)? | Is the journal agreed upon before writing? So, does the choice of the research topic, methodologies and content of the paper depend on that choice? | All |
| **Knowledge production - Doing research** | | What needs to be taken into consideration for designing a Methodologies Project Design? | All |
| | | Do you feel restricted in the research process? | All |
| | What are the effects on choices about the research process? (e.g. Effect on methodologies used?) | Have you heard about "responsible research methods"? And what's your stand towards it? | All |
| | Does the evaluation system foster collaboration or lead to competition? | How is collaboration organised in your project/institute/field? | All |
| **Knowledge production - Publishing research** | Is publication pressure a result of the evaluation system? And how does it influence the publications (e.g. premature publishing/ salami slicing)? | What are the most important factors in your field for deciding on when to publish research results? | All |

| Theme | Sub-question | Question | Target group |
|---|---|---|---|
| | | What are the most important factors in your field for deciding on what to publish [sexy results etc]? | All |
| | | Do you perceive publication pressure? | All |
| | | > Have you observed that people publish before the research has reached a more matured stage? | All |
| | | > Have you observed that people cut up your research just to produce more papers of it? | All |
| | | Do you feel like you need to concentrate more on quantity than quality of your work? | All |
| | Does the evaluation system influence content? | > Would one write up results differently if it weren't for the specific requirements measured by indicators such as impact factors and citation rates? | All |
| | | What do you define as 'failed' research? | All |
| | | Have research lines you have been engaged in ever failed? | All |
| | | > If yes, what were the consequences in terms of funding, publishing etc? | All |
| | | > If no, do you sometimes worry about not delivering the expected outcome due to a threat of not receiving further funds? | All |
| | How to deal with unexpected outcomes and "failures"? | Do you report "negative results"? Can they be published? Do astronomers/ you think that they should be published? | All |
| | How does the evaluation system influence replicability? | Do astronomers try to ensure that their published data is replicable or do you feel the necessity to keep information closed off? | All |
| | | | All |
| | Field: What is quality research in the field? | What is high quality research in your field? | All |
| | Institute: What is quality research in the institute? | How is high quality research defined in your institute? | All |
| | Researcher: What is quality research for the individual researcher? | What does high quality research mean to you? | All |
| | Researcher: What are motivational factors? | What drives you in your research? | All |
| | How does the funding system relate to good science quality as defined by the astronomer? | Does the funding system encourage good science? | All |
| **What is quality in astronomy (value, quality, excellence)** | How does the publication system relate to publication quality as defined by the astronomer? | How does the publication system reflect upon quality in science? | All |
| | | (Is the quantity of publications put above quality?) | All |
| | | | All |
| | Are there wishes/ways to improve the evaluation system? | What issues do you think need to be improved to guarantee better science? | All |
| | Consciousness about the evaluation system | Do you feel that you are given the chance to question how science is performed? | All |
| | | When did you have your first encounter with the way science is performed and assessed? How did that compare with your initial motivation to become a scientist? | Senior - Faculty |
| | How did the system change over time and what did senior researchers observe? | In your experience, did the definitions of value and academic quality change over time? | Senior - Faculty |
| | | When did you have your first encounter with the way science is performed and rewarded? How did that compare with your initial motivation to become a scientist? | Junior - Faculty, Postdoc, PhD, Master |
| **Improving research evaluation & Consciousness** | Do young researchers perceive that they need to adapt to the evaluation system? | Can you pick topics and methods yourself or do you feel like you'll only be free to do that once you reached tenure? | Junior - Faculty, Postdoc, PhD, Master |

*Table S-2:* These codes represent themes which emerged by combining sensitivity towards existing literature on constitutive effects of indicator use with insights from our data. The interviews were coded using this codes and Grounded Theory.

| Code | Explanation & Related Keywords |
| --- | --- |
| CAREER Clarity/ Expectations | Has the path been clear? What is expected in terms of career steps? Tenure. |
| Politics | |
| Prestige | |
| Output orientation | Both, in terms of output = basis of assessment & what output is expected. |
| Pressure | Publication/ Funding |
| Impact | |
| Competition | |
| Collaboration | |
| Riskiness | |
| Failure | |
| Negative results | Non-detections |
| Authorship | |
| Salami slicing | |
| Quality | |
| Curiosity | "Wanting to understand" |
| Referees | |
| Matthew effect | |
| Citation rates | |
| Publication rates | |
| Funding | |
| Gaming | Strategies, Targeting, "Sales men" |
| Replicability | |
| Epistemic Subculture | Topic of research, Instrumentation/ Observational/ Theoretician |
| Sexy topics | |
| Uncertainty (research) | |
| Uncertainty (career) | |
| Integrity | Fraud, Fake, Cheat |
| Luck | |
| Indicator | |